\documentclass[english,prl, floatfix,  twocolumn,  english, superscriptaddress, showpacs]{revtex4-1}
\usepackage[T1]{fontenc}
\usepackage[latin9]{inputenc}
\usepackage{amsmath}
\usepackage{amssymb}
\usepackage{graphicx}
\usepackage{wasysym}

\makeatletter

\providecommand{\tabularnewline}{\\}

\usepackage{xcolor}
\definecolor{darkblue}{rgb}{0.1,0.2,0.6} 
\definecolor{lightblue}{rgb}{0.1,0.1,1.0}
\definecolor{darkred}{rgb}{0.8,0.1,0.2}
\usepackage[colorlinks,citecolor=lightblue,linkcolor=darkblue,urlcolor=lightblue] {hyperref}
\renewcommand{\BibitemShut}[1]{}

\makeatother

\usepackage{babel}
\begin{document}
\global\long\def\E{\mathrm{e}}
\global\long\def\D{\mathrm{d}}
\global\long\def\I{\mathrm{i}}
\global\long\def\mat#1{\mathsf{#1}}
\global\long\def\vec#1{\mathsf{#1}}
\global\long\def\cf{\textit{cf.}}
\global\long\def\ie{\textit{i.e.}}
\global\long\def\eg{\textit{e.g.}}
\global\long\def\vs{\textit{vs.}}
 \global\long\def\ket#1{\left|#1\right\rangle }

\global\long\def\etal{\textit{et al.}}
\global\long\def\tr{\text{Tr}\,}
 \global\long\def\im{\text{Im}\,}
 \global\long\def\re{\text{Re}\,}
 \global\long\def\bra#1{\left\langle #1\right|}
 \global\long\def\braket#1#2{\left.\left\langle #1\right|#2\right\rangle }
 \global\long\def\obracket#1#2#3{\left\langle #1\right|#2\left|#3\right\rangle }
 \global\long\def\proj#1#2{\left.\left.\left|#1\right\rangle \right\langle #2\right|}

\title{The Ergodic Side of the Many-Body Localization Transition}

\author{David J. Luitz}

\affiliation{Department of Physics and Institute for Condensed Matter Theory,
University of Illinois at Urbana-Champaign, Urbana, Illinois 61801,
USA}
\email{dluitz@illinois.edu}

\author{Yevgeny Bar Lev}

\affiliation{Department of Chemistry, Columbia University, 3000 Broadway, New
York, New York 10027, USA}
\email{yb2296@columbia.edu}

\begin{abstract}
Recent studies point towards nontriviality of the ergodic phase in
systems exhibiting many-body localization (MBL), which shows subexponential
relaxation of local observables, subdiffusive transport and sublinear
spreading of the entanglement entropy. Here we review the dynamical
properties of this phase and the available numerically exact and approximate
methods for its study. We discuss in which sense this phase could
be considered ergodic and present possible phenomenological explanations
of its dynamical properties. We close by analyzing to which extent
the proposed explanations were verified by numerical studies and present
the open questions in this field.
\end{abstract}
\maketitle
\tableofcontents{}

\section{Introduction}

Boltzmann's ergodic hypothesis \textemdash{} central to classical
statistical mechanics \textemdash{} allows to derive most equilibrium
results. It states that a trajectory of a system with many degrees
of freedom will spend equal times in regions of equal phase-space
measure \cite{Boltzmann1884ergodic_hypo}. This implies that the infinite
time average of observables is equivalent to their ensemble average.
Attempts to generalize this definition of ergodicity to quantum systems
had begun with the works of von Neumann \cite{VonNeumann1929,von_neumann_proof_2010}
and substantial progress was made in the 1980ies both analytically
and numerically in pioneering works by Berry, Pechukas, Peres, Feingold,
Jensen and Shankar \cite{Berry1977,Pechukas1983,Pechukas1984,Feingold1984,Feingold1985,Feingold1986,Peres1984,Peres1984a,jensen_statistical_1985},
culminating in the contributions by Deutsch \cite{Deutsch1991} and
Srednicki \cite{Srednicki1994,Srednicki1995,Srednicki1999}. It was
realized early on that not all complex systems are ergodic, as in
particular classically or quantum integrable systems are nonergodic
almost by definition. These systems are however not generic since
integrability and thus nonergodicity is inherently unstable to the
addition of generic perturbations \cite{Deutsch1991}. Ergodicity
breaking in more generic systems occurs during thermodynamic phase
transitions, where a system spontaneously breaks a symmetry when it
orders \cite{Goldenfeld1992lectures}. A novel mechanism of ergodicity
breaking in \emph{generic} \emph{disordered quantum} systems was proposed
ten years ago in a seminal work by Basko, Aleiner and Altshuler, a
phenomenon now widely known as many-body localization (MBL) \cite{Basko2006a}.
This work established the stability of the nonergodic Anderson insulator
to the addition of weak interactions at sufficiently small but \emph{finite}
energy densities, and the stability of the (ergodic) metal for sufficiently
large energy densities. It therefore predicted the existence of a
critical energy density (the so called many-body mobility edge) which
demarcates the ergodic and the nonergodic phases. Unlike ergodicity
breaking at thermodynamic transitions, this transition relies on the
system being completely \emph{isolated} from the environment and has
no signatures in \emph{static} thermodynamic quantities. The existence
of a nonergodic phase was recently rigorously proved for one-dimensional
random spin chains under a few physically reasonable assumptions \cite{Imbrie2014,Imbrie2016}.
Since MBL requires isolation from the environment, its realization
in conventional condensed matter systems is challenging \cite{Basko2007a,Ovadia2014}.
However signatures of MBL were observed in ultracold atomic gases
on optical lattices both in one-dimensional \cite{Schreiber2015a,Bordia2015,Smith2015}
and two-dimensional systems \cite{Choi2016}.

Most works on MBL concentrated on the study of the nonergodic phase,
paying little attention to the ergodic phase \cite{Altman2014,Nandkishore2014,Vasseur2016}.
The reason for this ``injustice'' is that following the work of
Basko, Aleiner and Altshuler it was largely accepted that the ergodic
phase in systems exhibiting the MBL transition is a trivial metal,
namely it has a finite dc conductivity \cite{Basko2006a}. Systems
with unbounded energy density, which were considered in this work,
are essentially classical at sufficiently high energy densities and
therefore have a finite dc conductivity as can be shown using a self-consistency
argument \cite{Basko2006a}. For systems with bounded energy density
this is not the case since even at infinite temperature there are
examples of systems which are far from being classical \cite{Anderson1958b}.
First evidence of the nontriviality of the ergodic phase for systems
with bounded energy density was obtained by one of us \cite{BarLev2014}.
Using a combination of nonequilibrium perturbation technique and exact
diagonalization (see Section \ref{sec:Numerical-Methods} for a brief
description of these methods) a surprisingly slow relaxation of the
density autocorrelation function was observed on the ergodic side
of the MBL transition which was attributed to the existence of an
intermediate phase with impeded transport due to localized inclusions
\cite{BarLev2014}. In a subsequent work, an extensive study of spin
transport in a large portion of the parameter phase space was performed
using exact diagonalization (ED) and the time-dependent density matrix
renormalization group (tDMRG) \cite{Lev2014}. This study showed that
most of the ergodic phase is subdiffusive up to simulated times and
argued that the dc conductivity must vanish if subdiffusion persists
asymptotically in time. Similar results were obtained in the study
of ac conductivity, where also a phenomenological explanation of the
observed subdiffusion was suggested \cite{Agarwal2014}.

In this review we concentrate on the ergodic phase and refer the reader
who is interested in the nonergodic phase or the MBL transition to
Refs.~\cite{Altman2014,Nandkishore2014,Vasseur2016} as also to more
recent reviews to appear in the current issue \cite{Imbrie2016a,Parameswaran2016b,Agarwal2016_review,haldar_dynamical_2017,Abanin2017}.
We limit the discussion to models with quenched disorder and refer
the reader interested in systems with quasiperiodic potentials to
Ref.~\cite{Deng2016b} for a review. The structure of the review
is the following: in Sec.~\ref{sec:Models} we present the models
which will be used throughout the review, in Sec.~\ref{sec:Dynamical-Properties}
we survey the properties of the ergodic phase and discuss in which
sense this phase is ergodic. In Sec.~\ref{sec:Phenomenological-models}
we present the phenomenological theory of the ergodic phase. Finally,
we close the review by surveying the available numerical techniques
in Sec.~\ref{sec:Numerical-Methods} and discuss open questions in
Sec.~\ref{sec:Discussion-and-Open}.

\section{\label{sec:Models}Models}

In this section we will introduce the models which will be used throughout
the rest of the review. Currently the most studied model in the context
of many-body localization is the XXZ model,
\begin{align}
\hat{H} & =\frac{J_{xy}}{2}\sum_{i=1}^{L-1}\left(\hat{S}_{i}^{+}\hat{S}_{i+1}^{-}+\hat{S}_{i}^{-}\hat{S}_{i+1}^{+}\right)\label{eq:xxz}\\
 & +J_{z}\sum_{i=1}^{L-1}\hat{S}_{i}^{z}\hat{S}_{i+1}^{z}+\sum_{i=1}^{L}h_{i}\hat{S}_{i}^{z},\nonumber 
\end{align}
were $\hat{S}_{i}^{z},$ is the $z-$projection of the spin-$1/2$
operator, $\hat{S}_{i}^{\pm}$, are the corresponding lowering and
raising operators, $J_{xy}$ and $J_{z}$ are inter-spin couplings
and $h_{i}$ are random magnetic fields taken to be uniformly distributed
in the interval $h_{i}\in\left[-W,W\right].$ This model conserves
the $z-$projection of the total spin. Using the Jordan-Wigner transformation
\cite{Jordan1928},
\begin{align}
\hat{S}_{i}^{z} & \to\hat{n}_{i}-\frac{1}{2}\label{eq:jw}\\
\hat{S}_{i}^{+} & \to\left(-1\right)^{\sum_{k=1}^{i-1}n_{k}}\hat{c}_{i}^{\dagger}\nonumber \\
\hat{S}_{i}^{-} & \to\left(-1\right)^{\sum_{k=1}^{i-1}n_{k}}\hat{c}_{i},\nonumber 
\end{align}
it can be exactly mapped to a model of spinless electrons,
\begin{align}
\hat{H} & =-t\sum_{i=1}^{L-1}\left(\hat{c}_{i}^{\dagger}\hat{c}_{i+1}+\hat{c}_{i+1}^{\dagger}\hat{c}_{i}\right)\label{eq:spinless}\\
 & +U\sum_{i=1}^{L-1}\left(\hat{n}_{i}-\frac{1}{2}\right)\left(\hat{n}_{i+1}-\frac{1}{2}\right)+\sum_{i=1}^{L}h_{i}\hat{n}_{i},\nonumber 
\end{align}
where $\hat{c}_{i}^{\dagger}$ creates a spinless fermion on site
$i$ and $\hat{n}_{i}=\hat{c}_{i}^{\dagger}\hat{c}_{i}$ is the fermion
density. We dropped a constant term and set $t\equiv-J_{xy}/2$ and
$U\equiv J_{z}$ to have a more conventional notation for fermions.
The conservation of $z-$projection of the total spin translates to
the conservation of the total charge in the fermionic model. In most
studies either the hopping $t,$ or the in plane coupling, $J_{xy}$,
are set to be one. We will pursue the latter convention here. Thus,
unless otherwise specified, all times are measured in units of $J_{xy}^{-1}$.
For $J_{z}=1$ $\left(U/t=2\right)$ both models have an ergodic to
nonergodic transition for a disorder strength of $W=3.7\pm0.1$ \cite{Luitz2015}.
Since in this review we focus on the ergodic side of the transition,
we mostly consider $W\leq3.7$ here.

Another model which we will discuss is the Anderson-Hubbard model,
\begin{align}
\hat{H} & =-t\sum_{\sigma,i=1}^{L-1}\left(\hat{c}_{i\sigma}^{\dagger}\hat{c}_{i+1,\sigma}+\hat{c}_{i+1,\sigma}^{\dagger}\hat{c}_{i\sigma}\right)\label{eq:anderson_hubbard}\\
 & +U\sum_{i=1}^{L}\left(\hat{n}_{i\uparrow}-\frac{1}{2}\right)\left(\hat{n}_{i\downarrow}-\frac{1}{2}\right)+\sum_{i=1}^{L}h_{i\sigma}\hat{n}_{i\sigma},\nonumber 
\end{align}
where $\hat{c}_{i\sigma}^{\dagger}$ creates a spinful fermion on
site $i$ and $\hat{n}_{i\sigma}=\hat{c}_{i\sigma}^{\dagger}\hat{c}_{i\sigma}$
is the corresponding density. The disorder potential $h_{i\sigma}$
is taken to be different for the two species in order to explicitly
break $SU\left(2\right)$ symmetries in the charge and the spin sectors
thus avoiding possible complications \cite{Potter2016,Prelovsek2016b}.
While this model naturally appears in cold atoms experiments it is
less popular than the XXZ model, mostly because it has a larger local
Hilbert space dimension, which makes it more challenging for numerical
study.

Further models, which display an ergodic to nonergodic transition,
are periodically driven systems. In these models, also knowns as Floquet-MBL
models, the transition can be tuned by the frequency or the amplitude
of the drive. We have decided to exclude these systems from our review
due to scarcity of numerical results on their dynamics in the ergodic
phase. A reader interested in these topics is referred to the recent
literature \cite{DAlessio2013,Ponte2014a,lazarides_fate_2015,Abanin2015a,ponte_many-body_2015,abanin_theory_2016,Bordia2016,Rehn2016,Zhang2016b}
and the review Ref. \cite{haldar_dynamical_2017}.

\section{Properties of the Ergodic Phase}

In this section we survey the numerical results on the properties
of the ergodic phase. Since the XXZ model (\ref{eq:xxz}) is equivalent
to the spinless fermion model (\ref{eq:spinless}), in order to avoid
repetition we use the spin language in the rest of the review. Readers
who prefer to think in terms of fermions are referred to the mapping
(\ref{eq:jw}). To minimize the notational overhead we assume that
all the considered quantities are\emph{ implicitly averaged over disorder
realizations}, and therefore are translationally invariant on average.
In this section we will also uniformly use periodic boundary conditions
and average over the volume of the system, which we believe enhances
readability and allows to operate with more physically transparent
formulas. Readers who are interested in the technicalities and precise
implementations are referred to Section~\ref{sec:Numerical-Methods}
or to the original works.

\subsection{\label{subsec:ergodicty}Three flavors of ergodicity}

We have postponed the precise definition of ergodicity which we use
in this review to this subsection due to the involved subtleties.
While the definition for classical systems, via Boltzmann's ergodic
hypothesis, presented in the beginning of the introduction is very
precise, currently, there is no commonly accepted definition of ergodicity
of quantum systems \cite{Peres1984,Goldstein2010}. Some of the reasons
for this are that some concepts from classical physics like: microstates,
phase-space and chaos cannot be immediately carried over to quantum
systems. Here, we discuss three different notions of ergodicity in
quantum systems based on (i) the statistics of eigenvalues, (ii) the
statistics of eigenvectors and (iii) the validity of the eigenstate
thermalization hypothesis.

\subsubsection{Eigenvalue statistics}

For quantum systems which are chaotic in their classical limit it
was conjectured by Bohigas, Giannoni and Schmit that the eigenvalue
statistics follow the statistics of an ensemble of random matrices,
which depends on the symmetries of the Hamiltonian \cite{Bohigas1986}.
Using semi-classical field theory this conjecture was later justified
\cite{Andreev1996}. For systems without a proper classical limit,
such as fermionic lattice models or spin systems, a direct connection
between eigenvalue statistics and ergodicity is still lacking. Nevertheless,
it was empirically shown that many generic quantum systems do follow
the eigenvalue statistics of random matrices \cite{Montambaux1993,Poilblanc1993}.
To study the eigenvalue statistics, the eigenvalues of the systems
are calculated and ordered ascendantly, then, traditionally, ``unfolding''
of the spectrum is performed, which eliminates the dependence of the
statistics on the density of states. The distribution of the unfolded
spacings is then obtained and compared to the corresponding random
matrix distribution (Wigner-Dyson (WD) distribution). A system is
assumed to be ergodic if the distribution of its eigenvalue spacing
follows the WD distribution. The distribution of eigenvalue spacings
in disordered (Coulomb) interacting systems was studied a decade \emph{before}
MBL was established \cite{Berkovits1994,Berkovits1996,Jacquod1997a,georgeot_integrability_1998}.
In these early studies a crossover from a Poisson to a WD distribution
was observed. In later studies eigenvalue statistics for disordered
spin chains were also studied in the context of quantum chaos \cite{Avishai2002,Santos2004,Santos2004a}.
In the context of MBL eigenvalue statistics was first considered in
Ref.~\cite{oganesyan_localization_2007} which introduced a useful
metric for short-range correlations in the eigenvalues statistics,
effectively eliminating the arbitrariness which exists in the unfolding
procedure \cite{Guhr1998}. Instead of unfolding, the eigenvalue spacings
$\delta_{n}=E_{n+1}-E_{n}$ (where $E_{n}$are the ordered eigenvalues)
are normalized by their magnitude, $r_{n}=\mathrm{min}\left(\delta_{n}/\delta_{n+1},\delta_{n+1}/\delta_{n}\right)$
\cite{Oganesyan2009}. Ergodicity is assumed when the obtained probability
distribution of $r_{n}$ (or the unfolded $\delta_{n}$) matches the
one of the corresponding random matrix ensemble \cite{Atas2013}.
The fact that the phase, which is the subject of this review, is ergodic
in this sense was first established in Ref.~\cite{Oganesyan2009}
and then repeatedly in almost every work on MBL. 

The distribution of the eigenvalue spacings can be viewed as a stationary
distribution of a Brownian motion in a space of Hamiltonians, where
at each step a different disorder realization is drawn \cite{Dyson1962,chalker_fictitious_1996}.
In this approach, commonly known as the effective plasma model, unfolded
eigenvalues are thought of as particles with an effective two-body
interaction which is responsible for the eigenvalue repulsion. By
noting that in a second order expansion in the disordered potential
the effective interaction is well described by a power law, Serbyn
and Moore derive the corresponding limiting spacings distribution,

\begin{equation}
P\left(\delta_{n}\right)=C_{1}\delta_{n}^{\beta}\exp\left(-C_{2}\delta_{n}^{2-\gamma}\right),\label{eq:semi_poisson}
\end{equation}
where $C_{1,2}$ are constants, $0\leq\beta\leq1$ controls the level
repulsion and $\gamma$ controls the tail of the distribution \cite{Serbyn2015}.
This distribution interpolates between the Poisson distribution $\gamma=1$,
$\beta=0$ and the WD distribution $\gamma=0$, $\beta=1$. Motivated
by this form Serbyn and Moore numerically obtain $\gamma$ within
the whole ergodic phase, even outside the region of validity of the
effective plasma model $\left(W\apprge2\right)$. It is argued that
while for weak disorder $W\lesssim2$ the spacings distribution appears
to flow to the WD distribution, for stronger disorder close to the
MBL transition $2\lesssim W\lesssim3.7$ a region with intermediate
statistics is found. The corresponding distribution is similar to
the critical distributions obtained for Anderson transitions, it has
an exponential tail $\gamma=1$ and a \emph{finite} level repulsion
$\beta>0$ \cite{Evers2008a}. It is therefore argued that the MBL
transition has critical statistics similar to the critical Anderson
statistics \footnote{We note that the level statistics with {$\beta>0$ and $\gamma=1$}
was called a semi-Poisson statistics in Ref.~\cite{Serbyn2015}.
To eliminate the confusion with semi-Poisson statistics which was
introduced in Ref.~\cite{Bogomolny1999} and implies {$\beta=1$
and $\gamma=1$}, we have instead used the term ``critical statistics.''}. The effective model used by Serbyn and Moore was criticized in a
follow-up exact diagonalization work \cite{Bertrand2016}, which pointed
out that the eigenvalue statistics of the ergodic phase does not appear
to be scale invariant as the plasma model of Ref.~\cite{Serbyn2015}
suggests, moreover the critical eigenvalue statistics seems to better
agree with a Poisson distribution, similarly to critical statistics
of Anderson transition on a Bethe lattice.

\subsubsection{Eigenvector statistics, multifractality and the ``bad metal''}

The first proposal of an intermediate phase sandwiched between the
deeply ergodic and nonergodic (MBL) phases appeared almost 20 years
ago \cite{Altshuler1997}. This phase, colloquially dubbed by Altshuler
a ``bad metal'' \cite{Altshuler2010}, was first defined as a delocalized
yet nonergodic phase. The definition of ergodicity and delocalization
in this context is however quite different from what we have discussed
above, therefore to avoid confusion we will use a \textsf{sans serif}
font face to designate this kind of \textsf{ergodicity.}

The motivation behind this definition is best understood for the case
of a single particle. The moments of the eigenstates of the single
particle Hamiltonian $\psi_{\alpha}\left(x\right)$ written in the
position basis are given by,
\begin{equation}
I_{q}^{\alpha}=\sum_{x}\left|\psi_{\alpha}\left(x\right)\right|^{2q}.
\end{equation}
Delocalized single-particle eigenstates (for example eigenvectors
of a random matrix) scale as $\psi_{\alpha}\left(x\right)\sim V^{-1/2}$
where $V$ is the volume of the system, which yields $I_{q}^{\alpha}\propto V^{-\left(q-1\right)}$.
Localized eigenstates which decay exponentially with distance from
some localization center, yield $I_{q}^{\alpha}\approx\mathrm{const.}$
Since the infinite time average of the density autocorrelation function
is given by $I_{2}^{\alpha}$ (see derivation in Eq.~(\ref{eq:ipr})),
a natural definition of a delocalized (localized) state would be a
state with $I_{2}^{\alpha}\to0$ $\left(I_{q}^{\alpha}\to\text{const }\right)$
for $V\to\infty$. The participation ratio $1/I_{2}^{\alpha}$ quantifies
the number of sites that a eigenstate occupies in \emph{real} space.
When this number of sites is extensive, the system is defined to be
\textsf{ergodic}. On the contrary when eigenstates cover a subextensive
volume in real space $I_{2}^{\alpha}\sim V^{-D_{2}}$ with $0<D_{2}<1$,
not all sites in real space are ``available'' and the system is
therefore \textsf{nonergodic.} The eigenstate will be called multifractal
if the generalized dimensions $D_{q}$ depend on $q$. It occurs,
for example, at the critical point of the Anderson transition, where
all the moments $I_{q}^{\alpha}$ follow an anomalous scaling $I_{q}^{\alpha}\sim V^{-D_{q}\left(q-1\right)}$
\cite{Evers2008a}. We stress that the sparseness of the eigenstates
in real space does \emph{not} imply that a generic initial condition
will be locked to a region in space. In fact almost all initial conditions
will explore the whole volume of the lattice. The sparseness of the
eigenfunctions in real space has implications on the \emph{dynamics}
of the wavepackets, which will be subdiffusive with a dynamical exponent
which could be related to the generalized dimension $D_{2}$ \cite{Ketzmerick1997,Ohtsuki1997}.

The many-body problem is equivalent to a single-particle hopping on
a complicated graph, where the nodes of the graph represent many-body
states weighted by the diagonal part of the Hamiltonian and the hopping
rates are given by the offdiagonal part. The apparent simplicity of
this view is however misleading, since the disorder (many times taken
to sit on the diagonal part) will be highly correlated. The number
of return paths on this graph is exponentially small in their length,
therefore by neglecting the loops it could be approximately mapped
to a Cayley tree \cite{Altshuler1997} (see also review by Imbrie
\emph{et al. }\cite{Imbrie2016a}). Using this analogy one can carry
over the above definition of \textsf{ergodicity} to the many-body
case by substitution of the volume in real space by the total number
of many-body states, $\mathcal{N}$, 

\begin{equation}
I_{q}^{\alpha}=\sum_{n}\left|\left\langle \alpha|n\right\rangle \right|^{2q}\propto\mathcal{N}^{-D_{q}\left(q-1\right)},\label{eq:i_q}
\end{equation}
where $\ket{\alpha}$ are the eigenstates of the Hamiltonian computed
in some basis $\ket n$. We note in passing that this quantity is
closely related to the basis dependent Rényi ``participation'' entropies
of the wave function as considered in \cite{atas_multifractality_2012,luitz_universal_2014,Luitz2015},

\begin{equation}
S_{q}^{P,\alpha}=\frac{1}{1-q}\ln I_{q}^{\alpha}\propto D_{q}\ln\mathcal{N}.\label{eq:Sq}
\end{equation}
In the limit of $q\to1$ it reduces to the Shannon entropy $S_{1}^{P,\alpha}=-\sum_{n}|\langle\alpha|n\rangle|^{2}\ln|\langle\alpha|n\rangle|^{2}$,
and allows to define $D_{1}$ as of $S_{1}^{\alpha}/\ln\mathcal{N}$.
There are a few problems with this definition of \textsf{ergodicity}.
First, while the real space basis is a natural choice for the single
particle problem there is no obvious choice of the basis $\ket n$
in the many-body case. The second and more serious problem is the
lack of a direct connection between the spreading of the wavepacket
on a complicated graph or tree in the many-body Hilbert space and
the dynamics in real space (\emph{cf.} Eq.~(\ref{eq:survival_prob})
for one possible connection). In particular, it is not clear whether
the sparseness of the eigenfunctions in the many-body space $\left(D_{q}<1\right)$
has implications on the thermalization in \emph{finite} many-body
systems, or has a signature in local observables (for a discussion
see Ref.~\cite{Borgonovi2016}).

The existence of a stable delocalized but \textsf{nonergodic} phase
was tested in numerous numerical studies. Most studies of multifractality
are focused on either the Bethe lattice or random-regular graphs.
After almost a decade of study, this question is still largely open
\cite{Biroli2012,DeLuca2013,Luca,Kravtsov2015,Facoetti2016,Tikhonov2016,Tikhonov2016a,Garcia-Mata2016,Altshuler2016},
while most extensive numerical studies suggest that this phase disappears
in the thermodynamic limit \cite{Tikhonov2016,Garcia-Mata2016}. For
physical lattice models this question was considered in a study of
a random Josephson array \cite{Pino2015} and for the XXZ model \cite{Luitz2015,serbyn_thouless_2016,Torres-Herrera2016},
with a similar inconclusive outcome. While Ref.~\cite{Luitz2015}
suggests that $D_{1}=1$ below the MBL transition, Refs.~\cite{Pino2015,Torres-Herrera2016}
argue in favor of a stable intermediate phase with $D_{2}<1$. Furthermore,
Ref.~\cite{serbyn_thouless_2016} argues that this phase shrinks
to the MBL critical point in the thermodynamic limit. We would like
to point out that it is possible that this apparent discrepancy might
follow from a different basis used to calculate $I_{q}^{\alpha}$
in these works \cite{Luitz2015,serbyn_thouless_2016,Torres-Herrera2016}.

An attempt to connect the notion of \textsf{ergodicity} from eigenvector
statistics to ergodicity defined through eigenvalues statistics was
performed by Serbyn and Moore in the work described above \cite{Serbyn2015}.
By using a specific choice of the basis $\ket{\psi^{\beta}}\equiv2\hat{S}_{i}^{z}\ket{\beta}$
(where $|\beta\rangle$ are the eigenstates of the Hamiltonian) in
(\ref{eq:i_q}) one can write, 
\begin{equation}
I_{q}^{\alpha}=\sum_{\beta}\left|\left\langle \alpha|\psi^{\beta}\right\rangle \right|^{2q}\propto\mathcal{N}^{-D_{q}\left(q-1\right)},
\end{equation}
where the scaling with the size of Hilbert space is taken as an assumption
(only $I_{2}^{\alpha}$ was considered in Ref.~\cite{Serbyn2015}).
A similar scaling of the moments, was conjectured in Ref.~\cite{Monthus2016}
and was recently numerically verified \cite{serbyn_thouless_2016}.
Using heuristic arguments, Serbyn and Moore connected the exponent
$\gamma$ in (\ref{eq:semi_poisson}), which parametrizes the distribution
of the eigenvalue spacing to the generalized dimension $\gamma=1-D_{2}$
\footnote{We note that there appears to be a misprint in Ref.~\cite{Serbyn2015}.
Since for {$\gamma=1$, $d_{2}=1-\gamma=0$,} while from the authors'
definition of {$\mathcal{N}\sum_{\alpha}\left|\left\langle \alpha|\psi^{\beta}\right\rangle \right|^{4}\propto\mathcal{N}^{-d_{2}}$}
one gets that, {$I_{2}\propto\mathcal{N}^{-1}$,} which corresponds
to a WD distribution {$\left(\gamma=0\right)$.}}. This relation was however never verified numerically.

\subsubsection{Eigenstate thermalization hypothesis}

In this review we utilize yet another definition of ergodicity, which
is more similar to the Boltzmann ergodic hypothesis for classical
systems \cite{Boltzmann1884ergodic_hypo} as also to ideas by von
Neumann \cite{VonNeumann1929,von_neumann_proof_2010}. It is commonly
known as the \textit{eigenstate thermalization hypothesis} (ETH) and
it was mostly developed by Deutsch and Srednicki more than two decades
ago \cite{Deutsch1991,Srednicki1994,Srednicki1995,Srednicki1999,Rigol2008},
based on a multitude of theoretical and numerical works in quantum
chaos \cite{Pechukas1983,Pechukas1984,Peres1984,Peres1984a,Feingold1984,Feingold1985,jensen_statistical_1985,Feingold1986}
(for recent reviews, see \cite{DAlessio2015,Borgonovi2016,Gogolin2016}).
One can show that a sufficient condition for local quantum observables
$\hat{O}$ to decay to their microcanonical value (and to stay close
to this value for sufficiently long times) is the validity of the
ansatz,
\begin{equation}
\left\langle \alpha\left|\hat{O}\right|\beta\right\rangle =\bar{O}\left(E\right)\delta_{\alpha\beta}+e^{-S\left(E\right)/2}f\left(E,\omega\right)R_{\alpha\beta},\label{eq:ETH}
\end{equation}
where $\ket{\alpha}$, $\ket{\beta}$ are eigenstates of the Hamiltonian,
$S\left(E\right)$ is the microcanonical entropy, $\bar{O}\left(E\right)$,
$f\left(E,\omega\right)$ are smooth functions of their arguments
with $E\equiv\left(E_{\alpha}+E_{\beta}\right)/2$ and $\omega=E_{\beta}-E_{\alpha}$
and $R_{\alpha\beta}$ are random independent variables with zero
mean and a unit variance. 

The fact that the ergodic phase is indeed ergodic under this definition
was established for the diagonal elements in Refs.~\cite{Pal2010a,Luitz2016}
and for the offdiagonal elements in Ref.~\cite{Luitz2016b}. However
it was observed that the shape of the probability distributions of
local operators in the eigenbasis of the Hamiltonian according to
Eq.~(\ref{eq:ETH}) depends strongly on the value of the disorder
strength for both diagonal and off-diagonal matrix elements. In particular,
the distributions are perfectly Gaussian for weak disorder, while
for intermediate disorder strength the distribution of $R_{\alpha\beta}$
becomes strongly non-Gaussian even in the thermodynamic limit. Interestingly,
even in this case, the ETH ansatz remains valid, although in a generalized
form with a non-Gaussian noise term $R_{\alpha\beta}$. These non-Gaussian
probability distributions are accompanied with a slower decrease of
the standard deviation of the offdiagonal matrix elements $\left\langle \alpha\left|\hat{O}\right|\beta\right\rangle $
with the size of the system in the low frequency limit ($\omega=E_{\alpha}-E_{\beta}$).
This modified scaling was connected to the dynamical exponent of the
system \cite{Luitz2016b}. The dependence of the $\left\langle \alpha\left|\hat{O}\right|\beta\right\rangle $
matrix elements on $\omega$ was also studied in Ref.~\cite{serbyn_thouless_2016}.

\subsection{Entanglement Structure }

The ETH ansatz is commonly assumed to hold for few-body operators
which have a finite support on a small subsystem of the total isolated
system. This implies that even when the whole system is in an eigenstate
$|\alpha\rangle$, a sufficiently small subsystem $A$ is thermalized
by the rest of the system. Clearly, thermalization requires that the
entropy of the subsystem, \emph{i.e.} the von Neumann entanglement
entropy obtained by tracing out degrees of freedom that are not in
the subsystem $A$, \emph{(cf.} also the recent review in Ref.~\cite{laflorencie_quantum_2016})

\[
S_{A}^{\alpha}=-\mathrm{Tr}\left(\hat{\rho}_{A}\ln\hat{\rho}_{A}\right),\quad\mathrm{with}\quad\hat{\rho}_{A}=\mathrm{Tr}_{\bar{A}}|\alpha\rangle\langle\alpha|
\]
has to be extensive in the subsystem size $S_{A}^{\alpha}\propto L_{A}$,
which is usually referred to as a volume law scaling. In the MBL phase,
on the other hand, this is not true and due to the finite localization
length, the entanglement entropy of MBL eigenstates scales as the
surface area of the subsystem (which is constant in one dimension).
This difference in the entanglement scaling across the MBL transition
was first observed by Bauer and Nayak \cite{Bauer2013} numerically
and has subsequently become a popular measure to detect the MBL transition
\cite{Kjall2014,Vosk2014,Luitz2015,Potter2015,Yu2016}.

Kjäll \emph{et al. }numerically studied the critical region in which
the dominant scaling changes from a volume law to an area law \cite{Kjall2014}.
They discovered that close to the transition the \emph{variance} of
the entanglement entropy exhibits a maximum. A careful analysis of
the probability distributions of the entanglement entropy showed that
close to the transition, a \emph{mixture} of volume-law and area-law
states exists \cite{Luitz2016,Yu2016}. In Ref.~\cite{Yu2016}, it
was shown that in periodic, disordered, one dimensional systems the
average of the entanglement entropy $\bar{S}_{\mathrm{A}}$ over all
possible bipartitions with the same subsystem length is a smooth and
concave function of the subsystem length $L_{A}$. The derivative
$\partial\bar{S}_{A}/\partial L_{A}$ was argued to capture the dominant
entanglement scaling in the system for single eigenstates, and is
close to its maximal value ($\ln2$ for spin-$1/2$ chains) in the
case of a volume law scaling, and zero in the case of an area law
scaling. The probability distribution of $\partial\bar{S}_{A}/\partial L_{A}$
close to the MBL transition becomes strongly bimodal \emph{even for
single disorder realizations} \cite{Yu2016}, although the inter-sample
variance is observed to be smaller than the sample-to-sample variance
\cite{Yu2016,khemani_critical_2016}.

At weaker disorder, the probability distributions of the entanglement
entropy and its slope are sharply peaked at a large, volume law value,
with exponentially suppressed tails at lower entanglement, which neither
affect the mean nor the variance. The analysis of the entanglement
structure in Ref.~\cite{Yu2016} seems to exclude the possibility
of critical eigenvectors which have a volume law scaling with a suppressed
prefactor \cite{Grover2014}. Such a sub-thermal volume law scaling
holds only for the disorder averaged entanglement entropy and is caused
by the disorder average over a sharply peaked bimodal distribution.
This average corresponds to a part of the distribution with exponentially
low weight and is therefore physically meaningless. 

The study of the spatial entanglement structure is especially interesting
in light of the rare Griffiths regions picture, which was proposed
to explain the observed subdiffusion (see Sec.~\ref{sec:Phenomenological-models}).
In Ref.~\cite{Luitz2016}, the entanglement entropy was calculated
as a function of the cut position, showing qualitatively that in the
ergodic phase some cuts between the two subsystems have much lower
entanglement entropies compared to other cuts. Moreover these regions
can be identified in any eigenstate of the system. The correlation
between eigenstates in the spatial variation of entanglement was also
observed in the nearest neighbor concurrence as a local probe of entanglement
\cite{Bera2015a}. In an analysis of the probability distribution
of the change of the entanglement entropy $\Delta(\ell)=S(\ell+1)-S(\ell)$
if the subsystem is enlarged by one site it was shown that at intermediate
disorder there is an increasing probability (when disorder is increased)
of finding $\Delta(\ell)<0$, which can be seen as indirect evidence
for the existence of insulating inclusions in the system \cite{Yu2016}.
To make progress in this direction, it may prove useful to study more
local probes of entanglement, such as the mutual information $I(A,B)=S_{A}+S_{B}-S_{A\cup B}$,
which was recently proposed as a generic measure to extract the correlation
length \cite{DeTomasi2016}.

\subsection{\label{sec:Dynamical-Properties}Dynamical Properties and Transport}

In this section we will survey the different results on the dynamical
properties of the ergodic phase of systems exhibiting MBL transition.
To emphasize the similarity between MBL systems and classical glasses
through this section we will adopt the notation commonly used in the
glasses community. At the end of the section we present a summary
of the relations between the different dynamical quantities.

\subsubsection{Mean-square displacement and ac conductivity}

The XXZ model conserves both the total energy and the $z-$projection
of the total spin, which is equivalent to the conservation of the
total number of particles in the fermion language. For this model
one can therefore study either the transport of spin or energy. Following
the work of Basko, Aleiner and Altshuler \cite{Basko2006a} and first
numerical studies of the dc conductivity it was largely believed that
the ergodic phase is a metal \cite{Karahalios2009a,Berkelbach2010a,Barisic2010a},
namely that it has a finite dc conductivity (similarly to a normal
liquids in structural glasses). First evidence of the surprisingly
slow relaxation of local observables deep in the ergodic phase was
obtained using a self-consistent second Born approximation \cite{BarLev2014}.
A more extensive exploration of transport in the ergodic phase using
numerically exact methods was performed in Ref.~\cite{Lev2014}.
In this work the spin-spin correlation function 
\begin{equation}
G_{r}\left(t\right)=\text{Re }\frac{1}{L}\sum_{i}\text{Tr }\left[\hat{\rho}_{0}\hat{S}_{i+r}^{z}\left(t\right)\hat{S}_{i}^{z}\left(0\right)\right]\label{eq:c_ij}
\end{equation}
was calculated, where $\hat{\rho}_{0}$ is the density operator of
the initial state of the system which is typically taken to be proportional
to the identity operator (infinite temperature). This correlation
function is analogous to the van Hove correlation function in structural
glasses \cite{Binder2005}, and intuitively describes the evolution
of a spin excitation created at time $t=0$. To assess transport properties
one can evaluate the analog of the classical mean-square displacement,
\begin{equation}
x^{2}\left(t\right)=\sum_{r}r^{2}G_{r}\left(t\right),
\end{equation}
which for diffusive systems should asymptotically scale linearly with
time, $x^{2}\sim t$. It was found that even for the smallest studied
disorder strength $\left(W\approx1\right)$ and through most of the
ergodic phase transport is subdiffusive,
\begin{equation}
x^{2}\left(t\right)\sim t^{2/z}\qquad\mathrm{for}\qquad t<t_{*}\left(J_{z},W\right),\label{eq:x2}
\end{equation}
with a dynamical exponent, $z\left(J_{z},W\right)\geq2$ which depends
on the parameters of the system. The simulation time $t_{*}\approx L^{z}$
was chosen such that finite-size effects were eliminated up to a predefined
precision \cite{Lev2014}. This time scale could be considered as
a generalized Thouless time, namely the time it takes to transport
a particle across the system \cite{Edwards1972,Bertrand2016,Luitz2016b,serbyn_thouless_2016}.
An analogous calculation was performed using the self-consistent second
Born approximation for a \emph{two-dimensional} Anderson-Hubbard model
(\ref{eq:anderson_hubbard}), yielding similar results \cite{BarLev2015}.
The evaluation of (\ref{eq:c_ij}) and (\ref{eq:x2}) is valid for
any initial state and therefore does\emph{ not} require the system
to be within the linear response regime. However for a thermal initial
state $\hat{\rho}_{0}$ it is directly related to the frequency dependent
diffusion coefficient calculated from linear response theory \cite{Scher1973a}
(for a derivation of this relation for quantum systems see Appendix
\ref{sec:Appendix}),
\begin{equation}
D\left(\omega\right)=-\omega^{2}\int_{0}^{\infty}\mathrm{d}t\,\E^{\I\omega t}x^{2}\left(t\right),\label{eq:d_omega}
\end{equation}
which is proportional to the ac conductivity, 
\begin{equation}
\sigma\left(\omega\right)\propto\omega^{2}\int_{-\infty}^{\infty}\mathrm{d}t\,\E^{\I\omega t}|t|^{2/z}\propto|\omega|^{1-2/z}.\label{eq:sigma_w}
\end{equation}
The dependence of the ac conductivity on the frequency was numerically
calculated in Refs.~\cite{Agarwal2014,Gopalakrishnan2015}. We note
in passing that for infinite temperatures what is actually computed
is $D\left(\omega\right)\sim T\sigma\left(\omega\right)$ since $\sigma\left(\omega\right)$
vanishes in this limit. At the MBL transition the dynamical exponent
is expected to diverge, $z\to\infty$, and therefore the critical
ac conductivity is $\sigma\left(\omega\right)\propto\omega$ \cite{Gopalakrishnan2015}.
The ac conductivity was the first dynamical quantity which was studied
in the context of MBL. Within the linear response theory its real
part is given by the Kubo formula \cite{Kubo1957},
\begin{equation}
\text{Re }\sigma\left(\omega\right)=\frac{1}{\omega L}\tanh\left(\frac{1}{2}\beta\omega\right)\int_{-\infty}^{\infty}\mathrm{d}t\,e^{i\omega t}\text{Re }\left\langle \hat{J}\left(t\right)\hat{J}\left(0\right)\right\rangle _{\beta},\label{eq:kubo}
\end{equation}
where $\beta=1/T$, and we set the Boltzmann constant to be one, $\left\langle .\right\rangle _{\beta}$
is the thermal expectation value and $\hat{J}$ is the total current
density operator, 
\begin{equation}
\hat{J}=i\frac{J_{xy}}{2}\sum_{n}\left(\hat{S}_{n}^{+}\hat{S}_{n+1}^{-}-\hat{S}_{n+1}^{+}\hat{S}_{n}^{-}\right).
\end{equation}
The use of the Kubo formula above assumes the validity of linear response
theory. While the validity of (\ref{eq:kubo}) within the ergodic
phase was not directly tested, the response of the system for sufficiently
small driving fields was shown to be linear \cite{Kozarzewski2016a,Znidaric2016},
as also the heating of the system \cite{Gopalakrishnan2016}.

The first results on spin and heat ac conductivities were obtained
using exact diagonalization (ED) \cite{Karahalios2009a}. In this
work it was argued that,
\begin{equation}
\sigma\left(\omega\right)\backsimeq\sigma_{dc}\left(J_{z},W\right)+A\left|\omega\right|\label{eq:ac_linear}
\end{equation}
(and similarly for the heat conductivity) with $\sigma_{\mathrm{dc}}>0$
for most $J_{z}$ and $W$. The putative delocalization of the MBL
phase was later challenged in Ref.~\cite{Berkelbach2010a} and then
also in Ref.~\cite{Barisic2010a}. Recent large scale ED studies,
with systems as large as $L=28$, confirmed a \emph{linear} scaling
with frequency and \emph{a finite} dc conductivity in the \emph{ergodic}
phase \cite{Steinigeweg2015,Barisic2016,Prelovsek2016a}, in contrast
to the finding of a sublinear scaling (\ref{eq:sigma_w}) and \emph{zero}
dc conductivity in the same region of the phase diagram as argued
in Refs.~\cite{Agarwal2014,Gopalakrishnan2015}. These contradicting
results highlight the difficulty of extracting the low frequency behavior
from the Kubo formula (\ref{eq:kubo}); a difficulty, which was pointed
out already by Thouless and Kirkpatrick \cite{Thouless1981} and more
recently by Berkelbach and Reichman \cite{Berkelbach2010a}. The evaluation
of the ac conductivity for any \emph{finite} system requires a broadening
of the many-body levels with an artificial width $\eta$, which could
be attributed to either a residual coupling to the environment or
to the timescale over which the conductivity is measured \cite{Steinigeweg2015}.
This coupling results in $\sigma_{dc}\left(\eta\right)>0$ for any
finite system. In order to eliminate the dependence on $\eta$ it
is crucial to take the thermodynamic limit $L\to\infty$ before taking
$\eta\to0^{+}$ \cite{Thouless1981,Imry2008}. For systems \emph{known}
to be metallic $\left(\sigma_{dc}>0\right)$, this apparently formidable
task is actually feasible, since even for finite systems $\sigma\left(\omega\right)$
is almost independent of $\eta$, as long as $\eta>\Delta$ (where
$\Delta$ is the mean level spacing) \cite{Thouless1981}. This is
however \emph{not} the case when it is not known \emph{a priori} if
$\sigma_{dc}>0$, and the way the extrapolation to the thermodynamic
limit is performed is extremely important. The main technical difference
between Refs.~\cite{Karahalios2009a,Barisic2010a,Barisic2016,Steinigeweg2015,Prelovsek2016a}
and Refs.~\cite{Agarwal2014,Gopalakrishnan2015} is the functional
form which was used to fit the ac conductivity. While the former works
assume a \emph{finite} dc conductivity and the form (\ref{eq:ac_linear}),
the later assume that the dc conductivity \emph{vanishes} and the
form (\ref{eq:sigma_w}). An attempt to circumvent the inherent finite
size constraint of ED studies was performed in Ref.~\cite{Khait2016},
where a continued fraction expansion of dynamical correlations using
a variational extrapolation of recurrents was developed. This allowed
the authors of Ref.~\cite{Khait2016} to work essentially at the
infinite system limit. The results of this work are consistent with
a vanishing dc conductivity and the functional dependence (\ref{eq:sigma_w}).

\subsubsection{Autocorrelation function and the Edwards-Anderson parameter}

A different way of examining dynamical properties is the calculation
of the local autocorrelation function, which is a special case of
zero displacement ($r=0$) in (\ref{eq:c_ij}),

\begin{equation}
G_{0}\left(t\right)=\frac{1}{L}\text{Re }\sum_{i=1}^{L}\text{Tr }\hat{\rho}_{0}\hat{S}_{i}^{z}\left(t\right)\hat{S}_{i}^{z}\left(0\right).
\end{equation}
 Its infinite time average for thermal initial states, $\hat{\rho}_{0}=\exp\left[-\beta\hat{H}\right]/Z$
is given by,
\begin{equation}
\lim_{T\to\infty}\int_{0}^{T}G_{0}\left(\bar{t}\right)\D\bar{t}=\frac{1}{L}\sum_{i}\sum_{\alpha}e^{-\beta E_{\alpha}}\left|\left\langle \alpha\left|\hat{S}_{i}^{z}\right|\alpha\right\rangle \right|^{2}=q_{\text{EA}},
\end{equation}
where $\left|\alpha\right\rangle $ and $E_{\alpha}$ are the eigenvectors
and eigenvalues of the Hamiltonian and $q_{\text{EA }}$ is the Edwards-Anderson
(EA) parameter \cite{Edwards1976}. Similarly to the situation for
spin-glasses the EA parameter is zero in the ergodic phase and nonzero
in the nonergodic MBL phase, and could be used as an order parameter
of the MBL transition \footnote{For systems with no (or broken) spin reflection symmetry.}\cite{Huse2013a,Pekker2014,Kjall2014,Vasseur2015b,monthus_level_2016}.
While there is no direct connection between the decay of the autocorrelation
function and transport, many times the following relation between
the autocorrelation function and the mean-square displacement is assumed
to hold (see derivation for subdiffusive classical systems in Sec.~\ref{sec:Phenomenological-models}),

\begin{equation}
G_{0}\left(t\right)\propto\frac{1}{\sqrt{x^{2}\left(t\right)}}=t^{-1/z},\label{eq:autocorr}
\end{equation}
which relies on a scaling hypothesis, and allows to relate between
the exponents of the ac conductivity (\ref{eq:sigma_w}) and the autocorrelation
function (\ref{eq:autocorr}) \cite{Alexander1981}. This relation
was also derived in Refs.~\cite{Agarwal2014,Gopalakrishnan2015}.
The spectral density can be evaluated by taking the Fourier transform
of the autocorrelation function,
\begin{equation}
A\left(\omega\right)\equiv\int\mathrm{d}t\,\E^{\I\omega t}G_{0}\left(t\right)\propto\left|\omega\right|^{-\left(1-1/z\right)},\label{eq:spectral_density}
\end{equation}
which diverges at small frequency \cite{Gopalakrishnan2015a}. While
the autocorrelation function was already considered in Refs.~\cite{Berkelbach2010a,Iyer2013},
the surprisingly slow relaxation deep in the ergodic phase was noted
in Ref.~\cite{BarLev2014}, and was attributed to the possibility
of an intermediate phase. In fact, a \emph{direct} study of the functional
dependence of the dynamical exponent extracted from the autocorrelation
function (\ref{eq:autocorr}) was only performed quite recently \cite{Agarwal2014,Luitz2016b}.

\subsubsection{Survival or return probability}

For an initial state which is a projector on an eigenstate, $\hat{\rho}_{0}=\left|\alpha\right\rangle \left\langle \alpha\right|$
the autocorrelation function is closely related to the survival probability,
\begin{equation}
C\left(t\right)=\left|\left\langle \alpha\left|\delta\hat{S}_{i}^{z}\left(t\right)\delta\hat{S}_{i}^{z}\left(0\right)\right|\alpha\right\rangle \right|^{2}=\left|\left\langle \psi_{\alpha}\left|\E^{-\I\hat{H}t}\right|\psi_{\alpha}\right\rangle \right|^{2},\label{eq:survival_prob}
\end{equation}
where we have defined, $\left|\psi_{\alpha}\right\rangle \equiv\delta\hat{S}_{i}^{z}\left|\alpha\right\rangle $,
and $\delta\hat{S}_{i}^{z}=\hat{S}_{i}^{z}-\left\langle \alpha\left|\hat{S}_{i}^{z}\right|\alpha\right\rangle $,
to set the infinite time average of $\left\langle \alpha\left|\delta\hat{S}_{i}^{z}\left(t\right)\delta\hat{S}_{i}^{z}\left(0\right)\right|\alpha\right\rangle $
to zero. The decay of the survival probability therefore corresponds
to the decay of the \emph{fluctuations} of the autocorrelation function.
The infinite time average of the survival probability is given by,
\begin{equation}
I_{2}\equiv\lim_{T\to\infty}\frac{1}{T}\int_{0}^{T}\mathrm{d}t\left|\sum_{\beta}\left|C_{\alpha\beta}\right|^{2}\E^{-\I E_{\beta}t}\right|^{2}=\sum_{\beta}\left|C_{\alpha\beta}\right|^{4},\label{eq:ipr}
\end{equation}
where we defined $C_{\alpha\beta}=\left\langle \alpha\left|\delta\hat{S}_{i}^{z}\right|\beta\right\rangle $
and $I_{2}$ is the inverse participation ratio. The inverse participation
ratio scales as $I_{2}\propto\mathcal{N}^{-\tilde{D}_{2}},$ where
$\mathcal{N}$ is the Hilbert space dimension and $\tilde{D}_{2}$
is a generalized dimension. For delocalized systems (even non-interacting)
$\tilde{D}_{2}>0$ and $I_{2}\to0$ in the limit $L\to\infty$, while
for localized systems, $\tilde{D}_{2}=0$ and $I_{2}\to\text{const}$.
There is no direct connection between the decay of the survival probability
in many-body systems and transport, yet a power law relaxation was
obtained \footnote{In Ref.~\cite{Torres-Herrera2015a} a product state initial condition
was used, which is different from the initial condition we have used
in our definition (\ref{eq:survival_prob}). },
\begin{equation}
C\left(t\right)\sim t^{-\tilde{D}_{2}},\label{eq:survival_prob_decay}
\end{equation}
with disorder dependent exponent, $0\leq\tilde{D}_{2}\leq1$, which
is just the generalized dimension defined above \cite{Torres-Herrera2015a,Torres-Herrera2016}.\emph{ }

\subsubsection{Dynamical structure factor and the imbalance}

Instead of studying the decay of local excitations one can also consider
the relaxation of collective (spin-wave like) excitations,
\begin{equation}
F\left(q,t\right)=\text{Re }\text{Tr }\left[\hat{\rho}_{0}\hat{S}_{q}^{z}\left(t\right)\hat{S}_{-q}^{z}\left(0\right)\right],\label{eq:c_q_t}
\end{equation}
where $\hat{S}_{q}^{z}\left(t\right)=\sum_{n}\hat{S}_{n}^{z}\left(t\right)\exp\left[iqn\right]/\sqrt{L}$,
and $F\left(q,t\right)$ is the analog of the coherent intermediate
scattering function in structural glasses \cite{Binder2005}. It is
simply related to the van Hove correlation function (\ref{eq:c_ij})
calculated in Refs.~\cite{Lev2014,BarLev2015},
\begin{equation}
F\left(q,t\right)=\sum_{r}G\left(r,t\right)\E^{-\I qt},\label{eq:f_q_t}
\end{equation}
and was directly studied in the context of MBL in Ref.~\cite{Mierzejewski2016}.
For diffusive systems this quantity relaxes exponentially, $F\left(q,t\right)\sim\exp\left[-Dq^{2}t\right]$,
while for supercooled liquids the relaxation is characterized by a
Kohlrausch-Williams-Watts (KWW) law, $F\left(q,t\right)\sim\exp\left[-At^{\beta}\right]$,
and is also known as the $\beta-$relaxation \cite{Binder2005}. Taking
a Fourier transform of $F\left(q,t\right)$ with respect to time gives
the dynamical structure factor $S\left(q,\omega\right)$ which was
recently numerically studied in Ref.~\cite{Prelovsek2016a}. 

Due to the destructive nature of measurements in cold atoms experiments
it is hard to measure two-time correlation functions, however for
$q=\pi$ and a Néel state initial condition Eq.~(\ref{eq:c_q_t})
reduces to a one time-quantity, dubbed the imbalance,
\begin{equation}
I\left(t\right)=\sum_{n=1}^{L}\left(-1\right)^{n}\left\langle \hat{S}_{n}^{z}\left(t\right)\right\rangle ,
\end{equation}
which was successfully measured in a cold atoms experiment \cite{Schreiber2015a}.
An extensive ED study of the decay of the imbalance (generalized to
random product states) was performed by one of us in Ref.~\cite{Luitz2015a},
where it was found that for systems up to $L\leq28$ the imbalance
decays as a power-law superimposed on decaying oscillations,
\begin{equation}
I\left(t\right)\sim t^{-\zeta\left(W\right)},\label{eq:imbalance}
\end{equation}
moreover the dynamical exponent is subdiffusive, $\zeta\left(W\right)<1/2$
for disorder strengths $W>0.5$ and vanishes at the MBL transition.
The exponent $\zeta\left(W\right)$ is related to the dynamical exponent
as $\zeta\left(W\right)=1/z$ \footnote{This relation is not surprising since for $q=\pi$ the correlation
function (\ref{eq:c_q_t}) is very close in form to the local autocorrelation
function. For smaller $q$ it however should not be expected.}\cite{Luitz2016b}. 

\subsubsection{Entanglement entropy growth}

In Ref.~\cite{Luitz2015a} it was also demonstrated that after a
local quench the entanglement entropy grows only sublinearly with
time,
\begin{equation}
S\left(t\right)\sim t^{1/z_{\text{ent}}\left(W\right)},
\end{equation}
such that $z_{\text{ent}}\left(W\right)\geq1$. Similar results were
obtained using a light-cone tDMRG, which allows to obtain bulk transport
up to some finite time \cite{Enss2016}. This study used a binary
disorder distribution which allowed to \emph{exactly} average over
\emph{all} disorder realizations by utilizing the ancilla trick \cite{Paredes2005}. 

If one assumes that ``quasi-particles'' become entangled on ``first
encounter'' and cannot disentangle, then it is clear that entanglement
has to spread faster than transport of particles, $z_{\text{ent}}<z$
\cite{Kim2013}. Indeed, due to the conservation of the total spin,
in order to reduce the \emph{total} spin in some interval $l$, one
has to transport a ``quasi-particle'' through the interval $l$
times. On the other hand, by the assumption above, to entangle all
the ``quasi-particles'' in this interval, it is enough to transport
a ``quasi-particle'' through it only once. This implies that the
time it takes to induce entanglement in this interval is $l$ times
smaller than the transport time, $t_{\text{ent}}=t_{\text{tr}}/l$
or,
\begin{equation}
z_{\text{ent}}=z-1.\label{eq:ent_vs_trasnport}
\end{equation}
This heuristic argument establishing the connection between the dynamical
exponents $z$ and $z_{\mathrm{ent}}$ was introduced in Refs.~\cite{Vosk2014,Potter2015}.
While a microscopic derivation of the dynamical exponent $z_{\text{ent}}$
and its connection to the dynamical exponent $z$ is still missing,
using a novel diagrammatic technique a related quantity was calculated
by Aleiner \emph{et al. }\cite{Aleiner2016}. In this work an equation
of motion for the out-of-time order correlator was obtained which
is similar to equations of motion customary in the field of combustion.
The out-of-time order correlator measures the spread of disturbances
\cite{Larkin1969} and is related to the spread of entanglement \cite{Fan2016}.
Ref.~\cite{Aleiner2016} provides therefore a microscopic justification
to the ``entanglement on first encounter'' conjecture raised by
Kim and Huse \cite{Kim2013}. For a numerical verification of this
conjecture see Ref.~\cite{luitz_information_2017}. Finally, we note
that slow information transport was also observed in a sublinear power
law growth of the operator entropy of the time evolution operator
\cite{zhou_operator_2016}.

\subsubsection{\label{subsec:Transport-ness}Transport from nonequilibrium stationary
states}

Another initial condition which is useful for cold atoms experiments
is the domain wall initial condition, $\left|\uparrow\cdots\uparrow\downarrow\cdots\downarrow\right\rangle ,$
where one measures the decay of the magnetization imbalance between
two halves of the system \cite{Choi2016}. While for this initial
condition dynamics has not yet been studied in an experiment, it was
simulated using tDMRG for systems up to $L=60$ in Ref.~\cite{Hauschild2016}.
The transported magnetization across the domain wall is consistent
with a power law in accord with the scaling (\ref{eq:x2}), $M\left(t\right)\sim t^{1/z}$
\footnote{The authors actually find that a logarithmic time dependence describes
their data better, although algebraic growth fits almost equally well.}. The domain wall initial condition and the decay of the longest wavelength
excitation were also used in an ED study of energy transport \cite{Lerose2015}.
In this work, using a phenomenological diffusion equation and by extrapolating
to the thermodynamic limit, an energy diffusion coefficient was calculated.
It was argued that energy diffusion coefficient is nonzero through
a large portion of the ergodic phase \cite{Lerose2015}.

In condensed matter systems, where the real time dynamics is fast,
and therefore mostly inaccessible, transport is normally assessed
by the calculation of a stationary state current after the system
has been connected to a constant bias. Normal diffusive metals obey
Ohm's law with a stationary current which decreases as $L^{-1}$.
For ballistic metals (with mean-free path larger than the system size)
the current does not depend on the size of the system and for (perfect)
insulators the current decreases exponentially with system size. More
generally a relation between the dynamical exponent and the decay
of the current can be established using the following classical consideration
\cite{Li2003}: The time it takes for one spin to be transported from
one side of the system to the other is given by $t_{*}=L^{z}$, (which
is the generalized Thouless time defined below Eq.~(\ref{eq:x2})).
Since a fixed bias makes an extensive number of spins available for
transport, $N\propto L$, the stationary current is given by the ratio,
\begin{equation}
j\propto\frac{L}{t_{*}}=\frac{1}{L^{z-1}}.\label{eq:current}
\end{equation}
A power law dependence of the stationary current on the system size
was obtained in an open system tDMRG study of the ergodic phase \cite{Znidaric2016}.
A direct comparison to the dynamical exponent was not performed there,
but it was found that for $W>0.5$ spin transport in the system is
subdiffusive, while for $W<0.5$ it appears to be diffusive (see Section~\ref{sec:quenchmethods}
for a description of the method). On the right panel of Fig.~\ref{fig:Comparison-of-exponents.}
we present the dynamical exponent $1/z$ calculated from equation
(\ref{eq:current}) and using the data of Ref.~\cite{Znidaric2016}.
To highlight the importance of finite size effects we plot the same
data, restricting the system sizes to $L<100$. For $W<0.6$ and small
system sizes $\left(L<100\right)$ the transport appears to be faster
than diffusive, while for larger sizes $\left(L<400\right)$ the transport
slows down, yet remaining slightly faster than diffusive, even for
the largest system sizes which were used in Ref.~\cite{Znidaric2016}.
To estimate the \emph{minimal} system sizes for which the effects
of the disorder become important, Žnidari\v{c} \emph{et al.} calculate
the mean-free path in the system in second order perturbation theory
in the weak disorder. This length scales as $l\propto W^{-4/3}$\footnote{We note that this scaling is special for the XXZ model which is integrable
in the {$W=0$ } limit. For more generic nonintegrable models it
is supposed to scale as {$l\propto W^{-2}$}}, and for $W<0.6$ becomes larger $l\gg L$ then the system sizes
available in ED, making ED an inappropriate numerical tool for the
study of such a small disorder. While the results of Ref.~\cite{Znidaric2016}
are consistent with asymptotic diffusion for $W<0.6$, whether for
even larger system sizes transport slows down and eventually becomes
subdiffusive is still an open question. If this is indeed the case
it will suggest that another length scale $\tilde{l}\left(W,U\right)>l$
exists in this problem.

\subsubsection{Summary}

For the convenience of the reader we summarize all the results presented
in this subsection,
\begin{align}
x^{2}\left(t\right)\sim t^{2/z} & \qquad G_{0}\left(t\right)\sim I\left(t\right)\sim t^{-1/z}\nonumber \\
\sigma\left(\omega\right)\sim\omega^{1-2/z} & \qquad A\left(\omega\right)\sim\omega^{-\left(1-1/z\right)}\nonumber \\
C\left(t\right)\sim t^{-\tilde{D}_{2}} & \qquad I_{2}\sim\mathcal{N}^{-\tilde{D}_{2}}\nonumber \\
j\left(L\right)\sim L^{-\left(z-1\right)} & \qquad S\left(t\right)\sim t^{1/\left(z-1\right)}\label{eq:dynam_prop_summ}
\end{align}
where $2\leq z<\infty$ is the dynamical exponent, the mean-square
displacement $x^{2}\left(t\right)$ is defined in (\ref{eq:x2}),
the autocorrelation function $G_{0}\left(t\right)$ in (\ref{eq:autocorr}),
the imbalance $I\left(t\right)$ in (\ref{eq:imbalance}), the spectral
density $A\left(\omega\right)$ in (\ref{eq:spectral_density}), the
survival probability $C\left(t\right)$ in (\ref{eq:survival_prob}),
the inverse-participation ratio $I_{2}$ in (\ref{eq:ipr}) and $S\left(t\right)$
is the entanglement entropy. Here we have considered only disorder
averaged quantities. For the discussion of the corresponding typical
quantities we refer the reader to Ref.~\cite{Gopalakrishnan2015a}.

\begin{figure}
\includegraphics[width=1\columnwidth]{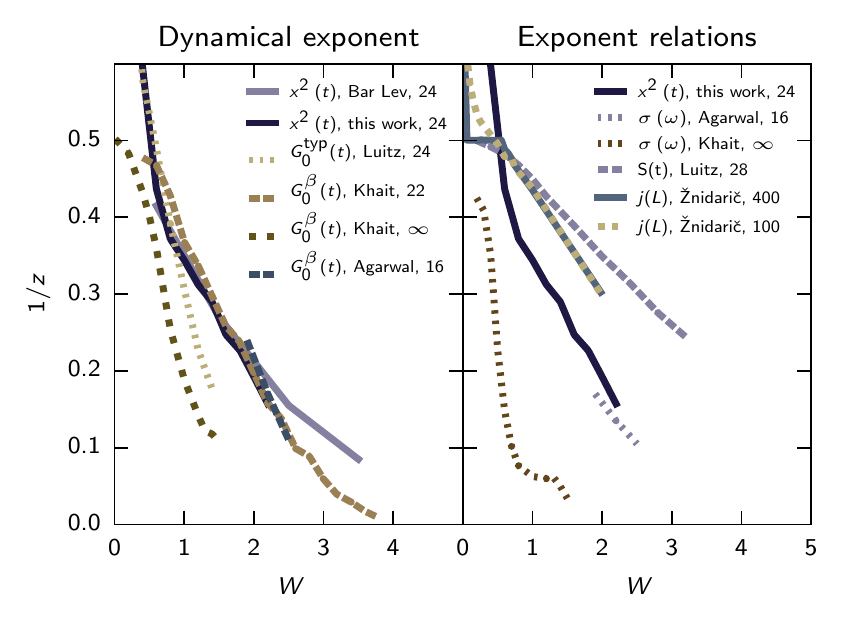}\caption{\label{fig:Comparison-of-exponents.}\textbf{Left panel:} Dynamical
exponent $1/z$. Here we show the finite size results from the Supp.
Mat. of Agarwal for $L=16$ $\protect\etal$ \cite{Agarwal2014} and
the ED result for $L=22$ Khait $\protect\etal$ \cite{Khait2016},
obtained from the decay of the infinite temperature correlation function
$G_{0}^{\beta=0}(t)$, as well as the result in the thermodynamic
limit using a variational extrapolation of recurrents (VER) by Khait
$\protect\etal$ \cite{Khait2016}. We also show our own result obtained
from the same quantity calculated for a (typical) pure state with
definite energy, where the functional form of the decay is fitted
to include the oscillations \cite{Luitz2016b}. Full lines are the
exponents as obtained from the width of the excitation $x^{2}(t)$
from Ref.~\cite{Lev2014} (infinite temperature) and calculated for
this work for the same typical pure state as mentioned before. \textbf{Right
panel: }Dynamical exponent estimated through the relations between
the exponents from Eq. (\ref{eq:dynam_prop_summ}), compared to the
best estimate from our calculation of $x^{2}(t)$ described in the
left panel. The exponent extracted from the ac conductivity $\sigma(\omega)$
calculated for a finite system size $\left(16\right)$ \cite{Agarwal2014},
we also show the thermodynamic limit result obtained from $\sigma(\omega)$
exponent calculated by Khait $\protect\etal$ using VER \cite{Khait2016}.
We include the best estimate of the exponent of the current scaling
with system size from Žnidari\v{c} $\protect\etal$ \cite{Znidaric2016}
as well as a our analysis of same data restricted to $L<100$, and
the exponent from the entanglement growth power law from Luitz $\protect\etal$
\cite{Luitz2015a}.}
\end{figure}

In Fig.~\ref{fig:Comparison-of-exponents.} we compare some of the
exponent relations which where discussed in this section and are summarized
above. We are skipping comparisons of trivial relations which follow
from a Fourier transform, such as the comparison between the exponents
of $G_{0}\left(t\right)$ and $A\left(\omega\right)$. One of the
most commonly used and assumed relations is the relation between the
decay of the autocorrelation function and the growth of the mean-square
displacement $G_{0}\left(t\right)\sim\left\langle x^{2}\left(t\right)\right\rangle ^{-1/2}$
{[}see Eq.~(\ref{eq:autocorr}){]}. While this equation clearly holds
for diffusive transport since the excitation profile $G_{r}\left(t\right)$
is asymptotically Gaussian, there is no reason why it should \emph{a
priori} hold for subdiffusive systems where asymptotic excitation
profiles can have heavy tails. This relation was indirectly tested
in Refs.~\cite{Agarwal2014,Khait2016} yielding not a very compelling
agreement. In the left panel we perform a direct test of this relation
using the dynamical exponent $1/z$ obtained from the decay of the
autocorrelation function in three different studies \cite{Agarwal2014,Khait2016,Luitz2016b},
compared to the dynamical exponent computed from the growth of the
mean-square displacement \cite{Lev2014}. We note that while the results
across the studies do agree qualitatively the quantitative discrepancy
is pretty large, sometimes as large as 100\%. We attribute this discrepancy
to the difficulty of fitting power laws to data on a limited time
domain and with superimposed oscillations, noting that the growth
of the mean-square displacement does not seem susceptible to such
problems. Another possible resolution could be that the extracted
power-laws are non asymptotic with different measures having different
sensitivity to the finite size effects. The relation between the two
exponents seems to hold well for $W<2$ with an increasing discrepancy
for stronger disorder, however the difficulty of reliably extracting
the autocorrelation exponent precludes from drawing strong conclusions.
On the right panel of Fig.~\ref{fig:Comparison-of-exponents.} we
compare the relation between the transport dynamical exponents extracted
from the mean-square displacement and the dynamical exponent obtained
from the ac conductivity, entanglement entropy and the decay of the
stationary current. Due to relation (\ref{eq:d_w}) the ac conductivity
exponents and the mean-square exponents have to agree asymptotically,
which is indeed what we observe. The exponent extracted from effectively
infinite systems is however dramatically different \cite{Khait2016}.
To the best of our knowledge relation (\ref{eq:ent_vs_trasnport})
was never explicitly verified for disordered systems. To verify this
scaling we plot the transport dynamical exponent $1/z$ and $1/\left(z_{\text{ent}}+1\right)$
as obtained in Ref.~\cite{Luitz2015a} on the right panel of Fig.~\ref{fig:Comparison-of-exponents.}.
It is clear that the relation holds only qualitatively with increasing
discrepancy for stronger disorder. Since finite size effects are negligible
for strong disorder it appears that the relation between the exponents
is more intricate than what is suggested by Eq.~(\ref{eq:ent_vs_trasnport}).
Interestingly, while there is a clear violation of the relation (\ref{eq:current})
the exponent extracted from the current coincides with the exponent
extracted from entanglement growth. We note in passing that since
entanglement cannot spread faster than particles there is an upper
bound of $1/2$ on the value of the $1/z$ exponent extracted from
entanglement growth. This means that the agreement for $W<1$ is in
some sense trivial, moreover in this regime as was pointed out in
the end of Sec.~\ref{subsec:Transport-ness}, ED results become increasingly
unreliable for such a weak disorder due to severe finite size effects.
The apparent violation of relation (\ref{eq:current}) for strong
disorder, where finite size effects are not pronounced has to be better
understood.

\section{\label{sec:Phenomenological-models}Phenomenological explanations}

\subsection{Griffiths effects}

Anomalous diffusion and subexponential relaxation of autocorrelation
functions are often associated with a failure of the central limit
theorem and the presence of heavy tailed distributions \cite{Metzler1999,Metzler2000,Sokolov2005}.
For example, for classical spin glasses a broad distribution of relaxation
times yields a subexponential relaxation of the magnetization and
spin autocorrelation functions \cite{Dhar1980,Palmer1984,Randeria1985,Kutasov1986},
and in the case of Lévy's flights, superdiffusion is a result of a
broad distribution of the hopping distances \cite{Levy1939}. A broad
distribution of relaxation times was also proposed as an explanation
for the observed subdiffusion in ergodic one-dimensional systems exhibiting
MBL \cite{Agarwal2014}. Microscopically the ``fat tail'' of the
distribution of the relaxation times follows from exponentially rare
inclusions which have exponentially long relaxation times and therefore
yield non-negligible contributions. Rare region effects on thermodynamical
phase transitions were first studied by Griffiths, who noted that
quenched disorder can make the free energy non-analytic in a finite
temperature interval \cite{Griffiths1969}. The importance of rare
spatial regions in quantum phase transitions and for dynamical properties
is even more dramatic, therefore rare region effects are overarchingly
called Griffiths effects \cite{Vojta2006}. It was proposed by Agarwal
\emph{et al. }that the subdiffusive ergodic phase, which was dubbed
the \emph{Griffiths phase}, could be effectively described by a one
dimensional random chain governed by the Master equation \cite{Agarwal2014},
\begin{equation}
\frac{\mathrm{d}P_{n}}{\mathrm{d}t}=W_{n,n-1}\left(P_{n-1}-P_{n}\right)+W_{n,n+1}\left(P_{n+1}-P_{n}\right),
\end{equation}
where $P_{n}$ is the probability to find a particle on site $n$
and $W_{n,n+1}=W_{n+1,n}>0$ are the corresponding transition rates,
which are taken to be independent random variables. This equation
had numerous appearances in various contexts. It was first considered
by Dyson, more than half a century ago, who calculated the density
of states of a random harmonic chain \cite{Dyson1953}. Replacing
$P_{n}$ by an electric potential on a node $n$ and $W_{n,n+1}$
by random conductances this model is equivalent to a random resistor
model, which was used in the hopping conductivity literature \cite{Miller1960,Kurkijarvi1973}.
It was also used as a phenomenological model to describe slow relaxation
in spin glasses \cite{Dhar1980,Palmer1984,Randeria1985,Kutasov1986}.
The properties of this model for various distributions of $W_{n}$
were extensively studied in the 80s by Alexander \cite{Alexander1981}.
It was established that the most important property of the distribution
$p\left(W\right)$ is whether its $\left\langle W^{-1}\right\rangle $
moment exists. If this moment is finite \footnote{In the language of the random resistors this means that the average
resistance is finite.} the random chain is diffusive with, $P_{0}\sim t^{-1/2}$ and $x^{2}\left(t\right)\sim t$.
Otherwise, the system is subdiffusive with an anomalous diffusion
which depends on the details of the distribution \cite{Alexander1981}.
For a power law distribution $p\left(W\right)\sim W^{-\alpha}$, which
was also the distribution considered in Ref.~\cite{Agarwal2014},
it was rigorously derived that the return probability asymptotically
scales as \cite{Bernasconi1978}, 
\begin{equation}
P_{0}\left(t\right)\sim t^{-\left(1-\alpha\right)/\left(2-\alpha\right)}\qquad\text{when}\qquad P_{n}\left(t=0\right)=\delta_{n0},
\end{equation}
with a Laplace transform, $\tilde{P}_{0}\left(\omega\right)\sim\omega^{-1/\left(2-\alpha\right)}$.
To derive the generalized diffusion coefficient one assumes the scaling
form \cite{Bernasconi1980},
\begin{equation}
\tilde{P}_{n}\left(\omega\right)\approx\tilde{P}_{0}\left(\omega\right)F\left(\frac{n}{\xi\left(\omega\right)}\right),\qquad\omega\to0,
\end{equation}
where $\xi\left(\omega\right)$ is some correlation length and $F\left(0\right)=1$.
Due to the normalization $\sum_{n}\tilde{P}_{n}\left(\omega\right)=\omega^{-1}$,
and
\begin{align}
\left(\omega\tilde{P}_{0}\left(\omega\right)\right)^{-1} & \approx\sum_{n}F\left(\frac{n}{\xi\left(\omega\right)}\right)\approx2\int_{0}^{\infty}\mathrm{d}xF\left(\frac{x}{\xi\left(\omega\right)}\right).
\end{align}
Changing the integration variables $x'=x/\xi\left(\omega\right)$
gives,
\begin{align}
\xi^{-1}\left(\omega\right)\approx & 2\omega\tilde{P}_{0}\left(\omega\right)\int_{0}^{\infty}\mathrm{d}x'F\left(x'\right).\label{eq:xi_omega}
\end{align}
Now using the relation (\ref{eq:d_omega}), one can write,
\begin{equation}
D\left(\omega\right)=\frac{1}{2}\omega^{2}\sum_{n}n^{2}\tilde{P}_{0}\left(\omega\right)=\frac{1}{2}\omega^{2}\tilde{P}_{0}\left(\omega\right)\sum_{n}n^{2}F\left(\frac{n}{\xi\left(\omega\right)}\right).
\end{equation}
Changing the variables again and using (\ref{eq:xi_omega}) yields,
\begin{align}
D\left(\omega\right) & =\frac{D_{0}}{\omega\tilde{P}_{0}^{2}\left(\omega\right)}\sim\omega^{\alpha/\left(2-\alpha\right)},\label{eq:d_w}
\end{align}
where $D_{0}=\int_{0}^{\infty}\mathrm{d}x'\,x'^{2}F\left(x'\right)/\left[8\left(\int_{0}^{\infty}\mathrm{d}x'F\left(x'\right)\right)^{3}\right]$
\cite{Bernasconi1980}. Similar relationships between the exponents
of the generalized diffusion coefficient (which is proportional to
the ac conductivity) and the return probability were obtained and
verified numerically for the XXZ model in Refs.~\cite{Agarwal2014,Gopalakrishnan2015a}.
In dimensions higher than one the random hopping model predicts asymptotically
diffusive transport, since contrary to the situation in one dimension,
links with low transition rates (high barriers) can be avoided \cite{Alexander1981a}.
Nevertheless, transport in a two-dimensional disordered Anderson-Hubbard
model (\ref{eq:anderson_hubbard}) was studied by one of us in Ref.~\cite{BarLev2015}
and found to be subdiffusive for a broad range of parameters and without
visible crossover to diffusion at the studied times.

A simplified explanation of Griffiths effects was presented by Gopalakrishnan
\emph{$\etal$} \cite{Gopalakrishnan2015a}. It assumed that the system
is composed of a collection of independently relaxing regions which
additively contribute to the decay of the autocorrelation function,
an approach familiar from the spin glass community \cite{Palmer1984}.
The autocorrelation function is taken to be,
\begin{equation}
C\left(t\right)=\left\langle \E^{-t/\tau}\right\rangle _{\tau}\equiv\int_{0}^{\infty}\mathrm{d}\tau\,p\left(\tau\right)\exp\left[-t/\tau\right],\label{eq:independent_times_assump}
\end{equation}
where $p\left(\tau\right)$ is the density of the regions with relaxation
time $\tau$. Instead of using an exponential cutoff, in Ref.~\cite{Gopalakrishnan2015a}
a sharp cutoff was assumed, namely,
\begin{equation}
C\left(t\right)=\int_{t}^{\infty}\mathrm{d}\tau\,p\left(\tau\right).\label{eq:sarang_c_t}
\end{equation}
Moreover it was assumed that the density of regions and their corresponding
relaxation rates are,
\begin{equation}
p\left(l\right)\sim\E^{-\gamma l^{d}}\qquad\tau\left(l\right)=\E^{\alpha l},\label{eq:rare_reg_dens_assump}
\end{equation}
where $\alpha$ and $\gamma$ are constants, $l$ is the linear dimension
of the region and $d$ is the dimension of the system. We note that
these assumptions are reasonable only for autocorrelation functions
which do \emph{not} decay to zero in the MBL phase, since only for
these correlation functions the relaxation time diverges at the transition.
An existence of rare \emph{spatial} regions created by rare local
realizations of the disordered potential is also implicitly assumed.
Therefore (\ref{eq:rare_reg_dens_assump}) is not expected to hold
for deterministic potentials such as the Aubry-André model \cite{Gopalakrishnan2015a}.
From (\ref{eq:rare_reg_dens_assump}) one can calculate the corresponding
distribution function of the relaxation times is,
\begin{equation}
p\left(\tau\right)\sim\frac{1}{\tau}\exp\left[-\frac{\gamma}{\alpha^{d}}\ln^{d}\left(\tau\right)\right].
\end{equation}
For $d=1$ the distribution of the relaxation times is given by a
power law and the integral in Eq.~(\ref{eq:sarang_c_t}) can be evaluated,
\begin{equation}
C\left(t\right)\sim t^{-\gamma/\alpha},
\end{equation}
which yields subdiffusive relaxation for $\gamma<2\alpha$, and a
superdiffusive relaxation otherwise. For $d\geq2$ approximating the
integral (\ref{eq:sarang_c_t}) by the largest integrand yields,
\begin{equation}
C\left(t\right)\approx\exp\left[-\frac{\gamma}{\alpha^{d}}\ln^{d}t\right],\label{eq:sarang_result}
\end{equation}
which relaxes faster than any power law, yet slower than exponential
\cite{Gopalakrishnan2015a}. The procedure above is somewhat arbitrary
since it strongly depends on the assumed distributions (\ref{eq:rare_reg_dens_assump}).
While they have a clear physical meaning, a more microscopic justification
would be preferable. Another problem with this approach is that it
neglects the dependence between the different regions. While this
is a reasonable approximation in higher dimensions, for one dimensional
systems it overestimates the relaxation rate since neighboring rare
regions should suppress the relaxation of their surrounding. For example,
naively calculating the typical autocorrelation function,
\begin{equation}
C\left(t\right)\sim\exp\left[-t/\left\langle \tau\right\rangle \right],
\end{equation}
yields exponential relaxation, since the average relaxation time $\left\langle \tau\right\rangle $
is finite even for one-dimensional systems (for $\alpha<\gamma<2\alpha$).
To correct for this discrepancy one has to take into account the dependence
between the regions, which was heuristically performed in Ref.~\cite{Gopalakrishnan2015a}.
For a more detailed discussion on the Griffiths effects we refer the
reader to Ref.~\cite{Agarwal2016_review}.

\subsection{Phenomenological Renormalization Group}

Several real space phenomenological renormalization group (RG) approaches
were developed to study the universal features of the MBL transition
\cite{Vosk2014,Potter2015,Zhang2016} (for a review see Ref.~\cite{Parameswaran2016b}).
A real space coarse-graining of the system is performed, accompanied
by a subdivision into ergodic and nonergodic regions. The main difference
between the approaches is the way in which these regions are identified
and combined during the RG steps. The simplified RG scheme presented
in Ref.~\cite{Zhang2016} starts from a random sequence of ergodic
and nonergodic regions of different lengths according to some initial
distribution. The RG step consists then of identifying the shortest
region and merging it with the two neighboring regions. The new region
will be ergodic or nonergodic according to a majority rule of the
three regions. Using these RG rules the critical distribution can
be derived as also the limiting distributions of the ergodic and nonergodic
phases. Interestingly, this RG procedure points to a fractal nature
of the nonergodic inclusions in the ergodic phase.

This procedure can be viewed as a maximally simplified version of
the more detailed RG approach proposed in Ref.~\cite{Vosk2014}.
In this work the regions were characterized by their many-body level
spacing $\Delta_{i}$ and an entanglement rate $\Gamma_{i}$, which
is inversely proportional to the time entanglement spreads across
the region. In addition, a set of two-region parameters $\Delta_{ij}$
and $\Gamma_{ij}$ is kept, which correspond to the parameters one
would obtain if two neighboring regions were merged. The RG step then
consists of merging two regions with the fastest (combined) entanglement
rate $\Gamma_{ij}$, after which the coupling $\Gamma_{k;ij}$ to
the neighboring region $k$ is renormalized. If the coupling between
the regions is effective, namely $\Gamma_{ij}\gg\Delta_{ij}$ and
$\Gamma_{jk}\gg\Delta_{jk}$ the rates are renormalized according
to (a) $\Gamma_{ij;k}^{-1}=\Gamma_{ij}^{-1}+\Gamma_{jk}^{-1}-\Gamma_{j}^{-1}$,
removing double counting of the transversal time of region $j$. On
the other hand, if the coupling is ineffective, the new entanglement
rate is obtained from second order perturbation theory via (b) $\Gamma_{ij;k}=\Gamma_{ij}\Gamma_{jk}/\Gamma_{j}$.
For the case when only one of the links is effective there is some
arbitrariness in the choice of the rules. If the effective link is
between two ergodic regions the rule (a) is used and when the effective
coupling is between an ergodic and nonergodic region rule (b) is used.
While this RG flow does not permit to directly obtain the entanglement
entropy between blocks, it was estimated from the lifetime of product
states $1/\Gamma_{ij}$ and the number of accessible states at a given
energy $1/\Delta_{ij}$, capturing correctly the transition from a
volume law scaling in the ergodic phase to an area law scaling in
the nonergodic phase and accompanied by a broad distribution of the
entanglement entropy close to the critical point. Transport properties
were studied by considering the scaling of the typical transport time
$l_{i}/\Gamma_{i}$ with the length of the region $l_{i}$. It was
found that transport in the ergodic phase in the vicinity of the critical
point is subdiffusive with a dynamical exponent smaller then $1/2$,
while the entanglement growth is sublinear in time.

In Ref.~\cite{Potter2015} a similar real space RG method was proposed.
Unlike the procedures discussed above, in this approach only resonant
regions are combined, and the nonresonant regions are left intact.
This removes the arbitrariness in RG rules when an ergodic and nonergodic
regions have to be combined. Each region has a length $l_{i}$ and
a bandwidth $\Lambda_{i}$, and all the regions are coupled using
a coupling strength $\Gamma_{i,j}$ which exponentially decreases
with the distance between the regions. After two regions $i$ and
$j$ of length $l_{i}$ ($l_{j}$) are merged, the coupling $\Gamma_{ij}$,
the bandwidth $\Lambda_{ij}=\Lambda_{i}+\Lambda_{j}+\Gamma_{ij}$
and the level spacing $\delta_{ij}=\Lambda_{ij}/(2^{l_{i}+l_{j}}-1)$
are renormalized. The coupling to other regions $\Gamma_{ij;k}$ has
to be updated too. This is the central step of the renormalization
procedure and involves analyzing all possible processes coupling the
regions $i$, $j$ and $k.$ It depends on the energy mismatch $\delta E_{ik}$
of the individually merged regions $i,k$ or $j,k$, which is defined
as the minimal energy difference in the spectrum of the merged regions.
If $\Gamma_{ik}\ll\delta E_{ik}$, then the renormalized coupling
can be computed in second order perturbation theory as $\Gamma_{ij;k}=\Gamma_{ik}\Gamma_{ij}/\delta E_{ik}$,
otherwise all three regions are strongly coupled and the coupling
is given by the addition of the two transport times $\Gamma_{ij;k}^{-1}=\Gamma_{ki}^{-1}+\Gamma_{ij}^{-1}$.
Out of all possible processes the largest coupling is retained. Iterating
this procedure until all resonant regions are exhausted generates
the largest resonant ``backbone'' in the system. If this backbone
percolates across the entire system the system will be ergodic, and
nonergodic otherwise.

Potter $\etal$ \cite{Potter2015} identify subdiffusive transport
in the ergodic phase from a broad power law distribution $p(\tau_{ij})\propto\tau_{ij}^{-\alpha}$
of the transport time scales $\tau_{ij}=1/\Gamma_{ij}$ with a divergent
mean ($1<\alpha<2$). The authors argue that transport can be viewed
as a random walk (with broadly distributed hopping rates) on the resonant
backbone which yields a dynamical exponent of transport of $z=\alpha/\left(\alpha-1\right)$.
In contrast to transport, entanglement is not a conserved quantity
and spreads deterministically across the chain, thus leading to a
different dynamical exponent $z_{\text{ent}}=1/\left(\alpha-1\right)=z-1$.

We emphasize that the RG procedures described above are completely
phenomenological and are not derived from any microscopic model. A
completely different real space RG method, which \emph{is} microscopically
based has been introduced in Refs.~\cite{Vosk2013a,Pekker2014},
generalizing the idea of strong disorder RG approaches for ground
state properties \cite{Dasgupta1980,Fisher1992,Fisher1994,Fisher1995}.
We refer the reader to the original works in Refs. \cite{Vosk2013a,Pekker2014,Agarwal2015}. 

\section{\label{sec:Numerical-Methods}Numerical Methods}

In this section we will describe some of the numerically exact and
approximate methods which can be used to study the many-body problem.
We note that this methods are not limited to the prototype model we
have considered in Sec.~\ref{sec:Models}. Through this section we
designate the Hilbert space dimension by $\mathcal{N}$ and note that
it scales exponentially with the system size, $L$, e.g. for spin-$\frac{1}{2}$
systems it grows like $\mathcal{N}=2^{L}$.

\subsection{Exact methods for nonequilibrium time evolution}

\label{sec:quenchmethods}

\subsubsection{Full diagonalization}

Studying the properties of strongly correlated quantum systems is
a formidable problem and an exact treatment of models is often possible
only numerically. In a typical nonequilibrium numerical experiment
the system is prepared in some initial state $\ket{\psi_{0}}$, which
is not an eigenstate of the Hamiltonian matrix $\mat H\in\mathbb{C}^{\mathcal{N}\times\mathcal{N}}$.
The propagation of the state in time can be performed by exactly diagonalizing
the Hamiltonian $\mat H=\mat{UDU}^{\dagger}$, where the matrix $\mat D=\mathrm{diag}\left(E_{0},\dots,E_{\mathcal{N}-1}\right)$
is diagonal and contains the eigenvalues $E_{n}$ of $\mat H$ while
the columns of $\mat U$ correspond to the orthonormal eigenvectors
$\ket n$ ($\cf$ Sec. \ref{sec:ethmethods} for more details). The
solution of the Schrödinger equation for the time dependent wave function
is given by $\ket{\psi(t)}=\sum_{n}\mathrm{e}^{-iE_{n}t}\ket n\left\langle n|\psi_{0}\right\rangle $,
where $\left\langle n|\psi_{0}\right\rangle $ are the coefficients
of the initial wave function in the eigenbasis of the Hamiltonian.
If the initial wavefunction is represented as a vector $\vec{x_{0}}\in\mathbb{C}^{\mathcal{N}}$
in the computational basis, then the wavefunction at time $t$ is
obtained by $\vec x(t)=\mat{U^{\dagger}\E^{-\I\mat Dt}\mat U\vec{x_{0}}}$,
where the matrix exponential of the diagonal matrix $\mat D$ is trivial.
While this method is able to access arbitrarily long times, it is
limited by the exponential growth of the Hilbert space with the size
of the system. The computational complexity of this method is about
$\mathcal{O}\left(\mathcal{N}^{3}\right)$, and the required memory
is $\mathcal{O}\left(\mathcal{N}^{2}\right)$, effectively limiting
the applicability of the method to lattice sizes of $\lesssim16$
(if the system has no additional symmetries).

\subsubsection{Krylov space time evolution}

Nautts and Wyatt realized in 1983 \cite{nauts_new_1983} that one
can avoid the full diagonalization of the Hamiltonian by using a Krylov
space method to calculate the exact time evolution $\ket{\psi\left(t+\Delta t\right)}=\mathrm{e}^{-\mathrm{i}\hat{H}\Delta t}\ket{\psi\left(t\right)}$.
Using the series expansion of the exponential, we obtain 
\begin{equation}
\mathrm{e}^{-\mathrm{i}\hat{H}\Delta t}\ket{\psi\left(t\right)}=\sum_{k=0}^{\infty}\frac{(-\mathrm{i}\Delta t)^{k}}{k!}\hat{H}^{k}\ket{\psi\left(t\right)},\label{eq:taylorexp}
\end{equation}
which for very small $\Delta t$ may be used directly, but is numerically
inherently unstable \cite{moler_nineteen_2003}. To obtain a more
stable expansion, it is useful to note that the wave function at time
$t+\Delta t$ is well approximated by a vector in the $m$ dimensional
Krylov space $\mathcal{K}_{m}=\mathrm{span}\left(\ket{\psi\left(t\right)},\hat{H}\ket{\psi\left(t\right)},\hat{H}^{2}\ket{\psi\left(t\right)},\dots,\hat{H}^{m-1}\ket{\psi\left(t\right)}\right)$.
Based on this observation, an orthonormal basis of the Krylov space
$\mathcal{K}_{m}$ is iteratively generated using the numerically
stable Arnoldi algorithm \cite{arnoldi_principle_1951} and the Hamiltonian
is projected into this subspace after $m$ iterations, yielding \cite{moler_nineteen_2003}
\begin{equation}
\mathrm{e}^{-\mathrm{i}\hat{H}\Delta t}\ket{\psi\left(t\right)}\approx\mat{V_{m}}\mathrm{e}^{-\mathrm{i}\mat{V_{m}^{\dagger}}\mat H\mat{V_{m}}\Delta t}\vec{e_{1}}.\label{eq:Krylov-timeevolution}
\end{equation}
Here the columns of the matrix $\mat{V_{m}}\in\mathbb{C}^{\mathcal{N}\times m}$
contain the orthonormal basis vectors of the Krylov space $\mathcal{K}_{m}$,
and $\vec{e_{1}}\in\mathbb{C}^{m}$ is the first unit vector (which
corresponds to $\ket{\psi\left(t\right)}$ in the new basis as this
is the first column of $\mat{V_{m}}$). Note that the matrix $\mat{V_{m}^{\dagger}}\mat H\mat{V_{m}}\in\mathbb{C}^{m\times m}$
is an upper Hessenberg matrix of \emph{small} dimension $m\ll\mathcal{N}$,
which can be readily exponentiated using standard methods, such as
a Padé approximation or a rotation to the eigenbasis. The dimension
of the Krylov space $m$ is continuously increased until the wavefunction
is converged to the desired precision. This method is very powerful
since it exploits the sparseness of the Hamiltonian and does not require
its full diagonalization. The memory requirements and the computational
complexity of this approach are much more favorable compared to exact
diagonalization (see Table~\ref{tab:dynamics}). This approach has
been used to study the nonequilibrium dynamics of spin chains with
lengths up to $L=28$ \cite{Luitz2015a,Lerose2015,Rehn2016}.

\subsubsection{tDMRG}

An independent approach to obtain the numerically exact time evolution
of the wave function after a quench employs a representation of the
wave function as a \emph{matrix product state} (MPS). For models for
which the Hamiltonian can be decomposed into terms which operate on
two adjacent sites, which we will call bond terms, the propagation
of the wavefunction in time is quite straightforward. In order to
calculate the wavefunction after a time step $\Delta t$, the Hamiltonian
is decomposed into two terms $\hat{H}=\hat{H}_{\mathrm{even}}+\hat{H}_{\mathrm{odd}}$,
where $\hat{H}_{\mathrm{even}}$ and $\hat{H}_{\mathrm{odd}}$ contain
even (odd) bond terms. While $\hat{H}_{\mathrm{even}}$ and $\hat{H}_{\mathrm{odd}}$
need not commute, all terms within $\hat{H}_{\mathrm{even}}$ $\left(\hat{H}_{\mathrm{odd}}\right)$
commute with each other. Therefore a Trotter decomposition of the
time evolution operator {[}\textit{cf.} Eq.~(\ref{eq:taylorexp}){]}
can be used to time evolve the state by $\Delta t$. The simplest
decomposition leads to an error of $\Delta t^{2}$,
\begin{equation}
\mathrm{e}^{-\mathrm{i}\hat{H}\Delta t}=\mathrm{e}^{-\mathrm{i}\hat{H}_{\mathrm{even}}\Delta t}\mathrm{e}^{-\mathrm{i}\hat{H}_{\mathrm{odd}}\Delta t}+\mathcal{O}\left(\Delta t^{2}\right),\label{eq:trotter1}
\end{equation}
however, higher order decompositions can be used (\textit{cf.} Refs.~\cite{Suzuki1990,Sornborger1999}.
Note that as the matrix exponentials on the right hand side of Eq.\ (\ref{eq:trotter1})
contain only commuting terms, they can be applied sequentially in
one DMRG sweep. During the application of the odd and even bond terms
to the MPS, the bond dimension of the MPS is adaptively truncated
such that the discarded weight, \textit{i.e.} the sum of the discarded
singular values does not exceed a certain threshold. As the number
of retained singular values directly limits the maximal entanglement
entropy that can be encoded by the MPS, it is clear that the bond
dimension of the MPS has to grow exponentially with the entanglement
entropy. In the ergodic phase the entanglement entropy after a quench
from a product state grows as a power law in time \cite{Luitz2015a},
thus leading to a stretched exponential growth of the bond dimension
with time and effectively limiting this method to short times. In
the MBL phase the situation is more favorable since the entanglement
entropy grows logarithmically in time \cite{Znidaric2008,bardarson_unbounded_2012,Serbyn2013b,Deng2016a},
leading to only a power law growth of the bond dimension. For details
on the method, we refer the reader to the original papers on this
adaptive method by Vidal \cite{vidal_efficient_2003,vidal_efficient_2004}
and to a review on DMRG \cite{schollwock_density-matrix_2005}.

\subsubsection{tDMRG for open systems}

The study of transport properties can be conveniently performed by
opening the system and attaching it to two (or more) leads with a
different chemical potential. For Markovian leads and under additional
approximations the evolution of the density matrix of the system $\hat{\rho}$
can be described using the Lindblad equation \cite{lindblad_generators_1976},

\begin{equation}
\frac{\mathrm{d}}{\mathrm{d}t}\hat{\rho}\equiv\mathcal{\hat{L}}\hat{\rho}=\I\left[\hat{\rho},\hat{H}\right]+\gamma\sum_{k}\left(\left[\hat{L}_{k}\hat{\rho},\hat{L}_{k}^{\dagger}\right]+\left[\hat{L}_{k},\hat{\rho}\hat{L}_{k}^{\dagger}\right]\right),\label{eq:lindblad}
\end{equation}
where the Lindblad operators $\hat{L}_{k}$, describe the coupling
between the system and the bath. The Lindlad equation can be numerically
solved using tDMRG \cite{prosen_matrix_2009,znidaric_dephasing-induced_2010}.
The evolution of the density matrix is performed by increasing the
size of the Hilbert space and considering the density matrix operator
$\hat{\rho}(t)$ as a vector in the enlarged space, whose time evolution
is governed by the Liouvillian $\mathcal{\hat{L}}$ (\emph{cf.}~(\ref{eq:lindblad}).
In this enlarged Hilbert space the tDMRG method described in the previous
section can be applied and it appears that in many cases the entanglement
entropy in the operator space, which governs the bond dimension and
therefore the efficiency of the method, grows slowly in time due to
decoherence effects caused by the Markovian bath. This favorable computational
complexity allows to reach the nonequilibrium steady state (NESS)
at long times, and to calculate the stationary magnetization and the
stationary current \cite{Znidaric2016}. If the bias between left
and right leads is small enough, the system is in the linear response
regime and the current in the NESS reveals the nature of the transport.
Žnidari\v{c} \emph{et al.} have used this method to study the dynamical
exponent in the random XXZ chain for system sizes up to $L=400$,
arguing in favor of a transition between a diffusive and a subdiffusive
regime at weak disorder strength \cite{Znidaric2016}. For strong
disorder this method becomes increasingly expensive since the time
it takes to reach the stationary state increases.

\subsubsection{Dynamical typicality}

Quantum typicality can be viewed as a geometrical concept that follows
from Lévy's Lemma. This lemma states that for a Lipschitz-continuous
function $f:S^{(2n-1)}\to\mathbb{R}$ defined on the surface of a
high dimensional sphere, any point $x\in S^{(2n-1)}$ drawn randomly
from a uniform distribution on the sphere will yield $f(x)$ exponentially
close to the average of $f$ over the surface of the sphere \cite{ledoux_concentration_2001}. 

Since any normalized quantum state in a finite dimensional Hilbert
space of dimension $\mathcal{N}$ can be represented as a point on
the surface of a $2\mathcal{N}$ dimensional unit hypersphere $S^{(2\mathcal{N}-1)}$,
and the trace of an operator $\mathrm{Tr}\,\hat{O}$ can be written
as the integral of $\hat{O}$ over the surface of this sphere with
respect to the Haar measure, it follows that the expectation value
of $\hat{O}$ for any random pure state $\ket{\psi}$ on the sphere
is exponentially close to the value of the trace, if the operator
can be represented as a Lipshitz-continuous function on the hypersphere.
This is typically the case for local operators. More precisely, the
probability to deviate from the trace by more than $\epsilon>0$ is
exponentially small, 
\begin{equation}
P\left[\left|\mathrm{Tr}\,\hat{O}-\bra{\psi}\hat{O}\ket{\psi}\right|\geq\epsilon\right]\leq a\E^{-b\mathcal{N}\epsilon^{2}},
\end{equation}
with positive constants $a$ and $b$. This means that the trace of
the operator $\hat{O}$ can be replaced by an expectation value obtained
from a random pure state $\ket{\psi}$ to a precision which improves
for larger Hilbert space dimension \cite{Popescu2006,goldstein_canonical_2006,Reimann2007,Bartsch2009,gelman_simulating_2003,sugiura_thermal_2012,sugiura_canonical_2013}.
To illustrate its application, we demonstrate how it can be used to
calculate a correlation function in the canonical ensemble:

\begin{equation}
C_{O}^{\beta}(t)=\frac{1}{Z}\mathrm{Tr}\left(\E^{-\beta\hat{H}}\hat{O}\left(t\right)\hat{O}\right)\approx\frac{1}{\left\langle \beta|\beta\right\rangle }\left\langle \beta\left|\hat{O}\left(t\right)\hat{O}\right|\beta\right\rangle ,
\end{equation}
where $\beta$ is the inverse temperature. Here we have used the cyclic
property of the trace, applied Lévy's lemma substituting the trace
by an expectation value of a random state $\ket{\psi}$ and finally
defined $\ket{\beta}\equiv\E^{-\frac{\beta}{2}\hat{H}t}\ket{\psi}$
($\cf$ \cite{sugiura_canonical_2013}). This state can be efficiently
calculated by imaginary time evolution of the pure state $\ket{\psi}$,
followed by real time evolution to obtain the correlation function.
This task can be performed either by integration of the Schrödinger
equation using the recently developed Runge-Kutta schemes \cite{elsayed_regression_2013,steinigeweg_pushing_2014},
or by utilizing the Krylov space technique discussed in the previous
section. All these approaches can be applied without full diagonalization
of the Hamiltonian and rely solely on the ability to calculate the
matrix vector product $\mat H\vec x$, which can be achieved even
without storing the sparse Hamiltonian matrix $\mat H$. The memory
requirement is thus reduced to the size of a few Hilbert space vectors.
We remark that in Ref.~\cite{Luitz2016b}, we have applied a simplified
version of this approach by creating a microcanonical typical state,
which we called ``energy squeezed state''. This state was constructed
by applying powers of $(\hat{H}-\sigma)^{2}$ to a random vector in
the Hilbert space to suppress contributions from eigenstates far away
from the target energy $\sigma$.

\subsection{Exact methods for eigenstates calculation}

\label{sec:ethmethods}

The absence of transport is the defining property that distinguishes
the MBL phase from the ergodic phase. Transport can be efficiently
studied using the numerical methods described in the previous section.
However, the MBL transition can also be viewed as an eigenstate phase
transition ($\cf$ Ref. \cite{Parameswaran2016b}), which is characterized
by strikingly different properties of the eigenstates of the Hamiltonian
in the ergodic and nonergodic phases, but also by different statistical
properties of the energy spectrum. To study this aspect numerically,
it is therefore important to be able to calculate some or all or the
eigenvalues and eigenstates of the Hamiltonian.

\subsubsection{Full diagonalization}

Clearly, the first choice to obtain exact high energy eigenstates
\footnote{This means typically states from the center of the spectrum where
the density of states is exponentially large.} is the full diagonalization of the Hamiltonian. This is typically
done using the standard protocol for dense matrices: First, the Hamiltonian
is brought to tridiagonal form by Householder reflections, the tridiagonal
matrix is then diagonalized by efficient algorithms, such as the divide
and conquer \cite{cuppen_divide_1981} approach or using multiple
relatively robust representations \cite{dhillon_new_1998}, and finally
the obtained eigenvectors are transformed back to the original basis
using the Householder transformations of the first step in inverse
order. This recipe is available in highly optimized \texttt{LAPACK}
implementations for many architectures, yielding high precision results.
Since these methods are based on dense matrices, they require $\mathcal{O}(\mathcal{N}^{2})$
memory to store the dense matrix (in addition to $\mathcal{O}(\mathcal{N}^{2})$
work space for the divide and conquer algorithm) as well as to store
all eigenvectors of the result. The computational complexity is dominated
by the Householder step and it scales as $\mathcal{O}(\mathcal{N}^{3})$.

\subsubsection{Subset diagonalization}

For some applications only a few eigenstates and eigenvalues within
some interval $[E_{-},E_{+}]$ are required. The \emph{shift-invert}
method is the current state-of-the-art method to tackle this problem.
It is closely related to inverse iteration and relies on the fact
that the extremal eigenvalues of $(\mat H-\sigma)^{-1}$ correspond
to the eigenvalues of $\mat H$ which are closest to the \emph{target
energy} $\sigma$. Furthermore, while the typical scaling of the level
spacing of the original problem is $\mathcal{N}^{-1}$, the level
spacing in the corresponding part of the transformed spectrum is $\mathcal{N}$.
Therefore standard Krylov space methods, such as the Lanczos algorithm
or the Arnoldi iteration can be efficiently used to obtain several
of the highest and lowest lying eigenvalues and eigenstates of the
transformed problem. The eigenvalues of the transformed problem are
trivially transformed back to the original problem and the eigenvectors
are invariant under this transformation and therefore are directly
obtained.

The hardest part of this procedure is the repeated calculation of
the action of $(\mat H-\sigma)^{-1}$ on vectors during the application
of Krylov space methods. This is typically done by first decomposing
the Hamiltonian into upper and lower triangular matrices (the LU decomposition)
such that $(\mat H-\sigma)=\mat{LU}$ using Gaussian elimination.
Subsequently $\mat{LU}\vec x=\vec b$ is solved, yielding $\vec x=(\mat H-\sigma)^{-1}\vec b$.
For the calculation of the $\mat{LU}$ decomposition, efficient implementations
that exploit the sparseness of $\mat H$ are available. For example,
for distributed memory machines the MUMPS \cite{amestoy_fully_2001,amestoy_hybrid_2006}
and SuperLU \cite{li_overview_2005} libraries can be used. The shift-invert
technique has been used to map the energy-disorder phase-diagram of
the XXZ model (\ref{eq:xxz}) by one of us \cite{Luitz2015}.

\subsubsection{Excited state DMRG}

While matrix product state methods are extremely successful for the
study of ground-state properties of one dimensional systems, they
were typically not employed to find matrix product state (MPS) representations
of highly excited eigenstates of the Hamiltonian, which only recently
became of interest in the context of MBL. Due to the area law entanglement
of eigenstates in the MBL phase it is natural to expect that such
a representation will be efficient. In the ergodic phase, on the other
hand it is highly inefficient due to the volume law scaling of entanglement.
Recently, several groups have developed methods to find highly excited
eigenstates with MPS based methods, which work well in the MBL phase.
Yu $\etal$ developed SIMPS (shift invert matrix product state method)
that relies on the idea of the shift-invert method \cite{yu_finding_2017},
trying to find an MPS which best approximates the eigenstate of $(\hat{H}-\sigma)^{-1}$
with the largest magnitude eigenvalue. For this purpose, Yu $\etal$
propose a method, which is used to iteratively apply $\left(\hat{H}-\sigma\right)^{-1}$
to an initial MPS $\ket{\psi_{0}}$. This iteration is converging
exponentially fast (at sufficiently large bond dimension) to an eigenstate
of $(\hat{H}-\sigma)^{-1}$ which corresponds to an excited state
of the Hamiltonian with an eigenvalue close to $\sigma.$ The key
insight of the method is that the next iteration $\ket{\psi_{k}}$
could be thought of as a solution of a variational minimization problem
$\left\Vert \left(\hat{H}-\sigma\right)\ket{\psi_{k}}-\ket{\psi_{k-1}}\right\Vert ^{2}$
where $\ket{\psi_{k-1}}$ corresponds to the previous iteration. The
solution of this problem is obtained using an adapted version of the
DMRG sweeping protocol. This method was subsequently used by Serbyn
\emph{et al.} to study the entanglement spectrum in the MBL phase
\cite{serbyn_power-law_2016}.

As SIMPS relies on a modification of DMRG, a simpler method was proposed
to obtain MPS representations of highly excited eigenstates in DMRG
\cite{yu_finding_2017,Khemani}. Instead of trying to invert the global
Hamiltonian, it is based on the local effective Hamiltonians appearing
during the DMRG sweep. The local matrices of the MPS are updated by
choosing an excited eigenstate of the local Hamiltonian and yielding
an eigenstate of the global Hamiltonian not necessarily close to a
target energy. The procedures of selecting the eigenstates of the
effective Hamiltonian either relies on choosing the eigenstate with
energy closest to the energy of the previous MPS \cite{yu_finding_2017},
or on the property of the MBL phase that the eigenstates are very
close to product states, such that an eigenstate which has the maximal
overlap with the MPS of the previous iteration is selected \cite{Khemani}.
This approach circumvents the general problem that for large system
sizes, the energy level spacing of the full spectrum becomes smaller
than machine precision. Other approaches exploit the idea of ``spectum
folding'', noting that the groundstate of $(\hat{H}-\sigma)^{2}$
corresponds to the eigenstate of $\hat{H}$ with an eigenvalue closest
to $\sigma$ \cite{Lim,Kennes2015}. All these methods are currently
employed only in the MBL phase and it is unclear whether they will
be useful to study the physics very close to the transition or in
the ergodic phase due to the presence of high entanglement entropy.
We note that in the fully MBL phase it has been argued that the complete
spectrum can be encoded in a single matrix product operator \cite{Pekker2014b,chandran_spectral_2015,Pollmann2016,Wahl2016}.

\subsubsection{Quantum Monte-Carlo}

Quantum Monte-Carlo (QMC) methods are extremely useful to study equilibrium
finite temperature physics as well as low temperature properties provided
that there is no sign problem. However, they are not able to resolve
single eigenstates which are not groundstates. Inglis and Pollet have
recently made progress in this direction by effectively shifting the
energies of the original Hamiltonian to make a highly excited eigenstate
the new groundstate \cite{inglis_accessing_2016}. This is achieved
by exploiting the fact that eigenstates of MBL systems can be labeled
by eigenvalues of on extensive number of conserved quasilocal quantities
\cite{Serbyn2013a,Huse2013}. This method is conceptually new and
very promising, although its current implementation relies on an approximate
construction of the quasilocal conserved operators with constraints
on their analytic form to make them compatible with the worm algorithm.
By construction, this method is only useful to study the MBL phase.

\begin{table}
\begin{tabular}{c|cccc}
\hline 
\textsf{\footnotesize{}Time evolution } & \textsf{\footnotesize{}memory } & \textsf{\footnotesize{}CPU } & \textsf{\footnotesize{}$L$ } & \textsf{\footnotesize{}time }\tabularnewline
\hline 
\textsf{\footnotesize{}ED } & \textsf{\footnotesize{}$\mathcal{O}(\mathcal{N}^{2})$ } & \textsf{\footnotesize{}$\mathcal{O}(\mathcal{N}^{3})$ } & \textsf{\footnotesize{}$\approx18$ } & \textsf{\footnotesize{}$\infty$ }\tabularnewline
\textsf{\footnotesize{}Krylov } & \textsf{\footnotesize{}$\mathcal{O}(m\mathcal{N})$ } & \textsf{\footnotesize{}$\mathcal{O}(LN_{t}\mathcal{N})$ } & \textsf{\footnotesize{}$\approx30$ } & \textsf{\footnotesize{}$t_{\mathrm{max}}$ }\tabularnewline
\textsf{\footnotesize{}tDMRG } & \textsf{\footnotesize{}$\mathcal{O}(L\chi^{2})$ } & \textsf{\footnotesize{}$\mathcal{O}(LN_{t}\chi^{3})$} & \textsf{\footnotesize{}$>100$} & \textsf{\footnotesize{}$\approx\mathcal{O}(\ln\chi)$ }\tabularnewline
\hline 
\end{tabular}

\caption{\label{tab:dynamics}Comparison of numerical methods for time evolution.
Here $m$ is the number of Krylov vectors, $N_{t}$ is the number
of time steps, $\chi$ is the bond dimension, $\mathcal{N}$ is the
Hilbert space dimension and $L$ is the system size.}
\end{table}

\begin{table}
\textsf{\footnotesize{}}%
\begin{tabular}{l|cccc}
\hline 
\textsf{\footnotesize{}Eigenstates } & \textsf{\footnotesize{}memory } & \textsf{\footnotesize{}CPU } & \textsf{\footnotesize{}$L$ } & \textsf{\footnotesize{}comment }\tabularnewline
\hline 
\textsf{\footnotesize{}ED } & \textsf{\footnotesize{}$\mathcal{O}(\mathcal{N}^{2})$ } & \textsf{\footnotesize{}$\mathcal{O}(\mathcal{N}^{3})$ } & \textsf{\footnotesize{}$\approx18$ } & \textsf{\footnotesize{}shared memory}\tabularnewline
\textsf{\footnotesize{}Shift-Invert } & \textsf{\footnotesize{}$\mathcal{O}(\mathcal{N}^{2})$ } & \textsf{\footnotesize{}$\mathcal{O}(\mathcal{N}^{3})$ } & \textsf{\footnotesize{}$\approx22$ } & \textsf{\footnotesize{}distributed memory}\tabularnewline
\textsf{\footnotesize{}ES-DMRG } & \textsf{\footnotesize{}$\mathcal{O}(L\chi^{2})$ } & \textsf{\footnotesize{}$\mathcal{O}(L\chi^{3})$} & \textsf{\footnotesize{}$\approx100$ } & \textsf{\footnotesize{}MBL only}\tabularnewline
\textsf{\footnotesize{}SI-DMRG } & \textsf{\footnotesize{}$\mathcal{O}(L\chi^{2})$ } & \textsf{\footnotesize{}$\mathcal{O}(L\chi^{4})$} & \textsf{\footnotesize{}$\approx100$ } & \textsf{\footnotesize{}MBL only}\tabularnewline
\textsf{\footnotesize{}QMC} & \textsf{\footnotesize{}$\mathcal{O}(L)$ } & \textsf{\footnotesize{}$\mathcal{O}\left(\frac{1}{\sigma_{\mathrm{MC}}^{2}}\right)$ } & \textsf{\footnotesize{}$\approx100$ } & \textsf{\footnotesize{}MBL only, approx.}\tabularnewline
\hline 
\end{tabular}\caption{\label{tab:statics}Comparison of numerical methods for the extraction
of high energy eigenstates. The notation is the same as in Table~\ref{tab:dynamics},
$\sigma_{MC}^{2}$ is the target variance of QMC result. The QMC method
mentioned here is currently not a strictly exact method \cite{inglis_accessing_2016}.}
\end{table}

\subsection{Approximate method: Perturbation theory\label{subsec:Perturbation-theory}}

All numerically exact methods have strong size or time constraints,
which result in finite size effects, especially in the limit of weak
disorder. These constraints are even more pronounced in higher dimensions,
where currently there are no efficient methods for an exact study
of nonequilibrium dynamics. Access to larger systems and longer times
can be gained by utilizing approximate methods. Below we survey one
such approach which was successfully applied for the study of transport
in one-dimensional \cite{BarLev2014} and a two dimensional system
by one of us \cite{BarLev2015}.

For the MBL problem this method was introduced in the work of Basko,
Aleiner and Altshuler \cite{Basko2006a}. The method is \emph{perturbative}
in the interaction strength and as was demonstrated by one of us,
is able to \emph{quantitatively} reproduce numerically exact results
in the limit of large disorder. For very weak disorder, the method
becomes increasingly unreliable and tends to overestimate the relaxation
in the system \cite{BarLev2014,BarLev2015}. We note that while perturbation
theory is clearly an analytical tool, its numerical implementation
requires the solution of certain numerical difficulties (for details
see \cite{Stan2009}). The quantities of interest are one particle
correlation functions,
\begin{eqnarray}
G_{ij}^{>}\left(t;t'\right) & = & -i\tr\left(\hat{\rho}_{0}\hat{c}_{i}\left(t\right)\hat{c}_{j}^{\dagger}\left(t'\right)\right)\label{eq:gtr_less_G}\\
G_{ij}^{<}\left(t;t'\right) & = & i\tr\left(\hat{\rho}_{0}\hat{c}_{j}^{\dagger}\left(t'\right)\hat{c}_{i}\left(t\right)\right),\nonumber 
\end{eqnarray}
where $\hat{\rho}_{0}$ is the initial density matrix and $\hat{c}_{j}^{\dagger}$
creates a spinless fermion at site $j$. For a noninteracting initial
density matrix, the Green's functions obey the Kadanoff\textendash Baym
equations of motion \cite{Kadanoff1994}, 
\begin{eqnarray}
i\partial_{t}G^{\gtrless}\left(t,t'\right) & = & \left(\hat{h}_{0}+\Sigma^{HF}\left(t\right)\right)G^{\gtrless}\left(t,t'\right)\nonumber \\
 & + & \int_{0}^{t}\Sigma^{R}\left(t,t_{2}\right)G^{\gtrless}\left(t_{2},t'\right)\mathrm{d}t_{2}\nonumber \\
 & + & \int_{0}^{t'}\Sigma^{\gtrless}\left(t,t_{2}\right)G^{A}\left(t_{2},t'\right)\mathrm{d}t_{2},\label{eq:KB_eq}
\end{eqnarray}
where spatial indices and summations are suppressed for clarity, $\hat{h}_{0,nm}$
is the one particle Hamiltonian, $\Sigma^{HF}\left(t\right)$, $\Sigma^{\gtrless}\left(t\right)$
are the Hartree-Fock, greater and lesser self-energies of the problem
respectively; and the superscripts 'R' and 'A' represent retarded
and advanced Green's functions and self-energies, which are defined
as 
\begin{eqnarray}
\Sigma^{R}\left(t,t_{2}\right) & = & \theta\left(t-t_{2}\right)\left(\Sigma^{>}\left(t,t_{2}\right)-\Sigma^{<}\left(t,t_{2}\right)\right)\\
G^{A}\left(t_{2},t'\right) & = & -\theta\left(t'-t_{2}\right)\left(G^{>}\left(t_{2},t'\right)-G^{<}\left(t_{2},t'\right)\right).\nonumber 
\end{eqnarray}
Since the exact form of the self-energies is normally unknown, they
are commonly approximated up to some order in the small parameter
of the problem. For the problem which is the subject of this review
the natural small parameter is $\lambda\equiv U/\delta$ where $U$
is the interaction strength and $\delta$ is the typical energy difference
of nearby localized single-particle states $\delta\equiv\Delta/\xi$.
Here $\Delta$ is the single-particle bandwidth and $\xi$ is the
single-particle localization length. To second order in $\lambda$
a particularly useful approximation is the self-consistent second-Born
approximation, 
\begin{eqnarray}
\Sigma_{ij}^{HF}\left(t\right) & = & -i\delta_{ij}\sum_{k}V_{ik}G_{kk}^{<}\left(t;t\right)+iV_{ij}G_{ij}^{<}\left(t;t\right)\nonumber \\
\Sigma_{ij}^{>}\left(t,t'\right) & = & \sum_{k,l}V_{il}V_{jk}G_{kl}^{<}\left(t',t\right)\times\label{eq:self-energies}\\
 &  & \left[G_{lk}^{>}\left(t,t'\right)G_{ij}^{>}\left(t,t'\right)-G_{lj}^{>}\left(t,t'\right)G_{ik}^{>}\left(t,t'\right)\right],\nonumber 
\end{eqnarray}
where $V_{ij}=V\left(\delta_{i,j+1}+\delta_{i,j-1}\right)$ is the
interaction potential. Using this approximation one can write a closed
form equation for the correlation functions (\ref{eq:gtr_less_G}),
which is then solved numerically. This method requires a memory which
scales like $\mathcal{O}\left(L^{2}N_{t}^{2}\right)$, where $N_{t}$
is the number of time steps required to solve (\ref{eq:KB_eq}) to
predetermined precision. The computational complexity of this method
scales like $\mathcal{O}\left(L^{3}N_{t}^{3}\right)$, and could be
further reduced to $\mathcal{O}\left(L^{3}N_{t}^{2}\right)$ by making
additional approximations \cite{Spicka2005,Spicka2005a,Latini2013}.
For more technical details on this method as also to detailed comparison
to exact methods the the reader is referred to Refs.~\cite{Latini2013,BarLev2014,BarLev2015}.

\section{\label{sec:Discussion-and-Open}Discussion and Open questions}

\begin{figure}[h]
\includegraphics[width=86mm]{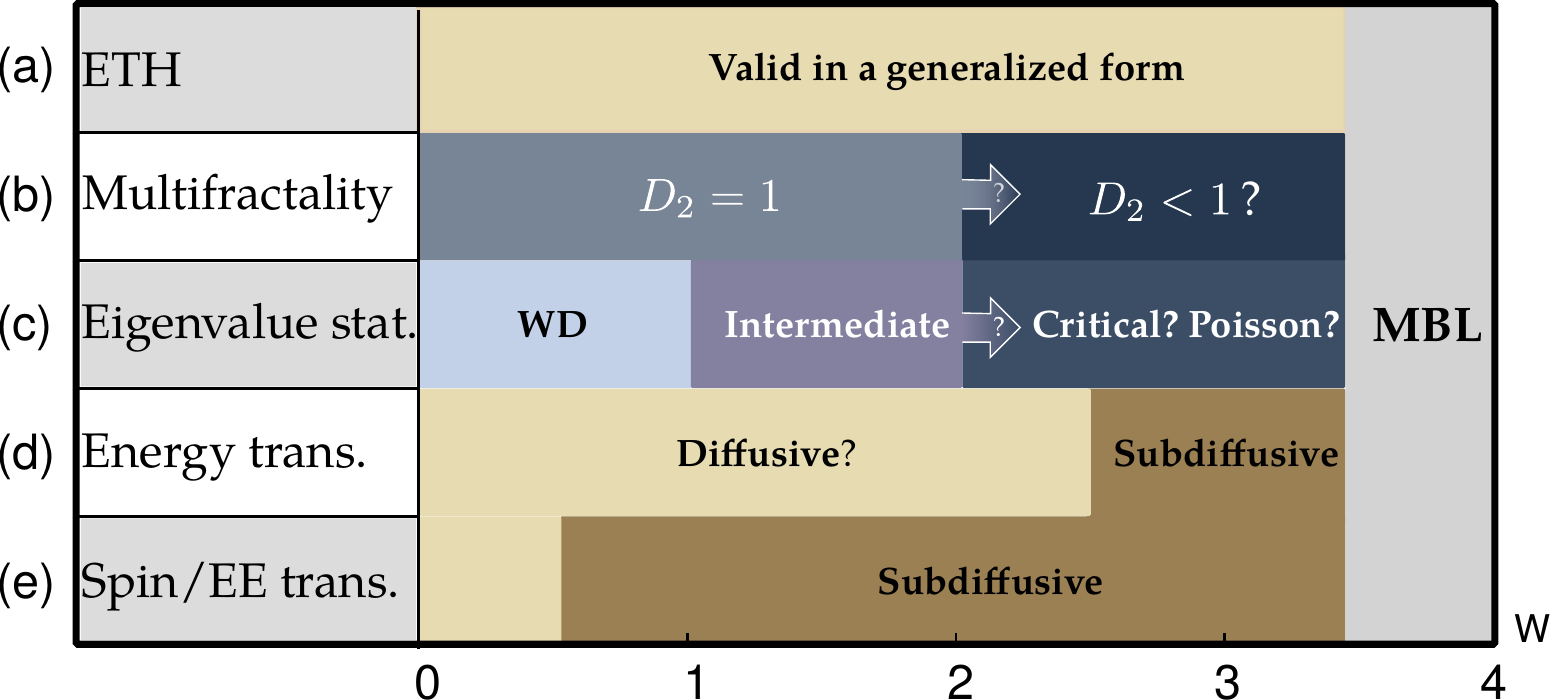}\caption{\label{fig:Visual-summary}A visual summary of current results and
open questions on the ergodic phase of the XXZ model (\ref{eq:xxz})
with $J_{z}=1$. Different color patches represent different phases
as suggested by various studies. The locations of transitions or crossovers
between the different phases are presented only approximately and
are displayed as sharp for better readability. Question marks represent
open questions and arrows indicate that some studies suggest that
the phase shrinks to the critical point $W_{c}\approx3.7$ in the
thermodynamic limit. \textbf{(a)} The validity of ETH was studied
in Refs.~\cite{Luitz2015a,Luitz2016b}, \textbf{(b)} the generalized
fractal dimensions $D_{1,2}$ were studied in Refs.~\cite{Luitz2015,serbyn_thouless_2016,Torres-Herrera2016},
\textbf{(c)} a detailed study of the eigenvalue statistics was done
in Refs.~\cite{Serbyn2015,Torres-Herrera2016}, \textbf{(d)} energy
transport was studied in Ref.~\cite{Lerose2015}, \textbf{(e)} spin
transport and entanglement spreading was studied in Refs.~\cite{Lev2014,Agarwal2014,Luitz2015a,Hauschild2016,Khait2016}.}
\end{figure}

In the previous sections we reviewed in detail the current knowledge
of the ergodic phase at weak disorder, preceding the MBL transition.
Here, we will identify some important open questions and discuss the
progress that has been made towards answering them. In Fig.~\ref{fig:Visual-summary}
we present a visual summary of the results as also some of the open
questions. It is apparent, that while recent works identified fascinating
possible scenarios for the rich physics of the ergodic phase, the
overall picture is not yet settled. Future works in this field will
have to clarify how the observed phenomenology evolves as a function
of system size to put existing contradictions into perspective.

\subsection{Subdiffusion and the subdiffusion to diffusion transition}

While the MBL phase can be defined by an absence of transport, the
nature of transport in the ergodic phase is not a priori clear. Many
numerical studies have addressed this question after first evidence
for subdiffusive transport in a one-dimensional XXZ model was found
\cite{Lev2014,Agarwal2014}. The results of most numerical studies
are consistent with the interpretation that at intermediate disorder
$1\lesssim W\lesssim3.7$ spin transport is subdiffusive and the entanglement
growth is sublinear \cite{Lev2014,Agarwal2014,Luitz2015a,Hauschild2016,Khait2016},
with a continuously varying dynamical exponent $z$, which diverges
at the MBL transition. In Ref.~\cite{Lerose2015} it was argued that
for $1\lesssim W\lesssim2.5$ energy transport is diffusive while
spin transport is subdiffusive. This study however is in contradiction
with Ref.~\cite{Luitz2015a}, since if true asymptotically in time,
it would suggest that energy was transported faster than information
(entanglement entropy). For very weak disorder, $W<1$, some studies
yet find subdiffusive spin transport \cite{Luitz2015a}, while others
argue in favor of a transition to diffusion \cite{Agarwal2014,Znidaric2016}.
Currently, the most compelling evidence stems from the study of an
\emph{open }XXZ chain with system sizes up to $L=400$ by Žnidari\v{c}
\emph{et al.} \cite{Znidaric2016}. This work argues in favor of a
transition between diffusive and subdiffusive behavior at $W\approx0.6$.
While this work cannot rule out weak subdiffusive transport for $W<0.6$,
which might occur for even larger system sizes (see right panel of
Fig.~\ref{fig:Comparison-of-exponents.}, and discussion at the end
of Sec.~\ref{subsec:Transport-ness}), it points out that ED studies
in this region of parameters are subject to severe finite size effects.
Interestingly, in this region, $W<0.6$ the fluctuations of local
operators in the eigenbasis of the Hamiltonian are perfectly Gaussian,
verifying exactly the ETH ansatz \cite{Luitz2016,Luitz2016b}. However
currently no direct connection between the nature of transport and
the shape of the probability distributions is known and we can only
speculate that such perfectly Gaussian distributions are a sign for
diffusion, while heavily tailed non-Gaussian distributions may signal
subdiffusion. The situation in dimensions higher than one is less
clear, since numerically addressing transport for $d\geq2$ remains
very challenging. Algorithmic progress and input from experiments
is required to clarify the nature of transport in this case. Currently
there is only one numerical study which points towards subdiffusion
in two dimensions based on perturbation theory ($\cf$ Sec. \ref{subsec:Perturbation-theory})
\cite{BarLev2015}. 

This seemingly clear picture of transport in one-dimensional systems
is disturbed if the numerical evidence is quantitatively compared.
We have presented a comparison of recent numerical estimates of the
dynamical exponent obtained by various methods. By looking on Fig.~\ref{fig:Comparison-of-exponents.}
it is obvious that the results match only qualitatively, moreover
some commonly used relations between the exponents {[}\emph{cf}. Eq.~(\ref{eq:dynam_prop_summ}){]}
do not hold. While this might indicate that some of these relations
should be reconsidered, the observed disagreement between the exponents
could also follow from the difficulty of extracting dynamical exponents
from numerical calculations on finite systems. In fact, a very recent
work evaluated the spread of spin perturbations starting from initial
conditions with fixed energy density \cite{Bera2016}. In this work
a convergent (with system size) dynamical exponent could not be obtained
and it was argued that the observed subdiffusion is a transient \cite{Bera2016}.
Asymptotic diffusion and finite dc conductivity in the whole ergodic
phase were also found in ac conductivity studies \cite{Steinigeweg2015,Barisic2016},
using system sizes up to $L=28$ similar to the studies which observe
subdiffusion. Future work will have to resolve this contradiction.

\subsection{Existence of the ``bad metal''}

The observation of intermediate level statistics as well as of multifractal
distributions of local operators at disorder strengths $W\gtrsim2$
\cite{Serbyn2015,serbyn_thouless_2016} is consistent with the prediction
of the existence of a delocalized, \textsf{nonergodic} phase (dubbed
``bad metal'' by Altshuler \cite{Altshuler2010}) for disorder strengths
below the MBL transition \cite{Altshuler1997}. However, the question
whether this phase shrinks in the thermodynamic limit to a critical
point or remains of finite extent in the parameter space is still
open \cite{serbyn_thouless_2016,Torres-Herrera2016}. Large scale
studies on random regular graphs (RRGs) seem to point out that this
phase disappears in the thermodynamic limit \cite{Tikhonov2016,Tikhonov2016a,Garcia-Mata2016},
although there is also no consensus here \cite{Biroli2012,DeLuca2013,Luca,Kravtsov2015,Facoetti2016,Tikhonov2016,Tikhonov2016a,Garcia-Mata2016,Altshuler2016}.
Moreover while both problems are related, it is not clear a priori
that results from RRGs apply for physical models.

The multifractal nature of this phase suggests that it might be related
to the observed subdiffusion, and could also explain the poor agreement
between the various dynamical exponents, as presented in the summary
of Sec.~\ref{sec:Dynamical-Properties} (see Fig.~\ref{fig:Comparison-of-exponents.}).
While a direct relation between multifractality in the many-body space
and real space subdiffusion was not established, a step in this direction
was performed in Refs.~\cite{Serbyn2015,Luitz2016b,serbyn_thouless_2016}.
There are however a few problems with this interpretation: (i) subdiffusion
appears to persist in the thermodynamic limit \cite{Znidaric2016}
while the ``bad metal'' seems to shrink in this limit \cite{serbyn_thouless_2016},
(ii) looking at Fig.~\ref{fig:Visual-summary} it is clear that the
extent of observed subdiffusion is much larger then the extent of
the observed multifractality (iii) in this region the definitions
of ergodicity via ETH and \textsf{ergodicity }(via eigenvector statistics)
do not agree, suggesting that the definitions of ergodicity as discussed
in Sec.~\ref{subsec:ergodicty} are not equivalent.

Another interesting question is whether there is a relation between
the ``bad metal'' and the Griffiths effects picture presented in
Sec.~\ref{sec:Phenomenological-models}.

\subsection{Mechanism of subdiffusive transport}

Even though the numerical evidence is not univocal, the existence
of a subdiffusive regime is certainly a valid scenario consistent
with many numerical studies. The proposed mechanism for subdiffusive
transport, dubbed Griffiths effects, is the existence of rare insulating
regions which serve as bottlenecks for the transport of particles
and entanglement \cite{Agarwal2014,Gopalakrishnan2015a}. While this
mechanism is difficult to test numerically, it seems consistent with
spatial variations in the entanglement structure as observed numerically
\cite{Bera2015a,Luitz2016,Yu2016}. However, other indirect tests
of the predictions of the Griffiths picture reveal several issues:
(i) it predicts asymptotic diffusion in dimensions higher then one
($\cf$ the review \cite{Agarwal2016_review}), which seems to contradict
the observation of subdiffusion in two dimensions \cite{BarLev2015}
(ii) it predicts diffusion for quasiperiodic potentials, although
sublinear entanglement growth was observed for the Aubry-André model
in Ref.~\cite{Naldesi2016a} as well as subdiffusive transport in
Refs.~\cite{luschen_evidence_2016,bar_lev_transport_2017} (iii)
Griffiths effects are expected to be subdominant far from the MBL
transition, which is in contrast to the extended subdiffusive phase
found in most studies. Due to the above, we feel that Griffiths picture
may have to be refined in future works. In particular its relation
to observed signatures of multifractality has to be better understood. 

\subsection{Relation to classical disordered models}

An important point that we did not discuss in this review is the connection
between the quantum and classical disordered models. The pertinent
question in the context of the ergodic phase is to which extent the
ergodic phase, which is the subject of this review, can be considered
classical. A pioneering study in this direction suggests that classical
disordered models are diffusive \cite{Oganesyan2009}. We refer the
reader on a recent review on this subject \cite{Huveneers2017}. 

The MBL transition is in many ways similar to the glass transition
\cite{Basko2006a}. In particular commonly used MBL models, such as
(\ref{eq:xxz}) and (\ref{eq:anderson_hubbard}), are superficially
reminiscent of spin-glass models. Both have quenched disorder and
a \emph{finite} temperature ergodic\textendash nonergodic transition.
However while much of the phenomenology is similar, there are important
differences. The spin glass transition is a thermodynamic phase transition
which occurs in a presence of an external heat bath \cite{Fischer1993spin},
while the MBL transition does not appear to have a thermodynamic signature
and occurs only for \emph{isolated} systems \cite{Basko2006a}. One
of the most interesting questions in this context, is whether the
ergodicity breaking mechanisms of spin glasses and MBL are somehow
related.

The relation to structural glasses is more remote, due to absence
of quenched disorder in the structural glasses models. Attempts to
find a stable MBL phase for disorder free, translationally invariant
models were unsuccessful \cite{Schiulaz2013,Yao2014,Schiulaz2014,Hickey2014,DeRoeck2014,DeRoeck2014a,Papic2015,Pino2015,Roeck2015a,Horssen2015,Garrison2016,Antipov2016}.
Notwithstanding, the ergodic phase which is the subject of this review,
shares many properties with supercooled liquids \cite{Binder2005}.
To point out these parallels, we have used notation borrowed from
the structural glasses community in our discussion of the dynamical
properties in Sec,~\ref{sec:Dynamical-Properties}. Similarly to
the supercooled liquids the ergodic phase thermalizes, and the relaxation
of density (spin) autocorrelation functions is subexponential with
a relaxation time which diverges at the transition. Interestingly,
the theory of the MBL transition as established in Ref.~\cite{Basko2006a}
is also reminiscent of the mode-coupling theory, which works remarkably
well for structural glasses \cite{Alamos1971}. Deeper connections
between the two fields should definitely be explored in future works.

\subsection{Summary}

In this review we surveyed the features of the ergodic phase, which
occurs in generic interacting systems with sufficiently weak quenched
disorder. We have explained in which sense this phase could be considered
ergodic, and elaborated on the different notions of ergodicity in
this context. We presented the peculiar static and dynamical properties
of this phase, which is characterized by intermediate eigenvalue statistics,
signatures of multifractality and the emergence of power-laws for
almost any dynamical property, including the growth of the entanglement
entropy. We have explained the phenomenological rare-region picture
(the Griffiths picture), and its predictions on the relations between
the different power-law exponents, as also the numerical verification
of these predictions. Finally, we presented all available numerically
exact and approximate methods for the exploration of this phase and
finished the review with discussion of some of the pertinent open
questions.
\begin{acknowledgments}
\textbf{\label{sec:ack}}
\end{acknowledgments}

We are grateful to Ilia Khait and Marko Žnidari\v{c} for sharing their
original data. YB acknowledges funding from the Simons Foundation
(\#454951, David R. Reichman). DJL was supported by the Gordon and
Betty Moore Foundation's EPiQS Initiative through Grant No. GBMF4305
at the University of Illinois. This research is part of the Blue Waters
sustained-petascale computing project, which is supported by the National
Science Foundation (awards OCI-0725070 and ACI-1238993) and the state
of Illinois. Blue Waters is a joint effort of the University of Illinois
at Urbana-Champaign and its National Center for Supercomputing Applications.

\appendix

\section{\label{sec:Appendix}Relation between the mean-square displacement
and the current-current correlation functions}

For convenience of the reader we derive a general relation between
the correlations of a conserved quantity and the correlations of the
corresponding current, which could be useful to applications well
beyond the scope of this review. Similar relations were derived previously,
see Refs.~\cite{Steinigeweg2009a,Yan2015,Steinigeweg2017}. 

We focus on  Hamiltonians which conserve the quantity, $\hat{Q}=\sum_{m=1}^{L}\hat{n}_{m}$,
namely $\left[\hat{H},\hat{Q}\right]=0$ , and for which the following
continuity equation applies,
\begin{equation}
\frac{\partial\hat{n}_{k}}{\partial t}=\Delta\hat{j}_{k},\label{eq:continuiuty}
\end{equation}
where $\Delta j_{k}\equiv j_{k}-j_{k-1}$ is the backward discrete
derivative. We define the means square displacement (MSD) of the excitation
in this density to be
\begin{equation}
x^{2}\left(t\right)=\frac{1}{L}\sum_{k=1}^{L}\sum_{l=1}^{L}\left(k-l\right)^{2}\text{Re }\left\langle \delta\hat{n}_{k}\left(t\right)\delta\hat{n}_{l}\right\rangle ,
\end{equation}
where $\left\langle .\right\rangle $ is the equilibrium average and
$\delta\hat{n}_{k}\left(t\right)\equiv\hat{n}_{k}\left(t\right)-\left\langle \hat{n}_{k}\right\rangle $.
We note the following identity,
\begin{align}
\left\langle \left(\hat{n}_{k}\left(t\right)-\hat{n}_{k}\right)\left(\hat{n}_{l}\left(t\right)-\hat{n}_{l}\right)\right\rangle  & =2\left\langle \delta\hat{n}_{k}\delta\hat{n}_{l}\right\rangle \\
 & -\left\langle \delta\hat{n}_{k}\left(t\right)\delta\hat{n}_{l}+\delta\hat{n}_{k}\delta\hat{n}_{l}\left(t\right)\right\rangle ,\nonumber 
\end{align}
which is true for expectation with respect to the equilibrium state.
To calculate MSD we multiply by $\left(k-l\right)^{2}$ and sum twice
over the lattice, which gives,
\begin{equation}
x^{2}\left(t\right)-x^{2}\left(0\right)=-\frac{1}{2L}\sum_{k,l=1}^{L}\left(k-l\right)^{2}\left\langle \left(\hat{n}_{k}\left(t\right)-\hat{n}_{k}\right)\left(\hat{n}_{l}\left(t\right)-\hat{n}_{l}\right)\right\rangle .
\end{equation}
Using the continuity equation we can write,
\begin{equation}
\hat{n}_{k}\left(t\right)-\hat{n}_{k}=\int_{0}^{t}\mathrm{d}\bar{t}\,\Delta\hat{j}_{k}\left(\bar{t}\right),
\end{equation}
such that
\begin{align}
x^{2}\left(t\right)-x^{2}\left(0\right) & =-\frac{1}{2L}\int_{0}^{t}\mathrm{d}t_{1}\int_{0}^{t}\mathrm{d}t_{2}\times\\
\times & \sum_{k,l=1}^{L}\left(k-l\right)^{2}\left\langle \Delta\hat{j}_{k}\left(t_{1}\right)\Delta\hat{j}_{l}\left(t_{2}\right)\right\rangle .\nonumber 
\end{align}
Taking a partial sum twice and assuming periodic boundary conditions
(or alternatively neglecting the boundary terms) gives,
\begin{equation}
x^{2}\left(t\right)-x^{2}\left(0\right)=\frac{1}{L}\int_{0}^{t}\mathrm{d}t_{1}\int_{0}^{t}\mathrm{d}t_{2}\left\langle \hat{J}\left(t_{1}\right)\hat{J}\left(t_{2}\right)\right\rangle ,
\end{equation}
where $\hat{J}\left(t\right)=\sum_{k=1}^{L}\hat{j}_{k}\left(t\right)$
is the total current. Since any correlation function in equilibrium
depends only on the time difference we change the variables to $\tau=t_{1}-t_{2}$,
and $t_{1}=t_{1}$ which has an unity Jacobian and the following transformation
of the integration boundaries,
\begin{equation}
x^{2}\left(t\right)-x^{2}\left(0\right)=\frac{2}{L}\int_{0}^{t}\mathrm{d}t_{1}\int_{0}^{t_{1}}\mathrm{d}\tau\left\langle \hat{J}\left(\tau\right)\hat{J}\left(0\right)\right\rangle ,
\end{equation}
which could also be written as,
\begin{equation}
\left\langle \hat{J}\left(t\right)\hat{J}\left(0\right)\right\rangle =\frac{\mathrm{d^{2}}}{\mathrm{d}t^{2}}x^{2}\left(t\right),
\end{equation}
which is very similar to its classical form. Taking the Fourier transform
we get the relation between the corresponding frequency dependent
diffusion coefficient and the MSD (Eq.~(\ref{eq:d_omega})),
\begin{equation}
D\left(\omega\right)=-\omega^{2}\int_{-\infty}^{\infty}\mathrm{d}t\,x^{2}\left(t\right)\E^{\I\omega t}.
\end{equation}
We note that this derivation assumes only the continuity equation
(\ref{eq:continuiuty}), periodic boundary conditions and an expectation
value with respect to the thermal state. It does not assume linear
response and any knowledge about the Hamiltonian except of the existence
of the conserved quantity. It also does not assume any specific form
of the conserved quantity or the corresponding current.

\bibliographystyle{apsrev4-1}
\bibliography{lib_david,lib_yevgeny}

\begin{thebibliography}{249}%
\makeatletter
\providecommand \@ifxundefined [1]{%
 \@ifx{#1\undefined}
}%
\providecommand \@ifnum [1]{%
 \ifnum #1\expandafter \@firstoftwo
 \else \expandafter \@secondoftwo
 \fi
}%
\providecommand \@ifx [1]{%
 \ifx #1\expandafter \@firstoftwo
 \else \expandafter \@secondoftwo
 \fi
}%
\providecommand \natexlab [1]{#1}%
\providecommand \enquote  [1]{``#1''}%
\providecommand \bibnamefont  [1]{#1}%
\providecommand \bibfnamefont [1]{#1}%
\providecommand \citenamefont [1]{#1}%
\providecommand \href@noop [0]{\@secondoftwo}%
\providecommand \href [0]{\begingroup \@sanitize@url \@href}%
\providecommand \@href[1]{\@@startlink{#1}\@@href}%
\providecommand \@@href[1]{\endgroup#1\@@endlink}%
\providecommand \@sanitize@url [0]{\catcode `\\12\catcode `\$12\catcode
  `\&12\catcode `\#12\catcode `\^12\catcode `\_12\catcode `\%12\relax}%
\providecommand \@@startlink[1]{}%
\providecommand \@@endlink[0]{}%
\providecommand \url  [0]{\begingroup\@sanitize@url \@url }%
\providecommand \@url [1]{\endgroup\@href {#1}{\urlprefix }}%
\providecommand \urlprefix  [0]{URL }%
\providecommand \Eprint [0]{\href }%
\providecommand \doibase [0]{http://dx.doi.org/}%
\providecommand \selectlanguage [0]{\@gobble}%
\providecommand \bibinfo  [0]{\@secondoftwo}%
\providecommand \bibfield  [0]{\@secondoftwo}%
\providecommand \translation [1]{[#1]}%
\providecommand \BibitemOpen [0]{}%
\providecommand \bibitemStop [0]{}%
\providecommand \bibitemNoStop [0]{.\EOS\space}%
\providecommand \EOS [0]{\spacefactor3000\relax}%
\providecommand \BibitemShut  [1]{\csname bibitem#1\endcsname}%
\let\auto@bib@innerbib\@empty
\bibitem [{\citenamefont {Boltzmann}(1884)}]{Boltzmann1884ergodic_hypo}%
  \BibitemOpen
  \bibfield  {author} {\bibinfo {author} {\bibfnamefont {L.}~\bibnamefont
  {Boltzmann}},\ }\href@noop {} {\bibfield  {journal} {\bibinfo  {journal}
  {Crelle's J.}\ }\textbf {\bibinfo {volume} {98}},\ \bibinfo {pages} {68}
  (\bibinfo {year} {1884})}\BibitemShut {NoStop}%
\bibitem [{\citenamefont {von Neumann}(1929)}]{VonNeumann1929}%
  \BibitemOpen
  \bibfield  {author} {\bibinfo {author} {\bibfnamefont {J.}~\bibnamefont {von
  Neumann}},\ }\href {\doibase 10.1007/BF01339852} {\bibfield  {journal}
  {\bibinfo  {journal} {Z. Phys.}\ }\textbf {\bibinfo {volume} {57}},\ \bibinfo
  {pages} {30} (\bibinfo {year} {1929})}\BibitemShut {NoStop}%
\bibitem [{\citenamefont {von Neumann}(2010)}]{von_neumann_proof_2010}%
  \BibitemOpen
  \bibfield  {author} {\bibinfo {author} {\bibfnamefont {J.}~\bibnamefont {von
  Neumann}},\ }\href {\doibase 10.1140/epjh/e2010-00008-5} {\bibfield
  {journal} {\bibinfo  {journal} {The European Physical Journal H}\ }\textbf
  {\bibinfo {volume} {35}},\ \bibinfo {pages} {201} (\bibinfo {year}
  {2010})}\BibitemShut {NoStop}%
\bibitem [{\citenamefont {Berry}(1977)}]{Berry1977}%
  \BibitemOpen
  \bibfield  {author} {\bibinfo {author} {\bibfnamefont {M.~V.}\ \bibnamefont
  {Berry}},\ }\href {\doibase 10.1088/0305-4470/10/12/016} {\bibfield
  {journal} {\bibinfo  {journal} {J. Phys. A. Math. Gen.}\ }\textbf {\bibinfo
  {volume} {10}},\ \bibinfo {pages} {2083} (\bibinfo {year}
  {1977})}\BibitemShut {NoStop}%
\bibitem [{\citenamefont {Pechukas}(1983)}]{Pechukas1983}%
  \BibitemOpen
  \bibfield  {author} {\bibinfo {author} {\bibfnamefont {P.}~\bibnamefont
  {Pechukas}},\ }\href {\doibase 10.1103/PhysRevLett.51.943} {\bibfield
  {journal} {\bibinfo  {journal} {Phys. Rev. Lett.}\ }\textbf {\bibinfo
  {volume} {51}},\ \bibinfo {pages} {943} (\bibinfo {year} {1983})}\BibitemShut
  {NoStop}%
\bibitem [{\citenamefont {Pechukas}(1984)}]{Pechukas1984}%
  \BibitemOpen
  \bibfield  {author} {\bibinfo {author} {\bibfnamefont {P.}~\bibnamefont
  {Pechukas}},\ }\href {\doibase 10.1021/j150665a006} {\bibfield  {journal}
  {\bibinfo  {journal} {J. Phys. Chem.}\ }\textbf {\bibinfo {volume} {88}},\
  \bibinfo {pages} {4823} (\bibinfo {year} {1984})}\BibitemShut {NoStop}%
\bibitem [{\citenamefont {Feingold}\ \emph {et~al.}(1984)\citenamefont
  {Feingold}, \citenamefont {Moiseyev},\ and\ \citenamefont
  {Peres}}]{Feingold1984}%
  \BibitemOpen
  \bibfield  {author} {\bibinfo {author} {\bibfnamefont {M.}~\bibnamefont
  {Feingold}}, \bibinfo {author} {\bibfnamefont {N.}~\bibnamefont {Moiseyev}},
  \ and\ \bibinfo {author} {\bibfnamefont {A.}~\bibnamefont {Peres}},\ }\href
  {\doibase 10.1103/PhysRevA.30.509} {\bibfield  {journal} {\bibinfo  {journal}
  {Phys. Rev. A}\ }\textbf {\bibinfo {volume} {30}},\ \bibinfo {pages} {509}
  (\bibinfo {year} {1984})}\BibitemShut {NoStop}%
\bibitem [{\citenamefont {Feingold}\ \emph {et~al.}(1985)\citenamefont
  {Feingold}, \citenamefont {Moiseyev},\ and\ \citenamefont
  {Peres}}]{Feingold1985}%
  \BibitemOpen
  \bibfield  {author} {\bibinfo {author} {\bibfnamefont {M.}~\bibnamefont
  {Feingold}}, \bibinfo {author} {\bibfnamefont {N.}~\bibnamefont {Moiseyev}},
  \ and\ \bibinfo {author} {\bibfnamefont {A.}~\bibnamefont {Peres}},\ }\href
  {\doibase 10.1016/0009-2614(85)85241-6} {\bibfield  {journal} {\bibinfo
  {journal} {Chem. Phys. Lett.}\ }\textbf {\bibinfo {volume} {117}},\ \bibinfo
  {pages} {344} (\bibinfo {year} {1985})}\BibitemShut {NoStop}%
\bibitem [{\citenamefont {Feingold}\ and\ \citenamefont
  {Peres}(1986)}]{Feingold1986}%
  \BibitemOpen
  \bibfield  {author} {\bibinfo {author} {\bibfnamefont {M.}~\bibnamefont
  {Feingold}}\ and\ \bibinfo {author} {\bibfnamefont {A.}~\bibnamefont
  {Peres}},\ }\href {\doibase 10.1103/PhysRevA.34.591} {\bibfield  {journal}
  {\bibinfo  {journal} {Phys. Rev. A}\ }\textbf {\bibinfo {volume} {34}},\
  \bibinfo {pages} {591} (\bibinfo {year} {1986})}\BibitemShut {NoStop}%
\bibitem [{\citenamefont {Peres}(1984{\natexlab{a}})}]{Peres1984}%
  \BibitemOpen
  \bibfield  {author} {\bibinfo {author} {\bibfnamefont {A.}~\bibnamefont
  {Peres}},\ }\href {\doibase 10.1103/PhysRevA.30.504} {\bibfield  {journal}
  {\bibinfo  {journal} {Phys. Rev. A}\ }\textbf {\bibinfo {volume} {30}},\
  \bibinfo {pages} {504} (\bibinfo {year} {1984}{\natexlab{a}})}\BibitemShut
  {NoStop}%
\bibitem [{\citenamefont {Peres}(1984{\natexlab{b}})}]{Peres1984a}%
  \BibitemOpen
  \bibfield  {author} {\bibinfo {author} {\bibfnamefont {A.}~\bibnamefont
  {Peres}},\ }\href {\doibase 10.1103/PhysRevA.30.1610} {\bibfield  {journal}
  {\bibinfo  {journal} {Phys. Rev. A}\ }\textbf {\bibinfo {volume} {30}},\
  \bibinfo {pages} {1610} (\bibinfo {year} {1984}{\natexlab{b}})}\BibitemShut
  {NoStop}%
\bibitem [{\citenamefont {Jensen}\ and\ \citenamefont
  {Shankar}(1985)}]{jensen_statistical_1985}%
  \BibitemOpen
  \bibfield  {author} {\bibinfo {author} {\bibfnamefont {R.~V.}\ \bibnamefont
  {Jensen}}\ and\ \bibinfo {author} {\bibfnamefont {R.}~\bibnamefont
  {Shankar}},\ }\href {\doibase 10.1103/PhysRevLett.54.1879} {\bibfield
  {journal} {\bibinfo  {journal} {Physical Review Letters}\ }\textbf {\bibinfo
  {volume} {54}},\ \bibinfo {pages} {1879} (\bibinfo {year}
  {1985})}\BibitemShut {NoStop}%
\bibitem [{\citenamefont {Deutsch}(1991)}]{Deutsch1991}%
  \BibitemOpen
  \bibfield  {author} {\bibinfo {author} {\bibfnamefont {J.~M.}\ \bibnamefont
  {Deutsch}},\ }\href {\doibase 10.1103/PhysRevA.43.2046} {\bibfield  {journal}
  {\bibinfo  {journal} {Phys. Rev. A}\ }\textbf {\bibinfo {volume} {43}},\
  \bibinfo {pages} {2046} (\bibinfo {year} {1991})}\BibitemShut {NoStop}%
\bibitem [{\citenamefont {Srednicki}(1994)}]{Srednicki1994}%
  \BibitemOpen
  \bibfield  {author} {\bibinfo {author} {\bibfnamefont {M.}~\bibnamefont
  {Srednicki}},\ }\href {\doibase 10.1103/PhysRevE.50.888} {\bibfield
  {journal} {\bibinfo  {journal} {Phys. Rev. E}\ }\textbf {\bibinfo {volume}
  {50}},\ \bibinfo {pages} {888} (\bibinfo {year} {1994})}\BibitemShut
  {NoStop}%
\bibitem [{\citenamefont {Srednicki}(1996)}]{Srednicki1995}%
  \BibitemOpen
  \bibfield  {author} {\bibinfo {author} {\bibfnamefont {M.}~\bibnamefont
  {Srednicki}},\ }\href {\doibase 10.1088/0305-4470/29/4/003} {\bibfield
  {journal} {\bibinfo  {journal} {J. Phys. A. Math. Gen.}\ }\textbf {\bibinfo
  {volume} {29}},\ \bibinfo {pages} {L75} (\bibinfo {year} {1996})}\BibitemShut
  {NoStop}%
\bibitem [{\citenamefont {Srednicki}(1999)}]{Srednicki1999}%
  \BibitemOpen
  \bibfield  {author} {\bibinfo {author} {\bibfnamefont {M.}~\bibnamefont
  {Srednicki}},\ }\href {\doibase 10.1088/0305-4470/32/7/007} {\bibfield
  {journal} {\bibinfo  {journal} {J. Phys. A. Math. Gen.}\ }\textbf {\bibinfo
  {volume} {32}},\ \bibinfo {pages} {1163} (\bibinfo {year}
  {1999})}\BibitemShut {NoStop}%
\bibitem [{\citenamefont {Goldenfeld}(1992)}]{Goldenfeld1992lectures}%
  \BibitemOpen
  \bibfield  {author} {\bibinfo {author} {\bibfnamefont {N.}~\bibnamefont
  {Goldenfeld}},\ }\href {https://books.google.com/books?id=DdB1{\_}{\_}nl7CYC}
  {\emph {\bibinfo {title} {{Lectures on Phase Transitions and the
  Renormalization Group}}}},\ Frontiers in physics\ (\bibinfo  {publisher}
  {Addison-Wesley, Advanced Book Program},\ \bibinfo {year} {1992})\BibitemShut
  {NoStop}%
\bibitem [{\citenamefont {Basko}\ \emph {et~al.}(2006)\citenamefont {Basko},
  \citenamefont {Aleiner},\ and\ \citenamefont {Altshuler}}]{Basko2006a}%
  \BibitemOpen
  \bibfield  {author} {\bibinfo {author} {\bibfnamefont {D.}~\bibnamefont
  {Basko}}, \bibinfo {author} {\bibfnamefont {I.~L.}\ \bibnamefont {Aleiner}},
  \ and\ \bibinfo {author} {\bibfnamefont {B.~L.}\ \bibnamefont {Altshuler}},\
  }\href {\doibase 10.1016/j.aop.2005.11.014} {\bibfield  {journal} {\bibinfo
  {journal} {Ann. Phys. (N. Y).}\ }\textbf {\bibinfo {volume} {321}},\ \bibinfo
  {pages} {1126} (\bibinfo {year} {2006})}\BibitemShut {NoStop}%
\bibitem [{\citenamefont {Imbrie}(2016{\natexlab{a}})}]{Imbrie2014}%
  \BibitemOpen
  \bibfield  {author} {\bibinfo {author} {\bibfnamefont {J.~Z.}\ \bibnamefont
  {Imbrie}},\ }\href {\doibase 10.1007/s10955-016-1508-x} {\bibfield  {journal}
  {\bibinfo  {journal} {J. Stat. Phys.}\ }\textbf {\bibinfo {volume} {163}},\
  \bibinfo {pages} {998} (\bibinfo {year} {2016}{\natexlab{a}})}\BibitemShut
  {NoStop}%
\bibitem [{\citenamefont {Imbrie}(2016{\natexlab{b}})}]{Imbrie2016}%
  \BibitemOpen
  \bibfield  {author} {\bibinfo {author} {\bibfnamefont {J.~Z.}\ \bibnamefont
  {Imbrie}},\ }\href {\doibase 10.1103/PhysRevLett.117.027201} {\bibfield
  {journal} {\bibinfo  {journal} {Phys. Rev. Lett.}\ }\textbf {\bibinfo
  {volume} {117}},\ \bibinfo {pages} {027201} (\bibinfo {year}
  {2016}{\natexlab{b}})}\BibitemShut {NoStop}%
\bibitem [{\citenamefont {Basko}\ \emph {et~al.}(2007)\citenamefont {Basko},
  \citenamefont {Aleiner},\ and\ \citenamefont {Altshuler}}]{Basko2007a}%
  \BibitemOpen
  \bibfield  {author} {\bibinfo {author} {\bibfnamefont {D.~M.}\ \bibnamefont
  {Basko}}, \bibinfo {author} {\bibfnamefont {I.~L.}\ \bibnamefont {Aleiner}},
  \ and\ \bibinfo {author} {\bibfnamefont {B.~L.}\ \bibnamefont {Altshuler}},\
  }\href {\doibase 10.1103/PhysRevB.76.052203} {\bibfield  {journal} {\bibinfo
  {journal} {Phys. Rev. B}\ }\textbf {\bibinfo {volume} {76}},\ \bibinfo
  {pages} {052203} (\bibinfo {year} {2007})}\BibitemShut {NoStop}%
\bibitem [{\citenamefont {Ovadia}\ \emph {et~al.}(2015)\citenamefont {Ovadia},
  \citenamefont {Kalok}, \citenamefont {Tamir}, \citenamefont {Mitra},
  \citenamefont {Sac{\'{e}}p{\'{e}}},\ and\ \citenamefont
  {Shahar}}]{Ovadia2014}%
  \BibitemOpen
  \bibfield  {author} {\bibinfo {author} {\bibfnamefont {M.}~\bibnamefont
  {Ovadia}}, \bibinfo {author} {\bibfnamefont {D.}~\bibnamefont {Kalok}},
  \bibinfo {author} {\bibfnamefont {I.}~\bibnamefont {Tamir}}, \bibinfo
  {author} {\bibfnamefont {S.}~\bibnamefont {Mitra}}, \bibinfo {author}
  {\bibfnamefont {B.}~\bibnamefont {Sac{\'{e}}p{\'{e}}}}, \ and\ \bibinfo
  {author} {\bibfnamefont {D.}~\bibnamefont {Shahar}},\ }\href {\doibase
  10.1038/srep13503} {\bibfield  {journal} {\bibinfo  {journal} {Sci. Rep.}\
  }\textbf {\bibinfo {volume} {5}},\ \bibinfo {pages} {13503} (\bibinfo {year}
  {2015})}\BibitemShut {NoStop}%
\bibitem [{\citenamefont {Schreiber}\ \emph {et~al.}(2015)\citenamefont
  {Schreiber}, \citenamefont {Hodgman}, \citenamefont {Bordia}, \citenamefont
  {Luschen}, \citenamefont {Fischer}, \citenamefont {Vosk}, \citenamefont
  {Altman}, \citenamefont {Schneider},\ and\ \citenamefont
  {Bloch}}]{Schreiber2015a}%
  \BibitemOpen
  \bibfield  {author} {\bibinfo {author} {\bibfnamefont {M.}~\bibnamefont
  {Schreiber}}, \bibinfo {author} {\bibfnamefont {S.~S.}\ \bibnamefont
  {Hodgman}}, \bibinfo {author} {\bibfnamefont {P.}~\bibnamefont {Bordia}},
  \bibinfo {author} {\bibfnamefont {H.~P.}\ \bibnamefont {Luschen}}, \bibinfo
  {author} {\bibfnamefont {M.~H.}\ \bibnamefont {Fischer}}, \bibinfo {author}
  {\bibfnamefont {R.}~\bibnamefont {Vosk}}, \bibinfo {author} {\bibfnamefont
  {E.}~\bibnamefont {Altman}}, \bibinfo {author} {\bibfnamefont
  {U.}~\bibnamefont {Schneider}}, \ and\ \bibinfo {author} {\bibfnamefont
  {I.}~\bibnamefont {Bloch}},\ }\href {\doibase 10.1126/science.aaa7432}
  {\bibfield  {journal} {\bibinfo  {journal} {Science}\ }\textbf {\bibinfo
  {volume} {349}},\ \bibinfo {pages} {842} (\bibinfo {year}
  {2015})}\BibitemShut {NoStop}%
\bibitem [{\citenamefont {Bordia}\ \emph
  {et~al.}(2016{\natexlab{a}})\citenamefont {Bordia}, \citenamefont
  {L{\"{u}}schen}, \citenamefont {Hodgman}, \citenamefont {Schreiber},
  \citenamefont {Bloch},\ and\ \citenamefont {Schneider}}]{Bordia2015}%
  \BibitemOpen
  \bibfield  {author} {\bibinfo {author} {\bibfnamefont {P.}~\bibnamefont
  {Bordia}}, \bibinfo {author} {\bibfnamefont {H.~P.}\ \bibnamefont
  {L{\"{u}}schen}}, \bibinfo {author} {\bibfnamefont {S.~S.}\ \bibnamefont
  {Hodgman}}, \bibinfo {author} {\bibfnamefont {M.}~\bibnamefont {Schreiber}},
  \bibinfo {author} {\bibfnamefont {I.}~\bibnamefont {Bloch}}, \ and\ \bibinfo
  {author} {\bibfnamefont {U.}~\bibnamefont {Schneider}},\ }\href {\doibase
  10.1103/PhysRevLett.116.140401} {\bibfield  {journal} {\bibinfo  {journal}
  {Phys. Rev. Lett.}\ }\textbf {\bibinfo {volume} {116}},\ \bibinfo {pages}
  {140401} (\bibinfo {year} {2016}{\natexlab{a}})}\BibitemShut {NoStop}%
\bibitem [{\citenamefont {Smith}\ \emph {et~al.}(2016)\citenamefont {Smith},
  \citenamefont {Lee}, \citenamefont {Richerme}, \citenamefont {Neyenhuis},
  \citenamefont {Hess}, \citenamefont {Hauke}, \citenamefont {Heyl},
  \citenamefont {Huse},\ and\ \citenamefont {Monroe}}]{Smith2015}%
  \BibitemOpen
  \bibfield  {author} {\bibinfo {author} {\bibfnamefont {J.}~\bibnamefont
  {Smith}}, \bibinfo {author} {\bibfnamefont {A.}~\bibnamefont {Lee}}, \bibinfo
  {author} {\bibfnamefont {P.}~\bibnamefont {Richerme}}, \bibinfo {author}
  {\bibfnamefont {B.}~\bibnamefont {Neyenhuis}}, \bibinfo {author}
  {\bibfnamefont {P.~W.}\ \bibnamefont {Hess}}, \bibinfo {author}
  {\bibfnamefont {P.}~\bibnamefont {Hauke}}, \bibinfo {author} {\bibfnamefont
  {M.}~\bibnamefont {Heyl}}, \bibinfo {author} {\bibfnamefont {D.~A.}\
  \bibnamefont {Huse}}, \ and\ \bibinfo {author} {\bibfnamefont
  {C.}~\bibnamefont {Monroe}},\ }\href {\doibase 10.1038/nphys3783} {\bibfield
  {journal} {\bibinfo  {journal} {Nat. Phys.}\ }\textbf {\bibinfo {volume}
  {12}},\ \bibinfo {pages} {907} (\bibinfo {year} {2016})}\BibitemShut
  {NoStop}%
\bibitem [{\citenamefont {Choi}\ \emph {et~al.}(2016)\citenamefont {Choi},
  \citenamefont {Hild}, \citenamefont {Zeiher}, \citenamefont {Schauss},
  \citenamefont {Rubio-Abadal}, \citenamefont {Yefsah}, \citenamefont
  {Khemani}, \citenamefont {Huse}, \citenamefont {Bloch},\ and\ \citenamefont
  {Gross}}]{Choi2016}%
  \BibitemOpen
  \bibfield  {author} {\bibinfo {author} {\bibfnamefont {J.-y.}\ \bibnamefont
  {Choi}}, \bibinfo {author} {\bibfnamefont {S.}~\bibnamefont {Hild}}, \bibinfo
  {author} {\bibfnamefont {J.}~\bibnamefont {Zeiher}}, \bibinfo {author}
  {\bibfnamefont {P.}~\bibnamefont {Schauss}}, \bibinfo {author} {\bibfnamefont
  {A.}~\bibnamefont {Rubio-Abadal}}, \bibinfo {author} {\bibfnamefont
  {T.}~\bibnamefont {Yefsah}}, \bibinfo {author} {\bibfnamefont
  {V.}~\bibnamefont {Khemani}}, \bibinfo {author} {\bibfnamefont {D.~A.}\
  \bibnamefont {Huse}}, \bibinfo {author} {\bibfnamefont {I.}~\bibnamefont
  {Bloch}}, \ and\ \bibinfo {author} {\bibfnamefont {C.}~\bibnamefont
  {Gross}},\ }\href {\doibase 10.1126/science.aaf8834} {\bibfield  {journal}
  {\bibinfo  {journal} {Science}\ }\textbf {\bibinfo {volume} {352}},\ \bibinfo
  {pages} {1547} (\bibinfo {year} {2016})}\BibitemShut {NoStop}%
\bibitem [{\citenamefont {Altman}\ and\ \citenamefont
  {Vosk}(2015)}]{Altman2014}%
  \BibitemOpen
  \bibfield  {author} {\bibinfo {author} {\bibfnamefont {E.}~\bibnamefont
  {Altman}}\ and\ \bibinfo {author} {\bibfnamefont {R.}~\bibnamefont {Vosk}},\
  }\href {\doibase 10.1146/annurev-conmatphys-031214-014701} {\bibfield
  {journal} {\bibinfo  {journal} {Annu. Rev. Condens. Matter Phys.}\ }\textbf
  {\bibinfo {volume} {6}},\ \bibinfo {pages} {383} (\bibinfo {year}
  {2015})}\BibitemShut {NoStop}%
\bibitem [{\citenamefont {Nandkishore}\ and\ \citenamefont
  {Huse}(2015)}]{Nandkishore2014}%
  \BibitemOpen
  \bibfield  {author} {\bibinfo {author} {\bibfnamefont {R.}~\bibnamefont
  {Nandkishore}}\ and\ \bibinfo {author} {\bibfnamefont {D.~A.}\ \bibnamefont
  {Huse}},\ }\href {\doibase 10.1146/annurev-conmatphys-031214-014726}
  {\bibfield  {journal} {\bibinfo  {journal} {Annu. Rev. Condens. Matter
  Phys.}\ }\textbf {\bibinfo {volume} {6}},\ \bibinfo {pages} {15} (\bibinfo
  {year} {2015})}\BibitemShut {NoStop}%
\bibitem [{\citenamefont {Vasseur}\ and\ \citenamefont
  {Moore}(2016)}]{Vasseur2016}%
  \BibitemOpen
  \bibfield  {author} {\bibinfo {author} {\bibfnamefont {R.}~\bibnamefont
  {Vasseur}}\ and\ \bibinfo {author} {\bibfnamefont {J.~E.}\ \bibnamefont
  {Moore}},\ }\href {\doibase 10.1088/1742-5468/2016/06/064010} {\bibfield
  {journal} {\bibinfo  {journal} {J. Stat. Mech. Theory Exp.}\ }\textbf
  {\bibinfo {volume} {2016}},\ \bibinfo {pages} {064010} (\bibinfo {year}
  {2016})}\BibitemShut {NoStop}%
\bibitem [{\citenamefont {Anderson}(1958)}]{Anderson1958b}%
  \BibitemOpen
  \bibfield  {author} {\bibinfo {author} {\bibfnamefont {P.~W.}\ \bibnamefont
  {Anderson}},\ }\href {\doibase 10.1103/PhysRev.109.1492} {\bibfield
  {journal} {\bibinfo  {journal} {Phys. Rev.}\ }\textbf {\bibinfo {volume}
  {109}},\ \bibinfo {pages} {1492} (\bibinfo {year} {1958})}\BibitemShut
  {NoStop}%
\bibitem [{\citenamefont {{Bar Lev}}\ and\ \citenamefont
  {Reichman}(2014)}]{BarLev2014}%
  \BibitemOpen
  \bibfield  {author} {\bibinfo {author} {\bibfnamefont {Y.}~\bibnamefont {{Bar
  Lev}}}\ and\ \bibinfo {author} {\bibfnamefont {D.~R.}\ \bibnamefont
  {Reichman}},\ }\href {\doibase 10.1103/PhysRevB.89.220201} {\bibfield
  {journal} {\bibinfo  {journal} {Phys. Rev. B}\ }\textbf {\bibinfo {volume}
  {89}},\ \bibinfo {pages} {220201} (\bibinfo {year} {2014})}\BibitemShut
  {NoStop}%
\bibitem [{\citenamefont {{Bar Lev}}\ \emph {et~al.}(2015)\citenamefont {{Bar
  Lev}}, \citenamefont {Cohen},\ and\ \citenamefont {Reichman}}]{Lev2014}%
  \BibitemOpen
  \bibfield  {author} {\bibinfo {author} {\bibfnamefont {Y.}~\bibnamefont {{Bar
  Lev}}}, \bibinfo {author} {\bibfnamefont {G.}~\bibnamefont {Cohen}}, \ and\
  \bibinfo {author} {\bibfnamefont {D.~R.}\ \bibnamefont {Reichman}},\ }\href
  {\doibase 10.1103/PhysRevLett.114.100601} {\bibfield  {journal} {\bibinfo
  {journal} {Phys. Rev. Lett.}\ }\textbf {\bibinfo {volume} {114}},\ \bibinfo
  {pages} {100601} (\bibinfo {year} {2015})}\BibitemShut {NoStop}%
\bibitem [{\citenamefont {Agarwal}\ \emph
  {et~al.}(2015{\natexlab{a}})\citenamefont {Agarwal}, \citenamefont
  {Gopalakrishnan}, \citenamefont {Knap}, \citenamefont {M{\"{u}}ller},\ and\
  \citenamefont {Demler}}]{Agarwal2014}%
  \BibitemOpen
  \bibfield  {author} {\bibinfo {author} {\bibfnamefont {K.}~\bibnamefont
  {Agarwal}}, \bibinfo {author} {\bibfnamefont {S.}~\bibnamefont
  {Gopalakrishnan}}, \bibinfo {author} {\bibfnamefont {M.}~\bibnamefont
  {Knap}}, \bibinfo {author} {\bibfnamefont {M.}~\bibnamefont {M{\"{u}}ller}},
  \ and\ \bibinfo {author} {\bibfnamefont {E.}~\bibnamefont {Demler}},\ }\href
  {\doibase 10.1103/PhysRevLett.114.160401} {\bibfield  {journal} {\bibinfo
  {journal} {Phys. Rev. Lett.}\ }\textbf {\bibinfo {volume} {114}},\ \bibinfo
  {pages} {160401} (\bibinfo {year} {2015}{\natexlab{a}})}\BibitemShut
  {NoStop}%
\bibitem [{\citenamefont {Imbrie}\ \emph {et~al.}(2016)\citenamefont {Imbrie},
  \citenamefont {Ros},\ and\ \citenamefont {Scardicchio}}]{Imbrie2016a}%
  \BibitemOpen
  \bibfield  {author} {\bibinfo {author} {\bibfnamefont {J.~Z.}\ \bibnamefont
  {Imbrie}}, \bibinfo {author} {\bibfnamefont {V.}~\bibnamefont {Ros}}, \ and\
  \bibinfo {author} {\bibfnamefont {A.}~\bibnamefont {Scardicchio}},\ }\href
  {http://arxiv.org/abs/1609.08076} {\  (\bibinfo {year} {2016})},\ \Eprint
  {http://arxiv.org/abs/1609.08076} {arXiv:1609.08076} \BibitemShut {NoStop}%
\bibitem [{\citenamefont {Parameswaran}\ \emph {et~al.}(2017)\citenamefont
  {Parameswaran}, \citenamefont {Potter},\ and\ \citenamefont
  {Vasseur}}]{Parameswaran2016b}%
  \BibitemOpen
  \bibfield  {author} {\bibinfo {author} {\bibfnamefont {S.~A.}\ \bibnamefont
  {Parameswaran}}, \bibinfo {author} {\bibfnamefont {A.~C.}\ \bibnamefont
  {Potter}}, \ and\ \bibinfo {author} {\bibfnamefont {R.}~\bibnamefont
  {Vasseur}},\ }\href {\doibase 10.1002/andp.201600302} {\bibfield  {journal}
  {\bibinfo  {journal} {Ann. Phys.}\ ,\ \bibinfo {pages} {1600302}} (\bibinfo
  {year} {2017})}\BibitemShut {NoStop}%
\bibitem [{\citenamefont {Agarwal}\ \emph {et~al.}(2017)\citenamefont
  {Agarwal}, \citenamefont {Altman}, \citenamefont {Demler}, \citenamefont
  {Gopalakrishnan}, \citenamefont {Huse},\ and\ \citenamefont
  {Knap}}]{Agarwal2016_review}%
  \BibitemOpen
  \bibfield  {author} {\bibinfo {author} {\bibfnamefont {K.}~\bibnamefont
  {Agarwal}}, \bibinfo {author} {\bibfnamefont {E.}~\bibnamefont {Altman}},
  \bibinfo {author} {\bibfnamefont {E.}~\bibnamefont {Demler}}, \bibinfo
  {author} {\bibfnamefont {S.}~\bibnamefont {Gopalakrishnan}}, \bibinfo
  {author} {\bibfnamefont {D.~A.}\ \bibnamefont {Huse}}, \ and\ \bibinfo
  {author} {\bibfnamefont {M.}~\bibnamefont {Knap}},\ }\href {\doibase
  10.1002/andp.201600326} {\bibfield  {journal} {\bibinfo  {journal} {Ann.
  Phys.}\ ,\ \bibinfo {pages} {1600326}} (\bibinfo {year} {2017})}\BibitemShut
  {NoStop}%
\bibitem [{\citenamefont {Haldar}\ and\ \citenamefont
  {Das}(2017)}]{haldar_dynamical_2017}%
  \BibitemOpen
  \bibfield  {author} {\bibinfo {author} {\bibfnamefont {A.}~\bibnamefont
  {Haldar}}\ and\ \bibinfo {author} {\bibfnamefont {A.}~\bibnamefont {Das}},\
  }\href@noop {} {\enquote {\bibinfo {title} {Dynamical {Many}-body
  {Localization} and {Delocalization} in {Periodically} {Driven} {Closed}
  {Quantum} {Systems}},}\ } (\bibinfo {year} {2017}),\ \Eprint
  {http://arxiv.org/abs/1702.03455} {arXiv:1702.03455} \BibitemShut {NoStop}%
\bibitem [{\citenamefont {Abanin}\ and\ \citenamefont
  {Papi{\'{c}}}(2017)}]{Abanin2017}%
  \BibitemOpen
  \bibfield  {author} {\bibinfo {author} {\bibfnamefont {D.~A.}\ \bibnamefont
  {Abanin}}\ and\ \bibinfo {author} {\bibfnamefont {Z.}~\bibnamefont
  {Papi{\'{c}}}},\ }\href {http://arxiv.org/abs/1705.09103} {\enquote {\bibinfo
  {title} {{Recent progress in many-body localization}},}\ } (\bibinfo {year}
  {2017}),\ \Eprint {http://arxiv.org/abs/1705.09103} {arXiv:1705.09103}
  \BibitemShut {NoStop}%
\bibitem [{\citenamefont {Deng}\ \emph {et~al.}(2016)\citenamefont {Deng},
  \citenamefont {Ganeshan}, \citenamefont {Li}, \citenamefont {Modak},
  \citenamefont {Mukerjee},\ and\ \citenamefont {Pixley}}]{Deng2016b}%
  \BibitemOpen
  \bibfield  {author} {\bibinfo {author} {\bibfnamefont {D.-L.}\ \bibnamefont
  {Deng}}, \bibinfo {author} {\bibfnamefont {S.}~\bibnamefont {Ganeshan}},
  \bibinfo {author} {\bibfnamefont {X.}~\bibnamefont {Li}}, \bibinfo {author}
  {\bibfnamefont {R.}~\bibnamefont {Modak}}, \bibinfo {author} {\bibfnamefont
  {S.}~\bibnamefont {Mukerjee}}, \ and\ \bibinfo {author} {\bibfnamefont
  {J.~H.}\ \bibnamefont {Pixley}},\ }\href {http://arxiv.org/abs/1612.00976} {\
   (\bibinfo {year} {2016})},\ \Eprint {http://arxiv.org/abs/1612.00976}
  {arXiv:1612.00976} \BibitemShut {NoStop}%
\bibitem [{\citenamefont {Jordan}\ and\ \citenamefont
  {Wigner}(1928)}]{Jordan1928}%
  \BibitemOpen
  \bibfield  {author} {\bibinfo {author} {\bibfnamefont {P.}~\bibnamefont
  {Jordan}}\ and\ \bibinfo {author} {\bibfnamefont {E.}~\bibnamefont
  {Wigner}},\ }\href {\doibase 10.1007/BF01331938} {\bibfield  {journal}
  {\bibinfo  {journal} {Z. Phys.}\ }\textbf {\bibinfo {volume} {47}},\ \bibinfo
  {pages} {631} (\bibinfo {year} {1928})}\BibitemShut {NoStop}%
\bibitem [{\citenamefont {Luitz}\ \emph {et~al.}(2015)\citenamefont {Luitz},
  \citenamefont {Laflorencie},\ and\ \citenamefont {Alet}}]{Luitz2015}%
  \BibitemOpen
  \bibfield  {author} {\bibinfo {author} {\bibfnamefont {D.~J.}\ \bibnamefont
  {Luitz}}, \bibinfo {author} {\bibfnamefont {N.}~\bibnamefont {Laflorencie}},
  \ and\ \bibinfo {author} {\bibfnamefont {F.}~\bibnamefont {Alet}},\ }\href
  {\doibase 10.1103/PhysRevB.91.081103} {\bibfield  {journal} {\bibinfo
  {journal} {Phys. Rev. B}\ }\textbf {\bibinfo {volume} {91}},\ \bibinfo
  {pages} {081103} (\bibinfo {year} {2015})}\BibitemShut {NoStop}%
\bibitem [{\citenamefont {Potter}\ and\ \citenamefont
  {Vasseur}(2016)}]{Potter2016}%
  \BibitemOpen
  \bibfield  {author} {\bibinfo {author} {\bibfnamefont {A.~C.}\ \bibnamefont
  {Potter}}\ and\ \bibinfo {author} {\bibfnamefont {R.}~\bibnamefont
  {Vasseur}},\ }\href {\doibase 10.1103/PhysRevB.94.224206} {\bibfield
  {journal} {\bibinfo  {journal} {Phys. Rev. B}\ }\textbf {\bibinfo {volume}
  {94}},\ \bibinfo {pages} {224206} (\bibinfo {year} {2016})}\BibitemShut
  {NoStop}%
\bibitem [{\citenamefont {Prelov{\v{s}}ek}\ \emph {et~al.}(2016)\citenamefont
  {Prelov{\v{s}}ek}, \citenamefont {Bari{\v{s}}i{\'{c}}},\ and\ \citenamefont
  {{\v{Z}}nidari{\v{c}}}}]{Prelovsek2016b}%
  \BibitemOpen
  \bibfield  {author} {\bibinfo {author} {\bibfnamefont {P.}~\bibnamefont
  {Prelov{\v{s}}ek}}, \bibinfo {author} {\bibfnamefont {O.~S.}\ \bibnamefont
  {Bari{\v{s}}i{\'{c}}}}, \ and\ \bibinfo {author} {\bibfnamefont
  {M.}~\bibnamefont {{\v{Z}}nidari{\v{c}}}},\ }\href {\doibase
  10.1103/PhysRevB.94.241104} {\bibfield  {journal} {\bibinfo  {journal} {Phys.
  Rev. B}\ }\textbf {\bibinfo {volume} {94}},\ \bibinfo {pages} {241104}
  (\bibinfo {year} {2016})}\BibitemShut {NoStop}%
\bibitem [{\citenamefont {D'Alessio}\ and\ \citenamefont
  {Polkovnikov}(2013)}]{DAlessio2013}%
  \BibitemOpen
  \bibfield  {author} {\bibinfo {author} {\bibfnamefont {L.}~\bibnamefont
  {D'Alessio}}\ and\ \bibinfo {author} {\bibfnamefont {A.}~\bibnamefont
  {Polkovnikov}},\ }\href {\doibase 10.1016/j.aop.2013.02.011} {\bibfield
  {journal} {\bibinfo  {journal} {Ann. Phys. (N. Y).}\ }\textbf {\bibinfo
  {volume} {333}},\ \bibinfo {pages} {19} (\bibinfo {year} {2013})}\BibitemShut
  {NoStop}%
\bibitem [{\citenamefont {Ponte}\ \emph
  {et~al.}(2015{\natexlab{a}})\citenamefont {Ponte}, \citenamefont {Chandran},
  \citenamefont {Papi{\'{c}}},\ and\ \citenamefont {Abanin}}]{Ponte2014a}%
  \BibitemOpen
  \bibfield  {author} {\bibinfo {author} {\bibfnamefont {P.}~\bibnamefont
  {Ponte}}, \bibinfo {author} {\bibfnamefont {A.}~\bibnamefont {Chandran}},
  \bibinfo {author} {\bibfnamefont {Z.}~\bibnamefont {Papi{\'{c}}}}, \ and\
  \bibinfo {author} {\bibfnamefont {D.~A.}\ \bibnamefont {Abanin}},\ }\href
  {\doibase 10.1016/j.aop.2014.11.008} {\bibfield  {journal} {\bibinfo
  {journal} {Ann. Phys. (N. Y).}\ }\textbf {\bibinfo {volume} {353}},\ \bibinfo
  {pages} {196} (\bibinfo {year} {2015}{\natexlab{a}})}\BibitemShut {NoStop}%
\bibitem [{\citenamefont {Lazarides}\ \emph {et~al.}(2015)\citenamefont
  {Lazarides}, \citenamefont {Das},\ and\ \citenamefont
  {Moessner}}]{lazarides_fate_2015}%
  \BibitemOpen
  \bibfield  {author} {\bibinfo {author} {\bibfnamefont {A.}~\bibnamefont
  {Lazarides}}, \bibinfo {author} {\bibfnamefont {A.}~\bibnamefont {Das}}, \
  and\ \bibinfo {author} {\bibfnamefont {R.}~\bibnamefont {Moessner}},\ }\href
  {\doibase 10.1103/PhysRevLett.115.030402} {\bibfield  {journal} {\bibinfo
  {journal} {Physical Review Letters}\ }\textbf {\bibinfo {volume} {115}},\
  \bibinfo {pages} {030402} (\bibinfo {year} {2015})}\BibitemShut {NoStop}%
\bibitem [{\citenamefont {Abanin}\ \emph {et~al.}(2015)\citenamefont {Abanin},
  \citenamefont {{De Roeck}},\ and\ \citenamefont {Huveneers}}]{Abanin2015a}%
  \BibitemOpen
  \bibfield  {author} {\bibinfo {author} {\bibfnamefont {D.~A.}\ \bibnamefont
  {Abanin}}, \bibinfo {author} {\bibfnamefont {W.}~\bibnamefont {{De Roeck}}},
  \ and\ \bibinfo {author} {\bibfnamefont {F.}~\bibnamefont {Huveneers}},\
  }\href {\doibase 10.1103/PhysRevLett.115.256803} {\bibfield  {journal}
  {\bibinfo  {journal} {Phys. Rev. Lett.}\ }\textbf {\bibinfo {volume} {115}},\
  \bibinfo {pages} {256803} (\bibinfo {year} {2015})}\BibitemShut {NoStop}%
\bibitem [{\citenamefont {Ponte}\ \emph
  {et~al.}(2015{\natexlab{b}})\citenamefont {Ponte}, \citenamefont {Papi{\'c}},
  \citenamefont {Huveneers},\ and\ \citenamefont
  {Abanin}}]{ponte_many-body_2015}%
  \BibitemOpen
  \bibfield  {author} {\bibinfo {author} {\bibfnamefont {P.}~\bibnamefont
  {Ponte}}, \bibinfo {author} {\bibfnamefont {Z.}~\bibnamefont {Papi{\'c}}},
  \bibinfo {author} {\bibfnamefont {F.}~\bibnamefont {Huveneers}}, \ and\
  \bibinfo {author} {\bibfnamefont {D.~A.}\ \bibnamefont {Abanin}},\ }\href
  {\doibase 10.1103/PhysRevLett.114.140401} {\bibfield  {journal} {\bibinfo
  {journal} {Physical Review Letters}\ }\textbf {\bibinfo {volume} {114}},\
  \bibinfo {pages} {140401} (\bibinfo {year} {2015}{\natexlab{b}})}\BibitemShut
  {NoStop}%
\bibitem [{\citenamefont {Abanin}\ \emph {et~al.}(2016)\citenamefont {Abanin},
  \citenamefont {De~Roeck},\ and\ \citenamefont
  {Huveneers}}]{abanin_theory_2016}%
  \BibitemOpen
  \bibfield  {author} {\bibinfo {author} {\bibfnamefont {D.~A.}\ \bibnamefont
  {Abanin}}, \bibinfo {author} {\bibfnamefont {W.}~\bibnamefont {De~Roeck}}, \
  and\ \bibinfo {author} {\bibfnamefont {F.}~\bibnamefont {Huveneers}},\ }\href
  {\doibase 10.1016/j.aop.2016.03.010} {\bibfield  {journal} {\bibinfo
  {journal} {Annals of Physics}\ }\textbf {\bibinfo {volume} {372}},\ \bibinfo
  {pages} {1} (\bibinfo {year} {2016})}\BibitemShut {NoStop}%
\bibitem [{\citenamefont {Bordia}\ \emph
  {et~al.}(2016{\natexlab{b}})\citenamefont {Bordia}, \citenamefont
  {L{\"{u}}schen}, \citenamefont {Schneider}, \citenamefont {Knap},\ and\
  \citenamefont {Bloch}}]{Bordia2016}%
  \BibitemOpen
  \bibfield  {author} {\bibinfo {author} {\bibfnamefont {P.}~\bibnamefont
  {Bordia}}, \bibinfo {author} {\bibfnamefont {H.}~\bibnamefont
  {L{\"{u}}schen}}, \bibinfo {author} {\bibfnamefont {U.}~\bibnamefont
  {Schneider}}, \bibinfo {author} {\bibfnamefont {M.}~\bibnamefont {Knap}}, \
  and\ \bibinfo {author} {\bibfnamefont {I.}~\bibnamefont {Bloch}},\ }\href
  {http://arxiv.org/abs/1607.07868} {\  (\bibinfo {year}
  {2016}{\natexlab{b}})},\ \Eprint {http://arxiv.org/abs/1607.07868}
  {arXiv:1607.07868} \BibitemShut {NoStop}%
\bibitem [{\citenamefont {Rehn}\ \emph {et~al.}(2016)\citenamefont {Rehn},
  \citenamefont {Lazarides}, \citenamefont {Pollmann},\ and\ \citenamefont
  {Moessner}}]{Rehn2016}%
  \BibitemOpen
  \bibfield  {author} {\bibinfo {author} {\bibfnamefont {J.}~\bibnamefont
  {Rehn}}, \bibinfo {author} {\bibfnamefont {A.}~\bibnamefont {Lazarides}},
  \bibinfo {author} {\bibfnamefont {F.}~\bibnamefont {Pollmann}}, \ and\
  \bibinfo {author} {\bibfnamefont {R.}~\bibnamefont {Moessner}},\ }\href
  {\doibase 10.1103/PhysRevB.94.020201} {\bibfield  {journal} {\bibinfo
  {journal} {Phys. Rev. B}\ }\textbf {\bibinfo {volume} {94}},\ \bibinfo
  {pages} {020201} (\bibinfo {year} {2016})}\BibitemShut {NoStop}%
\bibitem [{\citenamefont {Zhang}\ \emph
  {et~al.}(2016{\natexlab{a}})\citenamefont {Zhang}, \citenamefont {Khemani},\
  and\ \citenamefont {Huse}}]{Zhang2016b}%
  \BibitemOpen
  \bibfield  {author} {\bibinfo {author} {\bibfnamefont {L.}~\bibnamefont
  {Zhang}}, \bibinfo {author} {\bibfnamefont {V.}~\bibnamefont {Khemani}}, \
  and\ \bibinfo {author} {\bibfnamefont {D.~A.}\ \bibnamefont {Huse}},\ }\href
  {\doibase 10.1103/PhysRevB.94.224202} {\bibfield  {journal} {\bibinfo
  {journal} {Phys. Rev. B}\ }\textbf {\bibinfo {volume} {94}},\ \bibinfo
  {pages} {224202} (\bibinfo {year} {2016}{\natexlab{a}})}\BibitemShut
  {NoStop}%
\bibitem [{\citenamefont {Goldstein}\ \emph {et~al.}(2010)\citenamefont
  {Goldstein}, \citenamefont {Lebowitz}, \citenamefont {Tumulka},\ and\
  \citenamefont {Zangh{\`{i}}}}]{Goldstein2010}%
  \BibitemOpen
  \bibfield  {author} {\bibinfo {author} {\bibfnamefont {S.}~\bibnamefont
  {Goldstein}}, \bibinfo {author} {\bibfnamefont {J.~L.}\ \bibnamefont
  {Lebowitz}}, \bibinfo {author} {\bibfnamefont {R.}~\bibnamefont {Tumulka}}, \
  and\ \bibinfo {author} {\bibfnamefont {N.}~\bibnamefont {Zangh{\`{i}}}},\
  }\href {\doibase 10.1140/epjh/e2010-00007-7} {\bibfield  {journal} {\bibinfo
  {journal} {Eur. Phys. J. H}\ }\textbf {\bibinfo {volume} {35}},\ \bibinfo
  {pages} {173} (\bibinfo {year} {2010})}\BibitemShut {NoStop}%
\bibitem [{\citenamefont {Bohigas}\ \emph {et~al.}(1986)\citenamefont
  {Bohigas}, \citenamefont {Giannoni},\ and\ \citenamefont
  {Schmit}}]{Bohigas1986}%
  \BibitemOpen
  \bibfield  {author} {\bibinfo {author} {\bibfnamefont {O.}~\bibnamefont
  {Bohigas}}, \bibinfo {author} {\bibfnamefont {M.-J.}\ \bibnamefont
  {Giannoni}}, \ and\ \bibinfo {author} {\bibfnamefont {C.}~\bibnamefont
  {Schmit}},\ }in\ \href {\doibase 10.1007/3-540-17171-1_2} {\emph {\bibinfo
  {booktitle} {Quantum Chaos Stat. Nucl. Phys.}}}\ (\bibinfo {year} {1986})\
  pp.\ \bibinfo {pages} {18--40}\BibitemShut {NoStop}%
\bibitem [{\citenamefont {Andreev}\ \emph {et~al.}(1996)\citenamefont
  {Andreev}, \citenamefont {Agam}, \citenamefont {Simons},\ and\ \citenamefont
  {Altshuler}}]{Andreev1996}%
  \BibitemOpen
  \bibfield  {author} {\bibinfo {author} {\bibfnamefont {A.}~\bibnamefont
  {Andreev}}, \bibinfo {author} {\bibfnamefont {O.}~\bibnamefont {Agam}},
  \bibinfo {author} {\bibfnamefont {B.}~\bibnamefont {Simons}}, \ and\ \bibinfo
  {author} {\bibfnamefont {B.~L.}\ \bibnamefont {Altshuler}},\ }\href {\doibase
  10.1103/PhysRevLett.76.3947} {\bibfield  {journal} {\bibinfo  {journal}
  {Phys. Rev. Lett.}\ }\textbf {\bibinfo {volume} {76}},\ \bibinfo {pages}
  {3947} (\bibinfo {year} {1996})}\BibitemShut {NoStop}%
\bibitem [{\citenamefont {Montambaux}\ \emph {et~al.}(1993)\citenamefont
  {Montambaux}, \citenamefont {Poilblanc}, \citenamefont {Bellissard},\ and\
  \citenamefont {Sire}}]{Montambaux1993}%
  \BibitemOpen
  \bibfield  {author} {\bibinfo {author} {\bibfnamefont {G.}~\bibnamefont
  {Montambaux}}, \bibinfo {author} {\bibfnamefont {D.}~\bibnamefont
  {Poilblanc}}, \bibinfo {author} {\bibfnamefont {J.}~\bibnamefont
  {Bellissard}}, \ and\ \bibinfo {author} {\bibfnamefont {C.}~\bibnamefont
  {Sire}},\ }\href {\doibase 10.1103/PhysRevLett.70.497} {\bibfield  {journal}
  {\bibinfo  {journal} {Phys. Rev. Lett.}\ }\textbf {\bibinfo {volume} {70}},\
  \bibinfo {pages} {497} (\bibinfo {year} {1993})}\BibitemShut {NoStop}%
\bibitem [{\citenamefont {Poilblanc}\ \emph {et~al.}(1993)\citenamefont
  {Poilblanc}, \citenamefont {Ziman}, \citenamefont {Bellissard}, \citenamefont
  {Mila},\ and\ \citenamefont {Montambaux}}]{Poilblanc1993}%
  \BibitemOpen
  \bibfield  {author} {\bibinfo {author} {\bibfnamefont {D.}~\bibnamefont
  {Poilblanc}}, \bibinfo {author} {\bibfnamefont {T.}~\bibnamefont {Ziman}},
  \bibinfo {author} {\bibfnamefont {J.}~\bibnamefont {Bellissard}}, \bibinfo
  {author} {\bibfnamefont {F.}~\bibnamefont {Mila}}, \ and\ \bibinfo {author}
  {\bibfnamefont {G.}~\bibnamefont {Montambaux}},\ }\href {\doibase
  10.1209/0295-5075/22/7/010} {\bibfield  {journal} {\bibinfo  {journal} {EPL}\
  }\textbf {\bibinfo {volume} {22}},\ \bibinfo {pages} {537} (\bibinfo {year}
  {1993})}\BibitemShut {NoStop}%
\bibitem [{\citenamefont {Berkovits}(1994)}]{Berkovits1994}%
  \BibitemOpen
  \bibfield  {author} {\bibinfo {author} {\bibfnamefont {R.}~\bibnamefont
  {Berkovits}},\ }\href {\doibase 10.1209/0295-5075/25/9/008} {\bibfield
  {journal} {\bibinfo  {journal} {EPL}\ }\textbf {\bibinfo {volume} {25}},\
  \bibinfo {pages} {681} (\bibinfo {year} {1994})}\BibitemShut {NoStop}%
\bibitem [{\citenamefont {Berkovits}\ and\ \citenamefont
  {Avishai}(1996)}]{Berkovits1996}%
  \BibitemOpen
  \bibfield  {author} {\bibinfo {author} {\bibfnamefont {R.}~\bibnamefont
  {Berkovits}}\ and\ \bibinfo {author} {\bibfnamefont {Y.}~\bibnamefont
  {Avishai}},\ }\href {\doibase 10.1088/0953-8984/8/4/006} {\bibfield
  {journal} {\bibinfo  {journal} {J. Phys. Condens. Matter}\ }\textbf {\bibinfo
  {volume} {8}},\ \bibinfo {pages} {389} (\bibinfo {year} {1996})}\BibitemShut
  {NoStop}%
\bibitem [{\citenamefont {Jacquod}\ and\ \citenamefont
  {Shepelyansky}(1997)}]{Jacquod1997a}%
  \BibitemOpen
  \bibfield  {author} {\bibinfo {author} {\bibfnamefont {P.}~\bibnamefont
  {Jacquod}}\ and\ \bibinfo {author} {\bibfnamefont {D.~L.}\ \bibnamefont
  {Shepelyansky}},\ }\href {\doibase 10.1103/PhysRevLett.79.1837} {\bibfield
  {journal} {\bibinfo  {journal} {Phys. Rev. Lett.}\ }\textbf {\bibinfo
  {volume} {79}},\ \bibinfo {pages} {1837} (\bibinfo {year}
  {1997})}\BibitemShut {NoStop}%
\bibitem [{\citenamefont {Georgeot}\ and\ \citenamefont
  {Shepelyansky}(1998)}]{georgeot_integrability_1998}%
  \BibitemOpen
  \bibfield  {author} {\bibinfo {author} {\bibfnamefont {B.}~\bibnamefont
  {Georgeot}}\ and\ \bibinfo {author} {\bibfnamefont {D.~L.}\ \bibnamefont
  {Shepelyansky}},\ }\href {\doibase 10.1103/PhysRevLett.81.5129} {\bibfield
  {journal} {\bibinfo  {journal} {Physical Review Letters}\ }\textbf {\bibinfo
  {volume} {81}},\ \bibinfo {pages} {5129} (\bibinfo {year}
  {1998})}\BibitemShut {NoStop}%
\bibitem [{\citenamefont {Avishai}\ \emph {et~al.}(2002)\citenamefont
  {Avishai}, \citenamefont {Richert},\ and\ \citenamefont
  {Berkovits}}]{Avishai2002}%
  \BibitemOpen
  \bibfield  {author} {\bibinfo {author} {\bibfnamefont {Y.}~\bibnamefont
  {Avishai}}, \bibinfo {author} {\bibfnamefont {J.}~\bibnamefont {Richert}}, \
  and\ \bibinfo {author} {\bibfnamefont {R.}~\bibnamefont {Berkovits}},\ }\href
  {\doibase 10.1103/PhysRevB.66.052416} {\bibfield  {journal} {\bibinfo
  {journal} {Phys. Rev. B}\ }\textbf {\bibinfo {volume} {66}},\ \bibinfo
  {pages} {2} (\bibinfo {year} {2002})}\BibitemShut {NoStop}%
\bibitem [{\citenamefont {Santos}(2004)}]{Santos2004}%
  \BibitemOpen
  \bibfield  {author} {\bibinfo {author} {\bibfnamefont {L.~F.}\ \bibnamefont
  {Santos}},\ }\href {\doibase 10.1088/0305-4470/37/17/004} {\bibfield
  {journal} {\bibinfo  {journal} {J. Phys. A. Math. Gen.}\ }\textbf {\bibinfo
  {volume} {37}},\ \bibinfo {pages} {4723} (\bibinfo {year}
  {2004})}\BibitemShut {NoStop}%
\bibitem [{\citenamefont {Santos}\ \emph {et~al.}(2004)\citenamefont {Santos},
  \citenamefont {Rigolin},\ and\ \citenamefont {Escobar}}]{Santos2004a}%
  \BibitemOpen
  \bibfield  {author} {\bibinfo {author} {\bibfnamefont {L.~F.}\ \bibnamefont
  {Santos}}, \bibinfo {author} {\bibfnamefont {G.}~\bibnamefont {Rigolin}}, \
  and\ \bibinfo {author} {\bibfnamefont {C.~O.}\ \bibnamefont {Escobar}},\
  }\href {\doibase 10.1103/PhysRevA.69.042304} {\bibfield  {journal} {\bibinfo
  {journal} {Phys. Rev. A}\ }\textbf {\bibinfo {volume} {69}},\ \bibinfo
  {pages} {042304} (\bibinfo {year} {2004})}\BibitemShut {NoStop}%
\bibitem [{\citenamefont {Oganesyan}\ and\ \citenamefont
  {Huse}(2007)}]{oganesyan_localization_2007}%
  \BibitemOpen
  \bibfield  {author} {\bibinfo {author} {\bibfnamefont {V.}~\bibnamefont
  {Oganesyan}}\ and\ \bibinfo {author} {\bibfnamefont {D.~A.}\ \bibnamefont
  {Huse}},\ }\href {\doibase 10.1103/PhysRevB.75.155111} {\bibfield  {journal}
  {\bibinfo  {journal} {Phys. Rev. B}\ }\textbf {\bibinfo {volume} {75}},\
  \bibinfo {pages} {155111} (\bibinfo {year} {2007})}\BibitemShut {NoStop}%
\bibitem [{\citenamefont {Guhr}\ \emph {et~al.}(1998)\citenamefont {Guhr},
  \citenamefont {M{\"{u}}ller-Groeling},\ and\ \citenamefont
  {Weidenm{\"{u}}ller}}]{Guhr1998}%
  \BibitemOpen
  \bibfield  {author} {\bibinfo {author} {\bibfnamefont {T.}~\bibnamefont
  {Guhr}}, \bibinfo {author} {\bibfnamefont {A.}~\bibnamefont
  {M{\"{u}}ller-Groeling}}, \ and\ \bibinfo {author} {\bibfnamefont {H.~A.}\
  \bibnamefont {Weidenm{\"{u}}ller}},\ }\href {\doibase
  10.1016/S0370-1573(97)00088-4} {\bibfield  {journal} {\bibinfo  {journal}
  {Phys. Rep.}\ }\textbf {\bibinfo {volume} {299}},\ \bibinfo {pages} {189}
  (\bibinfo {year} {1998})}\BibitemShut {NoStop}%
\bibitem [{\citenamefont {Oganesyan}\ \emph {et~al.}(2009)\citenamefont
  {Oganesyan}, \citenamefont {Pal},\ and\ \citenamefont
  {Huse}}]{Oganesyan2009}%
  \BibitemOpen
  \bibfield  {author} {\bibinfo {author} {\bibfnamefont {V.}~\bibnamefont
  {Oganesyan}}, \bibinfo {author} {\bibfnamefont {A.}~\bibnamefont {Pal}}, \
  and\ \bibinfo {author} {\bibfnamefont {D.~A.}\ \bibnamefont {Huse}},\ }\href
  {\doibase 10.1103/PhysRevB.80.115104} {\bibfield  {journal} {\bibinfo
  {journal} {Phys. Rev. B}\ }\textbf {\bibinfo {volume} {80}},\ \bibinfo
  {pages} {115104} (\bibinfo {year} {2009})}\BibitemShut {NoStop}%
\bibitem [{\citenamefont {Atas}\ \emph {et~al.}(2013)\citenamefont {Atas},
  \citenamefont {Bogomolny}, \citenamefont {Giraud},\ and\ \citenamefont
  {Roux}}]{Atas2013}%
  \BibitemOpen
  \bibfield  {author} {\bibinfo {author} {\bibfnamefont {Y.~Y.}\ \bibnamefont
  {Atas}}, \bibinfo {author} {\bibfnamefont {E.}~\bibnamefont {Bogomolny}},
  \bibinfo {author} {\bibfnamefont {O.}~\bibnamefont {Giraud}}, \ and\ \bibinfo
  {author} {\bibfnamefont {G.}~\bibnamefont {Roux}},\ }\href {\doibase
  10.1103/PhysRevLett.110.084101} {\bibfield  {journal} {\bibinfo  {journal}
  {Phys. Rev. Lett.}\ }\textbf {\bibinfo {volume} {110}},\ \bibinfo {pages}
  {084101} (\bibinfo {year} {2013})}\BibitemShut {NoStop}%
\bibitem [{\citenamefont {Dyson}(1962)}]{Dyson1962}%
  \BibitemOpen
  \bibfield  {author} {\bibinfo {author} {\bibfnamefont {F.~J.}\ \bibnamefont
  {Dyson}},\ }\href {\doibase 10.1063/1.1703862} {\bibfield  {journal}
  {\bibinfo  {journal} {J. Math. Phys.}\ }\textbf {\bibinfo {volume} {3}},\
  \bibinfo {pages} {1191} (\bibinfo {year} {1962})}\BibitemShut {NoStop}%
\bibitem [{\citenamefont {Chalker}\ \emph {et~al.}(1996)\citenamefont
  {Chalker}, \citenamefont {Lerner},\ and\ \citenamefont
  {Smith}}]{chalker_fictitious_1996}%
  \BibitemOpen
  \bibfield  {author} {\bibinfo {author} {\bibfnamefont {J.~T.}\ \bibnamefont
  {Chalker}}, \bibinfo {author} {\bibfnamefont {I.~V.}\ \bibnamefont {Lerner}},
  \ and\ \bibinfo {author} {\bibfnamefont {R.~A.}\ \bibnamefont {Smith}},\
  }\href {\doibase 10.1063/1.531676} {\bibfield  {journal} {\bibinfo  {journal}
  {Journal of Mathematical Physics}\ }\textbf {\bibinfo {volume} {37}},\
  \bibinfo {pages} {5061} (\bibinfo {year} {1996})}\BibitemShut {NoStop}%
\bibitem [{\citenamefont {Serbyn}\ and\ \citenamefont
  {Moore}(2016)}]{Serbyn2015}%
  \BibitemOpen
  \bibfield  {author} {\bibinfo {author} {\bibfnamefont {M.}~\bibnamefont
  {Serbyn}}\ and\ \bibinfo {author} {\bibfnamefont {J.~E.}\ \bibnamefont
  {Moore}},\ }\href {\doibase 10.1103/PhysRevB.93.041424} {\bibfield  {journal}
  {\bibinfo  {journal} {Phys. Rev. B}\ }\textbf {\bibinfo {volume} {93}},\
  \bibinfo {pages} {041424} (\bibinfo {year} {2016})}\BibitemShut {NoStop}%
\bibitem [{\citenamefont {Evers}\ and\ \citenamefont
  {Mirlin}(2008)}]{Evers2008a}%
  \BibitemOpen
  \bibfield  {author} {\bibinfo {author} {\bibfnamefont {F.}~\bibnamefont
  {Evers}}\ and\ \bibinfo {author} {\bibfnamefont {A.}~\bibnamefont {Mirlin}},\
  }\href {\doibase 10.1103/RevModPhys.80.1355} {\bibfield  {journal} {\bibinfo
  {journal} {Rev. Mod. Phys.}\ }\textbf {\bibinfo {volume} {80}},\ \bibinfo
  {pages} {1355} (\bibinfo {year} {2008})}\BibitemShut {NoStop}%
\bibitem [{Note1()}]{Note1}%
  \BibitemOpen
  \bibinfo {note} {We note that the level statistics with {$\beta >0$ and
  $\gamma =1$} was called a semi-Poisson statistics in Ref.~\cite {Serbyn2015}.
  To eliminate the confusion with semi-Poisson statistics which was introduced
  in Ref.~\cite {Bogomolny1999} and implies {$\beta =1$ and $\gamma =1$}, we
  have instead used the term ``critical statistics.''}\BibitemShut {NoStop}%
\bibitem [{\citenamefont {Bertrand}\ and\ \citenamefont
  {Garc{\'{i}}a-Garc{\'{i}}a}(2016)}]{Bertrand2016}%
  \BibitemOpen
  \bibfield  {author} {\bibinfo {author} {\bibfnamefont {C.~L.}\ \bibnamefont
  {Bertrand}}\ and\ \bibinfo {author} {\bibfnamefont {A.~M.}\ \bibnamefont
  {Garc{\'{i}}a-Garc{\'{i}}a}},\ }\href {\doibase 10.1103/PhysRevB.94.144201}
  {\bibfield  {journal} {\bibinfo  {journal} {Phys. Rev. B}\ }\textbf {\bibinfo
  {volume} {94}},\ \bibinfo {pages} {144201} (\bibinfo {year}
  {2016})}\BibitemShut {NoStop}%
\bibitem [{\citenamefont {Altshuler}\ \emph {et~al.}(1997)\citenamefont
  {Altshuler}, \citenamefont {Gefen}, \citenamefont {Kamenev},\ and\
  \citenamefont {Levitov}}]{Altshuler1997}%
  \BibitemOpen
  \bibfield  {author} {\bibinfo {author} {\bibfnamefont {B.~L.}\ \bibnamefont
  {Altshuler}}, \bibinfo {author} {\bibfnamefont {Y.}~\bibnamefont {Gefen}},
  \bibinfo {author} {\bibfnamefont {A.}~\bibnamefont {Kamenev}}, \ and\
  \bibinfo {author} {\bibfnamefont {L.~S.}\ \bibnamefont {Levitov}},\ }\href
  {\doibase 10.1103/PhysRevLett.78.2803} {\bibfield  {journal} {\bibinfo
  {journal} {Phys. Rev. Lett.}\ }\textbf {\bibinfo {volume} {78}},\ \bibinfo
  {pages} {2803} (\bibinfo {year} {1997})}\BibitemShut {NoStop}%
\bibitem [{\citenamefont {Altshuler}(2010)}]{Altshuler2010}%
  \BibitemOpen
  \bibfield  {author} {\bibinfo {author} {\bibfnamefont {B.~L.}\ \bibnamefont
  {Altshuler}},\ }\href
  {http://www.lancaster.ac.uk/users/esqn/windsor10/lectures/Altshuler.pdf}
  {\enquote {\bibinfo {title} {{Many-Body Localization}},}\ } (\bibinfo {year}
  {2010})\BibitemShut {NoStop}%
\bibitem [{\citenamefont {Ketzmerick}\ \emph {et~al.}(1997)\citenamefont
  {Ketzmerick}, \citenamefont {Kruse}, \citenamefont {Kraut},\ and\
  \citenamefont {Geisel}}]{Ketzmerick1997}%
  \BibitemOpen
  \bibfield  {author} {\bibinfo {author} {\bibfnamefont {R.}~\bibnamefont
  {Ketzmerick}}, \bibinfo {author} {\bibfnamefont {K.}~\bibnamefont {Kruse}},
  \bibinfo {author} {\bibfnamefont {S.}~\bibnamefont {Kraut}}, \ and\ \bibinfo
  {author} {\bibfnamefont {T.}~\bibnamefont {Geisel}},\ }\href {\doibase
  10.1103/PhysRevLett.79.1959} {\bibfield  {journal} {\bibinfo  {journal}
  {Phys. Rev. Lett.}\ }\textbf {\bibinfo {volume} {79}},\ \bibinfo {pages}
  {1959} (\bibinfo {year} {1997})}\BibitemShut {NoStop}%
\bibitem [{\citenamefont {Ohtsuki}\ and\ \citenamefont
  {Kawarabayashi}(1997)}]{Ohtsuki1997}%
  \BibitemOpen
  \bibfield  {author} {\bibinfo {author} {\bibfnamefont {T.}~\bibnamefont
  {Ohtsuki}}\ and\ \bibinfo {author} {\bibfnamefont {T.}~\bibnamefont
  {Kawarabayashi}},\ }\href {\doibase 10.1143/JPSJ.66.314} {\bibfield
  {journal} {\bibinfo  {journal} {J. Phys. Soc. Japan}\ }\textbf {\bibinfo
  {volume} {66}},\ \bibinfo {pages} {314} (\bibinfo {year} {1997})}\BibitemShut
  {NoStop}%
\bibitem [{\citenamefont {Atas}\ and\ \citenamefont
  {Bogomolny}(2012)}]{atas_multifractality_2012}%
  \BibitemOpen
  \bibfield  {author} {\bibinfo {author} {\bibfnamefont {Y.~Y.}\ \bibnamefont
  {Atas}}\ and\ \bibinfo {author} {\bibfnamefont {E.}~\bibnamefont
  {Bogomolny}},\ }\href {\doibase 10.1103/PhysRevE.86.021104} {\bibfield
  {journal} {\bibinfo  {journal} {Physical Review E}\ }\textbf {\bibinfo
  {volume} {86}},\ \bibinfo {pages} {021104} (\bibinfo {year}
  {2012})}\BibitemShut {NoStop}%
\bibitem [{\citenamefont {Luitz}\ \emph {et~al.}(2014)\citenamefont {Luitz},
  \citenamefont {Alet},\ and\ \citenamefont
  {Laflorencie}}]{luitz_universal_2014}%
  \BibitemOpen
  \bibfield  {author} {\bibinfo {author} {\bibfnamefont {D.~J.}\ \bibnamefont
  {Luitz}}, \bibinfo {author} {\bibfnamefont {F.}~\bibnamefont {Alet}}, \ and\
  \bibinfo {author} {\bibfnamefont {N.}~\bibnamefont {Laflorencie}},\ }\href
  {\doibase 10.1103/PhysRevLett.112.057203} {\bibfield  {journal} {\bibinfo
  {journal} {Physical Review Letters}\ }\textbf {\bibinfo {volume} {112}},\
  \bibinfo {pages} {057203} (\bibinfo {year} {2014})}\BibitemShut {NoStop}%
\bibitem [{\citenamefont {Borgonovi}\ \emph {et~al.}(2016)\citenamefont
  {Borgonovi}, \citenamefont {Izrailev}, \citenamefont {Santos},\ and\
  \citenamefont {Zelevinsky}}]{Borgonovi2016}%
  \BibitemOpen
  \bibfield  {author} {\bibinfo {author} {\bibfnamefont {F.}~\bibnamefont
  {Borgonovi}}, \bibinfo {author} {\bibfnamefont {F.}~\bibnamefont {Izrailev}},
  \bibinfo {author} {\bibfnamefont {L.}~\bibnamefont {Santos}}, \ and\ \bibinfo
  {author} {\bibfnamefont {V.}~\bibnamefont {Zelevinsky}},\ }\href {\doibase
  10.1016/j.physrep.2016.02.005} {\bibfield  {journal} {\bibinfo  {journal}
  {Phys. Rep.}\ }\textbf {\bibinfo {volume} {626}},\ \bibinfo {pages} {1}
  (\bibinfo {year} {2016})}\BibitemShut {NoStop}%
\bibitem [{\citenamefont {Biroli}\ \emph {et~al.}(2012)\citenamefont {Biroli},
  \citenamefont {Ribeiro-Teixeira},\ and\ \citenamefont {Tarzia}}]{Biroli2012}%
  \BibitemOpen
  \bibfield  {author} {\bibinfo {author} {\bibfnamefont {G.}~\bibnamefont
  {Biroli}}, \bibinfo {author} {\bibfnamefont {A.~C.}\ \bibnamefont
  {Ribeiro-Teixeira}}, \ and\ \bibinfo {author} {\bibfnamefont
  {M.}~\bibnamefont {Tarzia}},\ }\href {http://arxiv.org/abs/1211.7334} {\
  (\bibinfo {year} {2012})},\ \Eprint {http://arxiv.org/abs/1211.7334}
  {arXiv:1211.7334} \BibitemShut {NoStop}%
\bibitem [{\citenamefont {{De Luca}}\ \emph {et~al.}(2013)\citenamefont {{De
  Luca}}, \citenamefont {Scardicchio}, \citenamefont {Kravtsov},\ and\
  \citenamefont {Altshuler}}]{DeLuca2013}%
  \BibitemOpen
  \bibfield  {author} {\bibinfo {author} {\bibfnamefont {A.}~\bibnamefont {{De
  Luca}}}, \bibinfo {author} {\bibfnamefont {A.}~\bibnamefont {Scardicchio}},
  \bibinfo {author} {\bibfnamefont {V.~E.}\ \bibnamefont {Kravtsov}}, \ and\
  \bibinfo {author} {\bibfnamefont {B.~L.}\ \bibnamefont {Altshuler}},\ }\href
  {http://arxiv.org/abs/1401.0019} {\  (\bibinfo {year} {2013})},\ \Eprint
  {http://arxiv.org/abs/1401.0019} {arXiv:1401.0019} \BibitemShut {NoStop}%
\bibitem [{\citenamefont {{De Luca}}\ \emph {et~al.}(2014)\citenamefont {{De
  Luca}}, \citenamefont {Altshuler}, \citenamefont {Kravtsov},\ and\
  \citenamefont {Scardicchio}}]{Luca}%
  \BibitemOpen
  \bibfield  {author} {\bibinfo {author} {\bibfnamefont {A.}~\bibnamefont {{De
  Luca}}}, \bibinfo {author} {\bibfnamefont {B.~L.}\ \bibnamefont {Altshuler}},
  \bibinfo {author} {\bibfnamefont {V.~E.}\ \bibnamefont {Kravtsov}}, \ and\
  \bibinfo {author} {\bibfnamefont {A.}~\bibnamefont {Scardicchio}},\ }\href
  {\doibase 10.1103/PhysRevLett.113.046806} {\bibfield  {journal} {\bibinfo
  {journal} {Phys. Rev. Lett.}\ }\textbf {\bibinfo {volume} {113}},\ \bibinfo
  {pages} {046806} (\bibinfo {year} {2014})}\BibitemShut {NoStop}%
\bibitem [{\citenamefont {Kravtsov}\ \emph {et~al.}(2015)\citenamefont
  {Kravtsov}, \citenamefont {Khaymovich}, \citenamefont {Cuevas},\ and\
  \citenamefont {Amini}}]{Kravtsov2015}%
  \BibitemOpen
  \bibfield  {author} {\bibinfo {author} {\bibfnamefont {V.~E.}\ \bibnamefont
  {Kravtsov}}, \bibinfo {author} {\bibfnamefont {I.~M.}\ \bibnamefont
  {Khaymovich}}, \bibinfo {author} {\bibfnamefont {E.}~\bibnamefont {Cuevas}},
  \ and\ \bibinfo {author} {\bibfnamefont {M.}~\bibnamefont {Amini}},\ }\href
  {\doibase 10.1088/1367-2630/17/12/122002} {\bibfield  {journal} {\bibinfo
  {journal} {New J. Phys.}\ }\textbf {\bibinfo {volume} {17}},\ \bibinfo
  {pages} {122002} (\bibinfo {year} {2015})}\BibitemShut {NoStop}%
\bibitem [{\citenamefont {Facoetti}\ \emph {et~al.}(2016)\citenamefont
  {Facoetti}, \citenamefont {Vivo},\ and\ \citenamefont
  {Biroli}}]{Facoetti2016}%
  \BibitemOpen
  \bibfield  {author} {\bibinfo {author} {\bibfnamefont {D.}~\bibnamefont
  {Facoetti}}, \bibinfo {author} {\bibfnamefont {P.}~\bibnamefont {Vivo}}, \
  and\ \bibinfo {author} {\bibfnamefont {G.}~\bibnamefont {Biroli}},\ }\href
  {\doibase 10.1209/0295-5075/115/47003} {\bibfield  {journal} {\bibinfo
  {journal} {EPL}\ }\textbf {\bibinfo {volume} {115}},\ \bibinfo {pages}
  {47003} (\bibinfo {year} {2016})}\BibitemShut {NoStop}%
\bibitem [{\citenamefont {Tikhonov}\ \emph {et~al.}(2016)\citenamefont
  {Tikhonov}, \citenamefont {Mirlin},\ and\ \citenamefont
  {Skvortsov}}]{Tikhonov2016}%
  \BibitemOpen
  \bibfield  {author} {\bibinfo {author} {\bibfnamefont {K.~S.}\ \bibnamefont
  {Tikhonov}}, \bibinfo {author} {\bibfnamefont {A.~D.}\ \bibnamefont
  {Mirlin}}, \ and\ \bibinfo {author} {\bibfnamefont {M.~A.}\ \bibnamefont
  {Skvortsov}},\ }\href {\doibase 10.1103/PhysRevB.94.220203} {\bibfield
  {journal} {\bibinfo  {journal} {Phys. Rev. B}\ }\textbf {\bibinfo {volume}
  {94}},\ \bibinfo {pages} {220203} (\bibinfo {year} {2016})}\BibitemShut
  {NoStop}%
\bibitem [{\citenamefont {Tikhonov}\ and\ \citenamefont
  {Mirlin}(2016)}]{Tikhonov2016a}%
  \BibitemOpen
  \bibfield  {author} {\bibinfo {author} {\bibfnamefont {K.~S.}\ \bibnamefont
  {Tikhonov}}\ and\ \bibinfo {author} {\bibfnamefont {A.~D.}\ \bibnamefont
  {Mirlin}},\ }\href {\doibase 10.1103/PhysRevB.94.184203} {\bibfield
  {journal} {\bibinfo  {journal} {Phys. Rev. B}\ }\textbf {\bibinfo {volume}
  {94}},\ \bibinfo {pages} {184203} (\bibinfo {year} {2016})}\BibitemShut
  {NoStop}%
\bibitem [{\citenamefont {Garc{\'{i}}a-Mata}\ \emph {et~al.}(2016)\citenamefont
  {Garc{\'{i}}a-Mata}, \citenamefont {Giraud}, \citenamefont {Georgeot},
  \citenamefont {Martin}, \citenamefont {Dubertrand},\ and\ \citenamefont
  {Lemari{\'{e}}}}]{Garcia-Mata2016}%
  \BibitemOpen
  \bibfield  {author} {\bibinfo {author} {\bibfnamefont {I.}~\bibnamefont
  {Garc{\'{i}}a-Mata}}, \bibinfo {author} {\bibfnamefont {O.}~\bibnamefont
  {Giraud}}, \bibinfo {author} {\bibfnamefont {B.}~\bibnamefont {Georgeot}},
  \bibinfo {author} {\bibfnamefont {J.}~\bibnamefont {Martin}}, \bibinfo
  {author} {\bibfnamefont {R.}~\bibnamefont {Dubertrand}}, \ and\ \bibinfo
  {author} {\bibfnamefont {G.}~\bibnamefont {Lemari{\'{e}}}},\ }\href
  {http://arxiv.org/abs/1609.05857} {\  (\bibinfo {year} {2016})},\ \Eprint
  {http://arxiv.org/abs/1609.05857} {arXiv:1609.05857} \BibitemShut {NoStop}%
\bibitem [{\citenamefont {Altshuler}\ \emph {et~al.}(2016)\citenamefont
  {Altshuler}, \citenamefont {Cuevas}, \citenamefont {Ioffe},\ and\
  \citenamefont {Kravtsov}}]{Altshuler2016}%
  \BibitemOpen
  \bibfield  {author} {\bibinfo {author} {\bibfnamefont {B.~L.}\ \bibnamefont
  {Altshuler}}, \bibinfo {author} {\bibfnamefont {E.}~\bibnamefont {Cuevas}},
  \bibinfo {author} {\bibfnamefont {L.~B.}\ \bibnamefont {Ioffe}}, \ and\
  \bibinfo {author} {\bibfnamefont {V.~E.}\ \bibnamefont {Kravtsov}},\ }\href
  {\doibase 10.1103/PhysRevLett.117.156601} {\bibfield  {journal} {\bibinfo
  {journal} {Phys. Rev. Lett.}\ }\textbf {\bibinfo {volume} {117}},\ \bibinfo
  {pages} {156601} (\bibinfo {year} {2016})}\BibitemShut {NoStop}%
\bibitem [{\citenamefont {Pino}\ \emph {et~al.}(2016)\citenamefont {Pino},
  \citenamefont {Ioffe},\ and\ \citenamefont {Altshuler}}]{Pino2015}%
  \BibitemOpen
  \bibfield  {author} {\bibinfo {author} {\bibfnamefont {M.}~\bibnamefont
  {Pino}}, \bibinfo {author} {\bibfnamefont {L.~B.}\ \bibnamefont {Ioffe}}, \
  and\ \bibinfo {author} {\bibfnamefont {B.~L.}\ \bibnamefont {Altshuler}},\
  }\href {\doibase 10.1073/pnas.1520033113} {\bibfield  {journal} {\bibinfo
  {journal} {Proc. Nat. Acad. Sci.}\ }\textbf {\bibinfo {volume} {113}},\
  \bibinfo {pages} {536} (\bibinfo {year} {2016})}\BibitemShut {NoStop}%
\bibitem [{\citenamefont {Serbyn}\ \emph
  {et~al.}(2016{\natexlab{a}})\citenamefont {Serbyn}, \citenamefont
  {Papi{\'c}},\ and\ \citenamefont {Abanin}}]{serbyn_thouless_2016}%
  \BibitemOpen
  \bibfield  {author} {\bibinfo {author} {\bibfnamefont {M.}~\bibnamefont
  {Serbyn}}, \bibinfo {author} {\bibfnamefont {Z.}~\bibnamefont {Papi{\'c}}}, \
  and\ \bibinfo {author} {\bibfnamefont {D.~A.}\ \bibnamefont {Abanin}},\
  }\href@noop {} {\enquote {\bibinfo {title} {Thouless energy and
  multifractality across the many-body localization transition},}\ } (\bibinfo
  {year} {2016}{\natexlab{a}}),\ \Eprint {http://arxiv.org/abs/1610.02389}
  {arXiv:1610.02389} \BibitemShut {NoStop}%
\bibitem [{\citenamefont {Torres-Herrera}\ and\ \citenamefont
  {Santos}(2017)}]{Torres-Herrera2016}%
  \BibitemOpen
  \bibfield  {author} {\bibinfo {author} {\bibfnamefont {E.~J.}\ \bibnamefont
  {Torres-Herrera}}\ and\ \bibinfo {author} {\bibfnamefont {L.~F.}\
  \bibnamefont {Santos}},\ }\href {\doibase 10.1002/andp.201600284} {\bibfield
  {journal} {\bibinfo  {journal} {Ann. Phys.}\ ,\ \bibinfo {pages} {1600284}}
  (\bibinfo {year} {2017})}\BibitemShut {NoStop}%
\bibitem [{\citenamefont {Monthus}(2016{\natexlab{a}})}]{Monthus2016}%
  \BibitemOpen
  \bibfield  {author} {\bibinfo {author} {\bibfnamefont {C.}~\bibnamefont
  {Monthus}},\ }\href {\doibase 10.1088/1742-5468/2016/07/073301} {\bibfield
  {journal} {\bibinfo  {journal} {J. Stat. Mech. Theory Exp.}\ }\textbf
  {\bibinfo {volume} {2016}},\ \bibinfo {pages} {073301} (\bibinfo {year}
  {2016}{\natexlab{a}})}\BibitemShut {NoStop}%
\bibitem [{Note2()}]{Note2}%
  \BibitemOpen
  \bibinfo {note} {We note that there appears to be a misprint in Ref.~\cite
  {Serbyn2015}. Since for {$\gamma =1$, $d_{2}=1-\gamma =0$,} while from the
  authors' definition of {$\protect \mathcal {N}\DOTSB \sum@ \slimits@ _{\alpha
  }\left |\left \delimiter "426830A \alpha |\psi ^{\beta }\right \delimiter
  "526930B \right |^{4}\propto \protect \mathcal {N}^{-d_{2}}$} one gets that,
  {$I_{2}\propto \protect \mathcal {N}^{-1}$,} which corresponds to a WD
  distribution {$\left (\gamma =0\right )$.}}\BibitemShut {Stop}%
\bibitem [{\citenamefont {Rigol}\ \emph {et~al.}(2008)\citenamefont {Rigol},
  \citenamefont {Dunjko},\ and\ \citenamefont {Olshanii}}]{Rigol2008}%
  \BibitemOpen
  \bibfield  {author} {\bibinfo {author} {\bibfnamefont {M.}~\bibnamefont
  {Rigol}}, \bibinfo {author} {\bibfnamefont {V.}~\bibnamefont {Dunjko}}, \
  and\ \bibinfo {author} {\bibfnamefont {M.}~\bibnamefont {Olshanii}},\ }\href
  {\doibase 10.1038/nature06838} {\bibfield  {journal} {\bibinfo  {journal}
  {Nature}\ }\textbf {\bibinfo {volume} {452}},\ \bibinfo {pages} {854}
  (\bibinfo {year} {2008})}\BibitemShut {NoStop}%
\bibitem [{\citenamefont {D'Alessio}\ \emph {et~al.}(2016)\citenamefont
  {D'Alessio}, \citenamefont {Kafri}, \citenamefont {Polkovnikov},\ and\
  \citenamefont {Rigol}}]{DAlessio2015}%
  \BibitemOpen
  \bibfield  {author} {\bibinfo {author} {\bibfnamefont {L.}~\bibnamefont
  {D'Alessio}}, \bibinfo {author} {\bibfnamefont {Y.}~\bibnamefont {Kafri}},
  \bibinfo {author} {\bibfnamefont {A.}~\bibnamefont {Polkovnikov}}, \ and\
  \bibinfo {author} {\bibfnamefont {M.}~\bibnamefont {Rigol}},\ }\href
  {\doibase 10.1080/00018732.2016.1198134} {\bibfield  {journal} {\bibinfo
  {journal} {Adv. Phys.}\ }\textbf {\bibinfo {volume} {65}},\ \bibinfo {pages}
  {239} (\bibinfo {year} {2016})}\BibitemShut {NoStop}%
\bibitem [{\citenamefont {Gogolin}\ and\ \citenamefont
  {Eisert}(2016)}]{Gogolin2016}%
  \BibitemOpen
  \bibfield  {author} {\bibinfo {author} {\bibfnamefont {C.}~\bibnamefont
  {Gogolin}}\ and\ \bibinfo {author} {\bibfnamefont {J.}~\bibnamefont
  {Eisert}},\ }\href {\doibase 10.1088/0034-4885/79/5/056001} {\bibfield
  {journal} {\bibinfo  {journal} {Reports Prog. Phys.}\ }\textbf {\bibinfo
  {volume} {79}},\ \bibinfo {pages} {056001} (\bibinfo {year}
  {2016})}\BibitemShut {NoStop}%
\bibitem [{\citenamefont {Pal}\ and\ \citenamefont {Huse}(2010)}]{Pal2010a}%
  \BibitemOpen
  \bibfield  {author} {\bibinfo {author} {\bibfnamefont {A.}~\bibnamefont
  {Pal}}\ and\ \bibinfo {author} {\bibfnamefont {D.~A.}\ \bibnamefont {Huse}},\
  }\href {\doibase 10.1103/PhysRevB.82.174411} {\bibfield  {journal} {\bibinfo
  {journal} {Phys. Rev. B}\ }\textbf {\bibinfo {volume} {82}},\ \bibinfo
  {pages} {174411} (\bibinfo {year} {2010})}\BibitemShut {NoStop}%
\bibitem [{\citenamefont {Luitz}(2016)}]{Luitz2016}%
  \BibitemOpen
  \bibfield  {author} {\bibinfo {author} {\bibfnamefont {D.~J.}\ \bibnamefont
  {Luitz}},\ }\href {\doibase 10.1103/PhysRevB.93.134201} {\bibfield  {journal}
  {\bibinfo  {journal} {Phys. Rev. B}\ }\textbf {\bibinfo {volume} {93}},\
  \bibinfo {pages} {134201} (\bibinfo {year} {2016})}\BibitemShut {NoStop}%
\bibitem [{\citenamefont {Luitz}\ and\ \citenamefont {{Bar
  Lev}}(2016)}]{Luitz2016b}%
  \BibitemOpen
  \bibfield  {author} {\bibinfo {author} {\bibfnamefont {D.~J.}\ \bibnamefont
  {Luitz}}\ and\ \bibinfo {author} {\bibfnamefont {Y.}~\bibnamefont {{Bar
  Lev}}},\ }\href {\doibase 10.1103/PhysRevLett.117.170404} {\bibfield
  {journal} {\bibinfo  {journal} {Phys. Rev. Lett.}\ }\textbf {\bibinfo
  {volume} {117}},\ \bibinfo {pages} {170404} (\bibinfo {year}
  {2016})}\BibitemShut {NoStop}%
\bibitem [{\citenamefont {Laflorencie}(2016)}]{laflorencie_quantum_2016}%
  \BibitemOpen
  \bibfield  {author} {\bibinfo {author} {\bibfnamefont {N.}~\bibnamefont
  {Laflorencie}},\ }\href {\doibase 10.1016/j.physrep.2016.06.008} {\bibfield
  {journal} {\bibinfo  {journal} {Physics Reports}\ }\bibinfo {series} {Quantum
  entanglement in condensed matter systems},\ \textbf {\bibinfo {volume}
  {646}},\ \bibinfo {pages} {1} (\bibinfo {year} {2016})}\BibitemShut {NoStop}%
\bibitem [{\citenamefont {Bauer}\ and\ \citenamefont
  {Nayak}(2013)}]{Bauer2013}%
  \BibitemOpen
  \bibfield  {author} {\bibinfo {author} {\bibfnamefont {B.}~\bibnamefont
  {Bauer}}\ and\ \bibinfo {author} {\bibfnamefont {C.}~\bibnamefont {Nayak}},\
  }\href {\doibase 10.1088/1742-5468/2013/09/P09005} {\bibfield  {journal}
  {\bibinfo  {journal} {J. Stat. Mech. Theory Exp.}\ }\textbf {\bibinfo
  {volume} {2013}},\ \bibinfo {pages} {P09005} (\bibinfo {year}
  {2013})}\BibitemShut {NoStop}%
\bibitem [{\citenamefont {Kj{\"{a}}ll}\ \emph {et~al.}(2014)\citenamefont
  {Kj{\"{a}}ll}, \citenamefont {Bardarson},\ and\ \citenamefont
  {Pollmann}}]{Kjall2014}%
  \BibitemOpen
  \bibfield  {author} {\bibinfo {author} {\bibfnamefont {J.~A.}\ \bibnamefont
  {Kj{\"{a}}ll}}, \bibinfo {author} {\bibfnamefont {J.~H.}\ \bibnamefont
  {Bardarson}}, \ and\ \bibinfo {author} {\bibfnamefont {F.}~\bibnamefont
  {Pollmann}},\ }\href {http://arxiv.org/abs/1403.1568
  http://link.aps.org/doi/10.1103/PhysRevLett.113.107204} {\bibfield  {journal}
  {\bibinfo  {journal} {Phys. Rev. Lett.}\ }\textbf {\bibinfo {volume} {113}},\
  \bibinfo {pages} {107204} (\bibinfo {year} {2014})}\BibitemShut {NoStop}%
\bibitem [{\citenamefont {Vosk}\ \emph {et~al.}(2015)\citenamefont {Vosk},
  \citenamefont {Huse},\ and\ \citenamefont {Altman}}]{Vosk2014}%
  \BibitemOpen
  \bibfield  {author} {\bibinfo {author} {\bibfnamefont {R.}~\bibnamefont
  {Vosk}}, \bibinfo {author} {\bibfnamefont {D.~A.}\ \bibnamefont {Huse}}, \
  and\ \bibinfo {author} {\bibfnamefont {E.}~\bibnamefont {Altman}},\ }\href
  {\doibase 10.1103/PhysRevX.5.031032} {\bibfield  {journal} {\bibinfo
  {journal} {Phys. Rev. X}\ }\textbf {\bibinfo {volume} {5}},\ \bibinfo {pages}
  {031032} (\bibinfo {year} {2015})}\BibitemShut {NoStop}%
\bibitem [{\citenamefont {Potter}\ \emph {et~al.}(2015)\citenamefont {Potter},
  \citenamefont {Vasseur},\ and\ \citenamefont {Parameswaran}}]{Potter2015}%
  \BibitemOpen
  \bibfield  {author} {\bibinfo {author} {\bibfnamefont {A.~C.}\ \bibnamefont
  {Potter}}, \bibinfo {author} {\bibfnamefont {R.}~\bibnamefont {Vasseur}}, \
  and\ \bibinfo {author} {\bibfnamefont {S.~A.}\ \bibnamefont {Parameswaran}},\
  }\href {\doibase 10.1103/PhysRevX.5.031033} {\bibfield  {journal} {\bibinfo
  {journal} {Phys. Rev. X}\ }\textbf {\bibinfo {volume} {5}},\ \bibinfo {pages}
  {031033} (\bibinfo {year} {2015})}\BibitemShut {NoStop}%
\bibitem [{\citenamefont {Yu}\ \emph {et~al.}(2016)\citenamefont {Yu},
  \citenamefont {Luitz},\ and\ \citenamefont {Clark}}]{Yu2016}%
  \BibitemOpen
  \bibfield  {author} {\bibinfo {author} {\bibfnamefont {X.}~\bibnamefont
  {Yu}}, \bibinfo {author} {\bibfnamefont {D.~J.}\ \bibnamefont {Luitz}}, \
  and\ \bibinfo {author} {\bibfnamefont {B.~K.}\ \bibnamefont {Clark}},\ }\href
  {\doibase 10.1103/PhysRevB.94.184202} {\bibfield  {journal} {\bibinfo
  {journal} {Phys. Rev. B}\ }\textbf {\bibinfo {volume} {94}},\ \bibinfo
  {pages} {184202} (\bibinfo {year} {2016})}\BibitemShut {NoStop}%
\bibitem [{\citenamefont {Khemani}\ \emph
  {et~al.}(2016{\natexlab{a}})\citenamefont {Khemani}, \citenamefont {Lim},
  \citenamefont {Sheng},\ and\ \citenamefont {Huse}}]{khemani_critical_2016}%
  \BibitemOpen
  \bibfield  {author} {\bibinfo {author} {\bibfnamefont {V.}~\bibnamefont
  {Khemani}}, \bibinfo {author} {\bibfnamefont {S.~P.}\ \bibnamefont {Lim}},
  \bibinfo {author} {\bibfnamefont {D.~N.}\ \bibnamefont {Sheng}}, \ and\
  \bibinfo {author} {\bibfnamefont {D.~A.}\ \bibnamefont {Huse}},\ }\href@noop
  {} {\enquote {\bibinfo {title} {Critical {Properties} of the {Many}-{Body}
  {Localization} {Transition}},}\ } (\bibinfo {year} {2016}{\natexlab{a}}),\
  \Eprint {http://arxiv.org/abs/1607.05756} {arXiv:1607.05756} \BibitemShut
  {NoStop}%
\bibitem [{\citenamefont {Grover}(2014)}]{Grover2014}%
  \BibitemOpen
  \bibfield  {author} {\bibinfo {author} {\bibfnamefont {T.}~\bibnamefont
  {Grover}},\ }\href {http://arxiv.org/abs/1405.1471} {\  (\bibinfo {year}
  {2014})},\ \Eprint {http://arxiv.org/abs/1405.1471} {arXiv:1405.1471}
  \BibitemShut {NoStop}%
\bibitem [{\citenamefont {Bera}\ and\ \citenamefont
  {Lakshminarayan}(2016)}]{Bera2015a}%
  \BibitemOpen
  \bibfield  {author} {\bibinfo {author} {\bibfnamefont {S.}~\bibnamefont
  {Bera}}\ and\ \bibinfo {author} {\bibfnamefont {A.}~\bibnamefont
  {Lakshminarayan}},\ }\href {\doibase 10.1103/PhysRevB.93.134204} {\bibfield
  {journal} {\bibinfo  {journal} {Phys. Rev. B}\ }\textbf {\bibinfo {volume}
  {93}},\ \bibinfo {pages} {134204} (\bibinfo {year} {2016})}\BibitemShut
  {NoStop}%
\bibitem [{\citenamefont {{De Tomasi}}\ \emph {et~al.}(2017)\citenamefont {{De
  Tomasi}}, \citenamefont {Bera}, \citenamefont {Bardarson},\ and\
  \citenamefont {Pollmann}}]{DeTomasi2016}%
  \BibitemOpen
  \bibfield  {author} {\bibinfo {author} {\bibfnamefont {G.}~\bibnamefont {{De
  Tomasi}}}, \bibinfo {author} {\bibfnamefont {S.}~\bibnamefont {Bera}},
  \bibinfo {author} {\bibfnamefont {J.~H.}\ \bibnamefont {Bardarson}}, \ and\
  \bibinfo {author} {\bibfnamefont {F.}~\bibnamefont {Pollmann}},\ }\href
  {\doibase 10.1103/PhysRevLett.118.016804} {\bibfield  {journal} {\bibinfo
  {journal} {Phys. Rev. Lett.}\ }\textbf {\bibinfo {volume} {118}},\ \bibinfo
  {pages} {016804} (\bibinfo {year} {2017})}\BibitemShut {NoStop}%
\bibitem [{\citenamefont {Karahalios}\ \emph {et~al.}(2009)\citenamefont
  {Karahalios}, \citenamefont {Metavitsiadis}, \citenamefont {Zotos},
  \citenamefont {Gorczyca},\ and\ \citenamefont
  {Prelov{\v{s}}ek}}]{Karahalios2009a}%
  \BibitemOpen
  \bibfield  {author} {\bibinfo {author} {\bibfnamefont {A.}~\bibnamefont
  {Karahalios}}, \bibinfo {author} {\bibfnamefont {A.}~\bibnamefont
  {Metavitsiadis}}, \bibinfo {author} {\bibfnamefont {X.}~\bibnamefont
  {Zotos}}, \bibinfo {author} {\bibfnamefont {A.}~\bibnamefont {Gorczyca}}, \
  and\ \bibinfo {author} {\bibfnamefont {P.}~\bibnamefont {Prelov{\v{s}}ek}},\
  }\href {\doibase 10.1103/PhysRevB.79.024425} {\bibfield  {journal} {\bibinfo
  {journal} {Phys. Rev. B}\ }\textbf {\bibinfo {volume} {79}},\ \bibinfo
  {pages} {024425} (\bibinfo {year} {2009})}\BibitemShut {NoStop}%
\bibitem [{\citenamefont {Berkelbach}\ and\ \citenamefont
  {Reichman}(2010)}]{Berkelbach2010a}%
  \BibitemOpen
  \bibfield  {author} {\bibinfo {author} {\bibfnamefont {T.~C.}\ \bibnamefont
  {Berkelbach}}\ and\ \bibinfo {author} {\bibfnamefont {D.~R.}\ \bibnamefont
  {Reichman}},\ }\href {\doibase 10.1103/PhysRevB.81.224429} {\bibfield
  {journal} {\bibinfo  {journal} {Phys. Rev. B}\ }\textbf {\bibinfo {volume}
  {81}},\ \bibinfo {pages} {224429} (\bibinfo {year} {2010})}\BibitemShut
  {NoStop}%
\bibitem [{\citenamefont {Bari{\v{s}}i{\'{c}}}\ and\ \citenamefont
  {Prelov{\v{s}}ek}(2010)}]{Barisic2010a}%
  \BibitemOpen
  \bibfield  {author} {\bibinfo {author} {\bibfnamefont {O.~S.}\ \bibnamefont
  {Bari{\v{s}}i{\'{c}}}}\ and\ \bibinfo {author} {\bibfnamefont
  {P.}~\bibnamefont {Prelov{\v{s}}ek}},\ }\href {\doibase
  10.1103/PhysRevB.82.161106} {\bibfield  {journal} {\bibinfo  {journal} {Phys.
  Rev. B}\ }\textbf {\bibinfo {volume} {82}},\ \bibinfo {pages} {161106}
  (\bibinfo {year} {2010})}\BibitemShut {NoStop}%
\bibitem [{\citenamefont {Binder}\ and\ \citenamefont
  {Kob}(2005)}]{Binder2005}%
  \BibitemOpen
  \bibfield  {author} {\bibinfo {author} {\bibfnamefont {K.}~\bibnamefont
  {Binder}}\ and\ \bibinfo {author} {\bibfnamefont {W.}~\bibnamefont {Kob}},\
  }\href
  {http://www.amazon.com/Glassy-Materials-Disordered-Solids-Introduction/dp/9812565108/ref=sr{\_}1{\_}1?s=books{\&}ie=UTF8{\&}qid=1405527473{\&}sr=1-1{\&}dpPl=1}
  {\emph {\bibinfo {title} {{Glassy Materials and Disordered Solids: An
  Introduction to Their Statistical Mechanics}}}}\ (\bibinfo  {publisher}
  {World Scientific Publishing Company},\ \bibinfo {year} {2005})\ p.\ \bibinfo
  {pages} {452}\BibitemShut {NoStop}%
\bibitem [{\citenamefont {Edwards}\ and\ \citenamefont
  {Thouless}(1972)}]{Edwards1972}%
  \BibitemOpen
  \bibfield  {author} {\bibinfo {author} {\bibfnamefont {J.~T.}\ \bibnamefont
  {Edwards}}\ and\ \bibinfo {author} {\bibfnamefont {D.}~\bibnamefont
  {Thouless}},\ }\href {\doibase 10.1088/0022-3719/5/8/007} {\bibfield
  {journal} {\bibinfo  {journal} {J. Phys. C Solid State Phys.}\ }\textbf
  {\bibinfo {volume} {5}},\ \bibinfo {pages} {807} (\bibinfo {year}
  {1972})}\BibitemShut {NoStop}%
\bibitem [{\citenamefont {{Bar Lev}}\ and\ \citenamefont
  {Reichman}(2016)}]{BarLev2015}%
  \BibitemOpen
  \bibfield  {author} {\bibinfo {author} {\bibfnamefont {Y.}~\bibnamefont {{Bar
  Lev}}}\ and\ \bibinfo {author} {\bibfnamefont {D.~R.}\ \bibnamefont
  {Reichman}},\ }\href {\doibase 10.1209/0295-5075/113/46001} {\bibfield
  {journal} {\bibinfo  {journal} {EPL}\ }\textbf {\bibinfo {volume} {113}},\
  \bibinfo {pages} {46001} (\bibinfo {year} {2016})}\BibitemShut {NoStop}%
\bibitem [{\citenamefont {Scher}\ and\ \citenamefont {Lax}(1973)}]{Scher1973a}%
  \BibitemOpen
  \bibfield  {author} {\bibinfo {author} {\bibfnamefont {H.}~\bibnamefont
  {Scher}}\ and\ \bibinfo {author} {\bibfnamefont {M.}~\bibnamefont {Lax}},\
  }\href {\doibase 10.1103/PhysRevB.7.4491} {\bibfield  {journal} {\bibinfo
  {journal} {Phys. Rev. B}\ }\textbf {\bibinfo {volume} {7}},\ \bibinfo {pages}
  {4491} (\bibinfo {year} {1973})}\BibitemShut {NoStop}%
\bibitem [{\citenamefont {Gopalakrishnan}\ \emph {et~al.}(2015)\citenamefont
  {Gopalakrishnan}, \citenamefont {M{\"{u}}ller}, \citenamefont {Khemani},
  \citenamefont {Knap}, \citenamefont {Demler},\ and\ \citenamefont
  {Huse}}]{Gopalakrishnan2015}%
  \BibitemOpen
  \bibfield  {author} {\bibinfo {author} {\bibfnamefont {S.}~\bibnamefont
  {Gopalakrishnan}}, \bibinfo {author} {\bibfnamefont {M.}~\bibnamefont
  {M{\"{u}}ller}}, \bibinfo {author} {\bibfnamefont {V.}~\bibnamefont
  {Khemani}}, \bibinfo {author} {\bibfnamefont {M.}~\bibnamefont {Knap}},
  \bibinfo {author} {\bibfnamefont {E.~A.}\ \bibnamefont {Demler}}, \ and\
  \bibinfo {author} {\bibfnamefont {D.~A.}\ \bibnamefont {Huse}},\ }\href
  {\doibase 10.1103/PhysRevB.92.104202} {\bibfield  {journal} {\bibinfo
  {journal} {Phys. Rev. B}\ }\textbf {\bibinfo {volume} {92}},\ \bibinfo
  {pages} {104202} (\bibinfo {year} {2015})}\BibitemShut {NoStop}%
\bibitem [{\citenamefont {Kubo}(1957)}]{Kubo1957}%
  \BibitemOpen
  \bibfield  {author} {\bibinfo {author} {\bibfnamefont {R.}~\bibnamefont
  {Kubo}},\ }\href {\doibase 10.1143/JPSJ.12.570} {\bibfield  {journal}
  {\bibinfo  {journal} {J. Phys. Soc. Japan}\ }\textbf {\bibinfo {volume}
  {12}},\ \bibinfo {pages} {570} (\bibinfo {year} {1957})}\BibitemShut
  {NoStop}%
\bibitem [{\citenamefont {Kozarzewski}\ \emph {et~al.}(2016)\citenamefont
  {Kozarzewski}, \citenamefont {Prelov{\v{s}}ek},\ and\ \citenamefont
  {Mierzejewski}}]{Kozarzewski2016a}%
  \BibitemOpen
  \bibfield  {author} {\bibinfo {author} {\bibfnamefont {M.}~\bibnamefont
  {Kozarzewski}}, \bibinfo {author} {\bibfnamefont {P.}~\bibnamefont
  {Prelov{\v{s}}ek}}, \ and\ \bibinfo {author} {\bibfnamefont {M.}~\bibnamefont
  {Mierzejewski}},\ }\href {\doibase 10.1103/PhysRevB.93.235151} {\bibfield
  {journal} {\bibinfo  {journal} {Phys. Rev. B}\ }\textbf {\bibinfo {volume}
  {93}},\ \bibinfo {pages} {235151} (\bibinfo {year} {2016})}\BibitemShut
  {NoStop}%
\bibitem [{\citenamefont {{\v{Z}}nidari{\v{c}}}\ \emph
  {et~al.}(2016)\citenamefont {{\v{Z}}nidari{\v{c}}}, \citenamefont
  {Scardicchio},\ and\ \citenamefont {Varma}}]{Znidaric2016}%
  \BibitemOpen
  \bibfield  {author} {\bibinfo {author} {\bibfnamefont {M.}~\bibnamefont
  {{\v{Z}}nidari{\v{c}}}}, \bibinfo {author} {\bibfnamefont {A.}~\bibnamefont
  {Scardicchio}}, \ and\ \bibinfo {author} {\bibfnamefont {V.~K.}\ \bibnamefont
  {Varma}},\ }\href {\doibase 10.1103/PhysRevLett.117.040601} {\bibfield
  {journal} {\bibinfo  {journal} {Phys. Rev. Lett.}\ }\textbf {\bibinfo
  {volume} {117}},\ \bibinfo {pages} {040601} (\bibinfo {year}
  {2016})}\BibitemShut {NoStop}%
\bibitem [{\citenamefont {Gopalakrishnan}\ \emph
  {et~al.}(2016{\natexlab{a}})\citenamefont {Gopalakrishnan}, \citenamefont
  {Knap},\ and\ \citenamefont {Demler}}]{Gopalakrishnan2016}%
  \BibitemOpen
  \bibfield  {author} {\bibinfo {author} {\bibfnamefont {S.}~\bibnamefont
  {Gopalakrishnan}}, \bibinfo {author} {\bibfnamefont {M.}~\bibnamefont
  {Knap}}, \ and\ \bibinfo {author} {\bibfnamefont {E.}~\bibnamefont
  {Demler}},\ }\href {\doibase 10.1103/PhysRevB.94.094201} {\bibfield
  {journal} {\bibinfo  {journal} {Phys. Rev. B}\ }\textbf {\bibinfo {volume}
  {94}},\ \bibinfo {pages} {094201} (\bibinfo {year}
  {2016}{\natexlab{a}})}\BibitemShut {NoStop}%
\bibitem [{\citenamefont {Steinigeweg}\ \emph {et~al.}(2016)\citenamefont
  {Steinigeweg}, \citenamefont {Herbrych}, \citenamefont {Pollmann},\ and\
  \citenamefont {Brenig}}]{Steinigeweg2015}%
  \BibitemOpen
  \bibfield  {author} {\bibinfo {author} {\bibfnamefont {R.}~\bibnamefont
  {Steinigeweg}}, \bibinfo {author} {\bibfnamefont {J.}~\bibnamefont
  {Herbrych}}, \bibinfo {author} {\bibfnamefont {F.}~\bibnamefont {Pollmann}},
  \ and\ \bibinfo {author} {\bibfnamefont {W.}~\bibnamefont {Brenig}},\ }\href
  {\doibase 10.1103/PhysRevB.94.180401} {\bibfield  {journal} {\bibinfo
  {journal} {Phys. Rev. B}\ }\textbf {\bibinfo {volume} {94}},\ \bibinfo
  {pages} {180401} (\bibinfo {year} {2016})}\BibitemShut {NoStop}%
\bibitem [{\citenamefont {Bari{\v{s}}i{\'{c}}}\ \emph
  {et~al.}(2016)\citenamefont {Bari{\v{s}}i{\'{c}}}, \citenamefont {Kokalj},
  \citenamefont {Balog},\ and\ \citenamefont {Prelov{\v{s}}ek}}]{Barisic2016}%
  \BibitemOpen
  \bibfield  {author} {\bibinfo {author} {\bibfnamefont {O.~S.}\ \bibnamefont
  {Bari{\v{s}}i{\'{c}}}}, \bibinfo {author} {\bibfnamefont {J.}~\bibnamefont
  {Kokalj}}, \bibinfo {author} {\bibfnamefont {I.}~\bibnamefont {Balog}}, \
  and\ \bibinfo {author} {\bibfnamefont {P.}~\bibnamefont {Prelov{\v{s}}ek}},\
  }\href {\doibase 10.1103/PhysRevB.94.045126} {\bibfield  {journal} {\bibinfo
  {journal} {Phys. Rev. B}\ }\textbf {\bibinfo {volume} {94}},\ \bibinfo
  {pages} {045126} (\bibinfo {year} {2016})}\BibitemShut {NoStop}%
\bibitem [{\citenamefont {Prelov{\v{s}}ek}\ and\ \citenamefont
  {Herbrych}(2016)}]{Prelovsek2016a}%
  \BibitemOpen
  \bibfield  {author} {\bibinfo {author} {\bibfnamefont {P.}~\bibnamefont
  {Prelov{\v{s}}ek}}\ and\ \bibinfo {author} {\bibfnamefont {J.}~\bibnamefont
  {Herbrych}},\ }\href {http://arxiv.org/abs/1609.05450} {\  (\bibinfo {year}
  {2016})},\ \Eprint {http://arxiv.org/abs/1609.05450} {arXiv:1609.05450}
  \BibitemShut {NoStop}%
\bibitem [{\citenamefont {Thouless}\ and\ \citenamefont
  {Kirkpatrick}(1981)}]{Thouless1981}%
  \BibitemOpen
  \bibfield  {author} {\bibinfo {author} {\bibfnamefont {D.}~\bibnamefont
  {Thouless}}\ and\ \bibinfo {author} {\bibfnamefont {S.}~\bibnamefont
  {Kirkpatrick}},\ }\href {\doibase 10.1088/0022-3719/14/3/007} {\bibfield
  {journal} {\bibinfo  {journal} {J. Phys. C Solid State Phys.}\ }\textbf
  {\bibinfo {volume} {14}},\ \bibinfo {pages} {235} (\bibinfo {year}
  {1981})}\BibitemShut {NoStop}%
\bibitem [{\citenamefont {Imry}(2008)}]{Imry2008}%
  \BibitemOpen
  \bibfield  {author} {\bibinfo {author} {\bibfnamefont {Y.}~\bibnamefont
  {Imry}},\ }\href
  {http://www.amazon.com/Introduction-Mesoscopic-Physics-Nanotechnology/dp/019955269X}
  {\emph {\bibinfo {title} {{Introduction to Mesoscopic Physics (Mesoscopic
  Physics and Nanotechnology)}}}}\ (\bibinfo  {publisher} {Oxford University
  Press, USA},\ \bibinfo {year} {2008})\ p.\ \bibinfo {pages} {252}\BibitemShut
  {NoStop}%
\bibitem [{\citenamefont {Khait}\ \emph {et~al.}(2016)\citenamefont {Khait},
  \citenamefont {Gazit}, \citenamefont {Yao},\ and\ \citenamefont
  {Auerbach}}]{Khait2016}%
  \BibitemOpen
  \bibfield  {author} {\bibinfo {author} {\bibfnamefont {I.}~\bibnamefont
  {Khait}}, \bibinfo {author} {\bibfnamefont {S.}~\bibnamefont {Gazit}},
  \bibinfo {author} {\bibfnamefont {N.~Y.}\ \bibnamefont {Yao}}, \ and\
  \bibinfo {author} {\bibfnamefont {A.}~\bibnamefont {Auerbach}},\ }\href
  {\doibase 10.1103/PhysRevB.93.224205} {\bibfield  {journal} {\bibinfo
  {journal} {Phys. Rev. B}\ }\textbf {\bibinfo {volume} {93}},\ \bibinfo
  {pages} {224205} (\bibinfo {year} {2016})}\BibitemShut {NoStop}%
\bibitem [{\citenamefont {Edwards}\ and\ \citenamefont
  {Anderson}(1975)}]{Edwards1976}%
  \BibitemOpen
  \bibfield  {author} {\bibinfo {author} {\bibfnamefont {S.~F.}\ \bibnamefont
  {Edwards}}\ and\ \bibinfo {author} {\bibfnamefont {P.~W.}\ \bibnamefont
  {Anderson}},\ }\href {\doibase 10.1088/0305-4608/5/5/017} {\bibfield
  {journal} {\bibinfo  {journal} {J. Phys. F Met. Phys.}\ }\textbf {\bibinfo
  {volume} {5}},\ \bibinfo {pages} {965} (\bibinfo {year} {1975})}\BibitemShut
  {NoStop}%
\bibitem [{Note3()}]{Note3}%
  \BibitemOpen
  \bibinfo {note} {For systems with no (or broken) spin reflection
  symmetry.}\BibitemShut {Stop}%
\bibitem [{\citenamefont {Huse}\ \emph {et~al.}(2013)\citenamefont {Huse},
  \citenamefont {Nandkishore}, \citenamefont {Oganesyan}, \citenamefont {Pal},\
  and\ \citenamefont {Sondhi}}]{Huse2013a}%
  \BibitemOpen
  \bibfield  {author} {\bibinfo {author} {\bibfnamefont {D.~A.}\ \bibnamefont
  {Huse}}, \bibinfo {author} {\bibfnamefont {R.}~\bibnamefont {Nandkishore}},
  \bibinfo {author} {\bibfnamefont {V.}~\bibnamefont {Oganesyan}}, \bibinfo
  {author} {\bibfnamefont {A.}~\bibnamefont {Pal}}, \ and\ \bibinfo {author}
  {\bibfnamefont {S.~L.}\ \bibnamefont {Sondhi}},\ }\href {\doibase
  10.1103/PhysRevB.88.014206} {\bibfield  {journal} {\bibinfo  {journal} {Phys.
  Rev. B}\ }\textbf {\bibinfo {volume} {88}},\ \bibinfo {pages} {014206}
  (\bibinfo {year} {2013})}\BibitemShut {NoStop}%
\bibitem [{\citenamefont {Pekker}\ \emph {et~al.}(2014)\citenamefont {Pekker},
  \citenamefont {Refael}, \citenamefont {Altman}, \citenamefont {Demler},\ and\
  \citenamefont {Oganesyan}}]{Pekker2014}%
  \BibitemOpen
  \bibfield  {author} {\bibinfo {author} {\bibfnamefont {D.}~\bibnamefont
  {Pekker}}, \bibinfo {author} {\bibfnamefont {G.}~\bibnamefont {Refael}},
  \bibinfo {author} {\bibfnamefont {E.}~\bibnamefont {Altman}}, \bibinfo
  {author} {\bibfnamefont {E.~A.}\ \bibnamefont {Demler}}, \ and\ \bibinfo
  {author} {\bibfnamefont {V.}~\bibnamefont {Oganesyan}},\ }\href {\doibase
  10.1103/PhysRevX.4.011052} {\bibfield  {journal} {\bibinfo  {journal} {Phys.
  Rev. X}\ }\textbf {\bibinfo {volume} {4}},\ \bibinfo {pages} {011052}
  (\bibinfo {year} {2014})}\BibitemShut {NoStop}%
\bibitem [{\citenamefont {Vasseur}\ \emph {et~al.}(2016)\citenamefont
  {Vasseur}, \citenamefont {Friedman}, \citenamefont {Parameswaran},\ and\
  \citenamefont {Potter}}]{Vasseur2015b}%
  \BibitemOpen
  \bibfield  {author} {\bibinfo {author} {\bibfnamefont {R.}~\bibnamefont
  {Vasseur}}, \bibinfo {author} {\bibfnamefont {A.~J.}\ \bibnamefont
  {Friedman}}, \bibinfo {author} {\bibfnamefont {S.~A.}\ \bibnamefont
  {Parameswaran}}, \ and\ \bibinfo {author} {\bibfnamefont {A.~C.}\
  \bibnamefont {Potter}},\ }\href {\doibase 10.1103/PhysRevB.93.134207}
  {\bibfield  {journal} {\bibinfo  {journal} {Phys. Rev. B}\ }\textbf {\bibinfo
  {volume} {93}},\ \bibinfo {pages} {134207} (\bibinfo {year}
  {2016})}\BibitemShut {NoStop}%
\bibitem [{\citenamefont {Monthus}(2016{\natexlab{b}})}]{monthus_level_2016}%
  \BibitemOpen
  \bibfield  {author} {\bibinfo {author} {\bibfnamefont {C.}~\bibnamefont
  {Monthus}},\ }\href {\doibase 10.1088/1742-5468/2016/03/033113} {\bibfield
  {journal} {\bibinfo  {journal} {Journal of Statistical Mechanics: Theory and
  Experiment}\ }\textbf {\bibinfo {volume} {2016}},\ \bibinfo {pages} {033113}
  (\bibinfo {year} {2016}{\natexlab{b}})}\BibitemShut {NoStop}%
\bibitem [{\citenamefont {Alexander}\ \emph {et~al.}(1981)\citenamefont
  {Alexander}, \citenamefont {Bernasconi}, \citenamefont {Schneider},\ and\
  \citenamefont {Orbach}}]{Alexander1981}%
  \BibitemOpen
  \bibfield  {author} {\bibinfo {author} {\bibfnamefont {S.}~\bibnamefont
  {Alexander}}, \bibinfo {author} {\bibfnamefont {J.}~\bibnamefont
  {Bernasconi}}, \bibinfo {author} {\bibfnamefont {W.~R.}\ \bibnamefont
  {Schneider}}, \ and\ \bibinfo {author} {\bibfnamefont {R.}~\bibnamefont
  {Orbach}},\ }\href {\doibase 10.1103/RevModPhys.53.175} {\bibfield  {journal}
  {\bibinfo  {journal} {Rev. Mod. Phys.}\ }\textbf {\bibinfo {volume} {53}},\
  \bibinfo {pages} {175} (\bibinfo {year} {1981})}\BibitemShut {NoStop}%
\bibitem [{\citenamefont {Gopalakrishnan}\ \emph
  {et~al.}(2016{\natexlab{b}})\citenamefont {Gopalakrishnan}, \citenamefont
  {Agarwal}, \citenamefont {Demler}, \citenamefont {Huse},\ and\ \citenamefont
  {Knap}}]{Gopalakrishnan2015a}%
  \BibitemOpen
  \bibfield  {author} {\bibinfo {author} {\bibfnamefont {S.}~\bibnamefont
  {Gopalakrishnan}}, \bibinfo {author} {\bibfnamefont {K.}~\bibnamefont
  {Agarwal}}, \bibinfo {author} {\bibfnamefont {E.~A.}\ \bibnamefont {Demler}},
  \bibinfo {author} {\bibfnamefont {D.~A.}\ \bibnamefont {Huse}}, \ and\
  \bibinfo {author} {\bibfnamefont {M.}~\bibnamefont {Knap}},\ }\href {\doibase
  10.1103/PhysRevB.93.134206} {\bibfield  {journal} {\bibinfo  {journal} {Phys.
  Rev. B}\ }\textbf {\bibinfo {volume} {93}},\ \bibinfo {pages} {134206}
  (\bibinfo {year} {2016}{\natexlab{b}})}\BibitemShut {NoStop}%
\bibitem [{\citenamefont {Iyer}\ \emph {et~al.}(2013)\citenamefont {Iyer},
  \citenamefont {Oganesyan}, \citenamefont {Refael},\ and\ \citenamefont
  {Huse}}]{Iyer2013}%
  \BibitemOpen
  \bibfield  {author} {\bibinfo {author} {\bibfnamefont {S.}~\bibnamefont
  {Iyer}}, \bibinfo {author} {\bibfnamefont {V.}~\bibnamefont {Oganesyan}},
  \bibinfo {author} {\bibfnamefont {G.}~\bibnamefont {Refael}}, \ and\ \bibinfo
  {author} {\bibfnamefont {D.~A.}\ \bibnamefont {Huse}},\ }\href {\doibase
  10.1103/PhysRevB.87.134202} {\bibfield  {journal} {\bibinfo  {journal} {Phys.
  Rev. B}\ }\textbf {\bibinfo {volume} {87}},\ \bibinfo {pages} {134202}
  (\bibinfo {year} {2013})}\BibitemShut {NoStop}%
\bibitem [{Note4()}]{Note4}%
  \BibitemOpen
  \bibinfo {note} {In Ref.~\cite {Torres-Herrera2015a} a product state initial
  condition was used, which is different from the initial condition we have
  used in our definition (\ref {eq:survival_prob}).}\BibitemShut {Stop}%
\bibitem [{\citenamefont {Torres-Herrera}\ and\ \citenamefont
  {Santos}(2015)}]{Torres-Herrera2015a}%
  \BibitemOpen
  \bibfield  {author} {\bibinfo {author} {\bibfnamefont {E.~J.}\ \bibnamefont
  {Torres-Herrera}}\ and\ \bibinfo {author} {\bibfnamefont {L.~F.}\
  \bibnamefont {Santos}},\ }\href {\doibase 10.1103/PhysRevB.92.014208}
  {\bibfield  {journal} {\bibinfo  {journal} {Phys. Rev. B}\ }\textbf {\bibinfo
  {volume} {92}},\ \bibinfo {pages} {014208} (\bibinfo {year}
  {2015})}\BibitemShut {NoStop}%
\bibitem [{\citenamefont {Mierzejewski}\ \emph {et~al.}(2016)\citenamefont
  {Mierzejewski}, \citenamefont {Herbrych},\ and\ \citenamefont
  {Prelov{\v{s}}ek}}]{Mierzejewski2016}%
  \BibitemOpen
  \bibfield  {author} {\bibinfo {author} {\bibfnamefont {M.}~\bibnamefont
  {Mierzejewski}}, \bibinfo {author} {\bibfnamefont {J.}~\bibnamefont
  {Herbrych}}, \ and\ \bibinfo {author} {\bibfnamefont {P.}~\bibnamefont
  {Prelov{\v{s}}ek}},\ }\href {\doibase 10.1103/PhysRevB.94.224207} {\bibfield
  {journal} {\bibinfo  {journal} {Phys. Rev. B}\ }\textbf {\bibinfo {volume}
  {94}},\ \bibinfo {pages} {224207} (\bibinfo {year} {2016})}\BibitemShut
  {NoStop}%
\bibitem [{\citenamefont {Luitz}\ \emph {et~al.}(2016)\citenamefont {Luitz},
  \citenamefont {Laflorencie},\ and\ \citenamefont {Alet}}]{Luitz2015a}%
  \BibitemOpen
  \bibfield  {author} {\bibinfo {author} {\bibfnamefont {D.~J.}\ \bibnamefont
  {Luitz}}, \bibinfo {author} {\bibfnamefont {N.}~\bibnamefont {Laflorencie}},
  \ and\ \bibinfo {author} {\bibfnamefont {F.}~\bibnamefont {Alet}},\ }\href
  {\doibase 10.1103/PhysRevB.93.060201} {\bibfield  {journal} {\bibinfo
  {journal} {Phys. Rev. B}\ }\textbf {\bibinfo {volume} {93}},\ \bibinfo
  {pages} {060201} (\bibinfo {year} {2016})}\BibitemShut {NoStop}%
\bibitem [{Note5()}]{Note5}%
  \BibitemOpen
  \bibinfo {note} {This relation is not surprising since for $q=\pi $ the
  correlation function (\ref {eq:c_q_t}) is very close in form to the local
  autocorrelation function. For smaller $q$ it however should not be
  expected.}\BibitemShut {Stop}%
\bibitem [{\citenamefont {Enss}\ \emph {et~al.}(2017)\citenamefont {Enss},
  \citenamefont {Andraschko},\ and\ \citenamefont {Sirker}}]{Enss2016}%
  \BibitemOpen
  \bibfield  {author} {\bibinfo {author} {\bibfnamefont {T.}~\bibnamefont
  {Enss}}, \bibinfo {author} {\bibfnamefont {F.}~\bibnamefont {Andraschko}}, \
  and\ \bibinfo {author} {\bibfnamefont {J.}~\bibnamefont {Sirker}},\ }\href
  {\doibase 10.1103/PhysRevB.95.045121} {\bibfield  {journal} {\bibinfo
  {journal} {Phys. Rev. B}\ }\textbf {\bibinfo {volume} {95}},\ \bibinfo
  {pages} {045121} (\bibinfo {year} {2017})}\BibitemShut {NoStop}%
\bibitem [{\citenamefont {Paredes}\ \emph {et~al.}(2005)\citenamefont
  {Paredes}, \citenamefont {Verstraete},\ and\ \citenamefont
  {Cirac}}]{Paredes2005}%
  \BibitemOpen
  \bibfield  {author} {\bibinfo {author} {\bibfnamefont {B.}~\bibnamefont
  {Paredes}}, \bibinfo {author} {\bibfnamefont {F.}~\bibnamefont {Verstraete}},
  \ and\ \bibinfo {author} {\bibfnamefont {J.~I.}\ \bibnamefont {Cirac}},\
  }\href {\doibase 10.1103/PhysRevLett.95.140501} {\bibfield  {journal}
  {\bibinfo  {journal} {Phys. Rev. Lett.}\ }\textbf {\bibinfo {volume} {95}},\
  \bibinfo {pages} {140501} (\bibinfo {year} {2005})}\BibitemShut {NoStop}%
\bibitem [{\citenamefont {Kim}\ and\ \citenamefont {Huse}(2013)}]{Kim2013}%
  \BibitemOpen
  \bibfield  {author} {\bibinfo {author} {\bibfnamefont {H.}~\bibnamefont
  {Kim}}\ and\ \bibinfo {author} {\bibfnamefont {D.~A.}\ \bibnamefont {Huse}},\
  }\href {\doibase 10.1103/PhysRevLett.111.127205} {\bibfield  {journal}
  {\bibinfo  {journal} {Phys. Rev. Lett.}\ }\textbf {\bibinfo {volume} {111}},\
  \bibinfo {pages} {127205} (\bibinfo {year} {2013})}\BibitemShut {NoStop}%
\bibitem [{\citenamefont {Aleiner}\ \emph {et~al.}(2016)\citenamefont
  {Aleiner}, \citenamefont {Faoro},\ and\ \citenamefont {Ioffe}}]{Aleiner2016}%
  \BibitemOpen
  \bibfield  {author} {\bibinfo {author} {\bibfnamefont {I.~L.}\ \bibnamefont
  {Aleiner}}, \bibinfo {author} {\bibfnamefont {L.}~\bibnamefont {Faoro}}, \
  and\ \bibinfo {author} {\bibfnamefont {L.~B.}\ \bibnamefont {Ioffe}},\ }\href
  {\doibase 10.1016/j.aop.2016.09.006} {\bibfield  {journal} {\bibinfo
  {journal} {Ann. Phys. (N. Y).}\ }\textbf {\bibinfo {volume} {375}},\ \bibinfo
  {pages} {378} (\bibinfo {year} {2016})}\BibitemShut {NoStop}%
\bibitem [{\citenamefont {Larkin}\ and\ \citenamefont
  {Ovchinnikov}(1969)}]{Larkin1969}%
  \BibitemOpen
  \bibfield  {author} {\bibinfo {author} {\bibfnamefont {A.~I.}\ \bibnamefont
  {Larkin}}\ and\ \bibinfo {author} {\bibfnamefont {Y.~N.}\ \bibnamefont
  {Ovchinnikov}},\ }\href@noop {} {\bibfield  {journal} {\bibinfo  {journal}
  {Jetp}\ }\textbf {\bibinfo {volume} {28}},\ \bibinfo {pages} {1200} (\bibinfo
  {year} {1969})}\BibitemShut {NoStop}%
\bibitem [{\citenamefont {Fan}\ \emph {et~al.}(2016)\citenamefont {Fan},
  \citenamefont {Zhang}, \citenamefont {Shen},\ and\ \citenamefont
  {Zhai}}]{Fan2016}%
  \BibitemOpen
  \bibfield  {author} {\bibinfo {author} {\bibfnamefont {R.}~\bibnamefont
  {Fan}}, \bibinfo {author} {\bibfnamefont {P.}~\bibnamefont {Zhang}}, \bibinfo
  {author} {\bibfnamefont {H.}~\bibnamefont {Shen}}, \ and\ \bibinfo {author}
  {\bibfnamefont {H.}~\bibnamefont {Zhai}},\ }\href
  {http://arxiv.org/abs/1608.01914} {\  (\bibinfo {year} {2016})},\ \Eprint
  {http://arxiv.org/abs/1608.01914} {arXiv:1608.01914} \BibitemShut {NoStop}%
\bibitem [{\citenamefont {Luitz}\ and\ \citenamefont
  {Bar~Lev}(2017)}]{luitz_information_2017}%
  \BibitemOpen
  \bibfield  {author} {\bibinfo {author} {\bibfnamefont {D.~J.}\ \bibnamefont
  {Luitz}}\ and\ \bibinfo {author} {\bibfnamefont {Y.}~\bibnamefont
  {Bar~Lev}},\ }\href@noop {} {\enquote {\bibinfo {title} {Information
  propagation in isolated quantum systems},}\ } (\bibinfo {year} {2017}),\
  \Eprint {http://arxiv.org/abs/1702.03929} {arXiv:1702.03929} \BibitemShut
  {NoStop}%
\bibitem [{\citenamefont {Zhou}\ and\ \citenamefont
  {Luitz}(2016)}]{zhou_operator_2016}%
  \BibitemOpen
  \bibfield  {author} {\bibinfo {author} {\bibfnamefont {T.}~\bibnamefont
  {Zhou}}\ and\ \bibinfo {author} {\bibfnamefont {D.~J.}\ \bibnamefont
  {Luitz}},\ }\href@noop {} {\enquote {\bibinfo {title} {Operator entanglement
  entropy of the time evolution operator in chaotic systems},}\ } (\bibinfo
  {year} {2016}),\ \Eprint {http://arxiv.org/abs/1612.07327} {arXiv:1612.07327}
  \BibitemShut {NoStop}%
\bibitem [{\citenamefont {Hauschild}\ \emph {et~al.}(2016)\citenamefont
  {Hauschild}, \citenamefont {Heidrich-Meisner},\ and\ \citenamefont
  {Pollmann}}]{Hauschild2016}%
  \BibitemOpen
  \bibfield  {author} {\bibinfo {author} {\bibfnamefont {J.}~\bibnamefont
  {Hauschild}}, \bibinfo {author} {\bibfnamefont {F.}~\bibnamefont
  {Heidrich-Meisner}}, \ and\ \bibinfo {author} {\bibfnamefont
  {F.}~\bibnamefont {Pollmann}},\ }\href {\doibase 10.1103/PhysRevB.94.161109}
  {\bibfield  {journal} {\bibinfo  {journal} {Phys. Rev. B}\ }\textbf {\bibinfo
  {volume} {94}},\ \bibinfo {pages} {161109} (\bibinfo {year}
  {2016})}\BibitemShut {NoStop}%
\bibitem [{Note6()}]{Note6}%
  \BibitemOpen
  \bibinfo {note} {The authors actually find that a logarithmic time dependence
  describes their data better, although algebraic growth fits almost equally
  well.}\BibitemShut {Stop}%
\bibitem [{\citenamefont {Varma}\ \emph {et~al.}(2015)\citenamefont {Varma},
  \citenamefont {Lerose}, \citenamefont {Pietracaprina}, \citenamefont
  {Goold},\ and\ \citenamefont {Scardicchio}}]{Lerose2015}%
  \BibitemOpen
  \bibfield  {author} {\bibinfo {author} {\bibfnamefont {V.~K.}\ \bibnamefont
  {Varma}}, \bibinfo {author} {\bibfnamefont {A.}~\bibnamefont {Lerose}},
  \bibinfo {author} {\bibfnamefont {F.}~\bibnamefont {Pietracaprina}}, \bibinfo
  {author} {\bibfnamefont {J.}~\bibnamefont {Goold}}, \ and\ \bibinfo {author}
  {\bibfnamefont {A.}~\bibnamefont {Scardicchio}},\ }\href
  {http://arxiv.org/abs/1511.09144} {\  (\bibinfo {year} {2015})},\ \Eprint
  {http://arxiv.org/abs/1511.09144} {arXiv:1511.09144} \BibitemShut {NoStop}%
\bibitem [{\citenamefont {Li}\ and\ \citenamefont {Wang}(2003)}]{Li2003}%
  \BibitemOpen
  \bibfield  {author} {\bibinfo {author} {\bibfnamefont {B.}~\bibnamefont
  {Li}}\ and\ \bibinfo {author} {\bibfnamefont {J.}~\bibnamefont {Wang}},\
  }\href {\doibase 10.1103/PhysRevLett.91.044301} {\bibfield  {journal}
  {\bibinfo  {journal} {Phys. Rev. Lett.}\ }\textbf {\bibinfo {volume} {91}},\
  \bibinfo {pages} {044301} (\bibinfo {year} {2003})}\BibitemShut {NoStop}%
\bibitem [{Note7()}]{Note7}%
  \BibitemOpen
  \bibinfo {note} {We note that this scaling is special for the XXZ model which
  is integrable in the {$W=0$ } limit. For more generic nonintegrable models it
  is supposed to scale as {$l\propto W^{-2}$}}\BibitemShut {NoStop}%
\bibitem [{\citenamefont {Metzler}\ \emph {et~al.}(1999)\citenamefont
  {Metzler}, \citenamefont {Barkai},\ and\ \citenamefont
  {Klafter}}]{Metzler1999}%
  \BibitemOpen
  \bibfield  {author} {\bibinfo {author} {\bibfnamefont {R.}~\bibnamefont
  {Metzler}}, \bibinfo {author} {\bibfnamefont {E.}~\bibnamefont {Barkai}}, \
  and\ \bibinfo {author} {\bibfnamefont {J.}~\bibnamefont {Klafter}},\ }\href
  {\doibase 10.1103/PhysRevLett.82.3563} {\bibfield  {journal} {\bibinfo
  {journal} {Phys. Rev. Lett.}\ }\textbf {\bibinfo {volume} {82}},\ \bibinfo
  {pages} {3563} (\bibinfo {year} {1999})}\BibitemShut {NoStop}%
\bibitem [{\citenamefont {Metzler}\ and\ \citenamefont
  {Klafter}(2000)}]{Metzler2000}%
  \BibitemOpen
  \bibfield  {author} {\bibinfo {author} {\bibfnamefont {R.}~\bibnamefont
  {Metzler}}\ and\ \bibinfo {author} {\bibfnamefont {J.}~\bibnamefont
  {Klafter}},\ }\href {\doibase 10.1016/S0370-1573(00)00070-3} {\bibfield
  {journal} {\bibinfo  {journal} {Phys. Rep.}\ }\textbf {\bibinfo {volume}
  {339}},\ \bibinfo {pages} {1} (\bibinfo {year} {2000})}\BibitemShut {NoStop}%
\bibitem [{\citenamefont {Sokolov}\ and\ \citenamefont
  {Klafter}(2005)}]{Sokolov2005}%
  \BibitemOpen
  \bibfield  {author} {\bibinfo {author} {\bibfnamefont {I.~M.}\ \bibnamefont
  {Sokolov}}\ and\ \bibinfo {author} {\bibfnamefont {J.}~\bibnamefont
  {Klafter}},\ }\href {\doibase 10.1063/1.1860472} {\bibfield  {journal}
  {\bibinfo  {journal} {Chaos}\ }\textbf {\bibinfo {volume} {15}},\ \bibinfo
  {pages} {26103} (\bibinfo {year} {2005})}\BibitemShut {NoStop}%
\bibitem [{\citenamefont {Dhar}\ and\ \citenamefont {Barma}(1980)}]{Dhar1980}%
  \BibitemOpen
  \bibfield  {author} {\bibinfo {author} {\bibfnamefont {D.}~\bibnamefont
  {Dhar}}\ and\ \bibinfo {author} {\bibfnamefont {M.}~\bibnamefont {Barma}},\
  }\href {\doibase 10.1007/BF01008051} {\bibfield  {journal} {\bibinfo
  {journal} {J. Stat. Phys.}\ }\textbf {\bibinfo {volume} {22}},\ \bibinfo
  {pages} {259} (\bibinfo {year} {1980})}\BibitemShut {NoStop}%
\bibitem [{\citenamefont {Palmer}\ \emph {et~al.}(1984)\citenamefont {Palmer},
  \citenamefont {Stein}, \citenamefont {Abrahams},\ and\ \citenamefont
  {Anderson}}]{Palmer1984}%
  \BibitemOpen
  \bibfield  {author} {\bibinfo {author} {\bibfnamefont {R.~G.}\ \bibnamefont
  {Palmer}}, \bibinfo {author} {\bibfnamefont {D.~L.}\ \bibnamefont {Stein}},
  \bibinfo {author} {\bibfnamefont {E.}~\bibnamefont {Abrahams}}, \ and\
  \bibinfo {author} {\bibfnamefont {P.~W.}\ \bibnamefont {Anderson}},\ }\href
  {\doibase 10.1103/PhysRevLett.53.958} {\bibfield  {journal} {\bibinfo
  {journal} {Phys. Rev. Lett.}\ }\textbf {\bibinfo {volume} {53}},\ \bibinfo
  {pages} {958} (\bibinfo {year} {1984})}\BibitemShut {NoStop}%
\bibitem [{\citenamefont {Randeria}\ \emph {et~al.}(1985)\citenamefont
  {Randeria}, \citenamefont {Sethna},\ and\ \citenamefont
  {Palmer}}]{Randeria1985}%
  \BibitemOpen
  \bibfield  {author} {\bibinfo {author} {\bibfnamefont {M.}~\bibnamefont
  {Randeria}}, \bibinfo {author} {\bibfnamefont {J.~P.}\ \bibnamefont
  {Sethna}}, \ and\ \bibinfo {author} {\bibfnamefont {R.~G.}\ \bibnamefont
  {Palmer}},\ }\href {\doibase 10.1103/PhysRevLett.54.1321} {\bibfield
  {journal} {\bibinfo  {journal} {Phys. Rev. Lett.}\ }\textbf {\bibinfo
  {volume} {54}},\ \bibinfo {pages} {1321} (\bibinfo {year}
  {1985})}\BibitemShut {NoStop}%
\bibitem [{\citenamefont {Kutasov}\ \emph {et~al.}(1986)\citenamefont
  {Kutasov}, \citenamefont {Aharony}, \citenamefont {Domany},\ and\
  \citenamefont {Kinzel}}]{Kutasov1986}%
  \BibitemOpen
  \bibfield  {author} {\bibinfo {author} {\bibfnamefont {D.}~\bibnamefont
  {Kutasov}}, \bibinfo {author} {\bibfnamefont {A.}~\bibnamefont {Aharony}},
  \bibinfo {author} {\bibfnamefont {E.}~\bibnamefont {Domany}}, \ and\ \bibinfo
  {author} {\bibfnamefont {W.}~\bibnamefont {Kinzel}},\ }\href {\doibase
  10.1103/PhysRevLett.56.2229} {\bibfield  {journal} {\bibinfo  {journal}
  {Phys. Rev. Lett.}\ }\textbf {\bibinfo {volume} {56}},\ \bibinfo {pages}
  {2229} (\bibinfo {year} {1986})}\BibitemShut {NoStop}%
\bibitem [{\citenamefont {Levy}(1939)}]{Levy1939}%
  \BibitemOpen
  \bibfield  {author} {\bibinfo {author} {\bibfnamefont {P.}~\bibnamefont
  {Levy}},\ }\href {http://eudml.org/doc/86716} {\bibfield  {journal} {\bibinfo
   {journal} {Bull. la Soci{\'{e}}t{\'{e}} Math{\'{e}}matique Fr.}\ }\textbf
  {\bibinfo {volume} {67}},\ \bibinfo {pages} {1} (\bibinfo {year}
  {1939})}\BibitemShut {NoStop}%
\bibitem [{\citenamefont {Griffiths}(1969)}]{Griffiths1969}%
  \BibitemOpen
  \bibfield  {author} {\bibinfo {author} {\bibfnamefont {R.~B.}\ \bibnamefont
  {Griffiths}},\ }\href {\doibase 10.1103/PhysRevLett.23.17} {\bibfield
  {journal} {\bibinfo  {journal} {Phys. Rev. Lett.}\ }\textbf {\bibinfo
  {volume} {23}},\ \bibinfo {pages} {17} (\bibinfo {year} {1969})}\BibitemShut
  {NoStop}%
\bibitem [{\citenamefont {Vojta}(2006)}]{Vojta2006}%
  \BibitemOpen
  \bibfield  {author} {\bibinfo {author} {\bibfnamefont {T.}~\bibnamefont
  {Vojta}},\ }\href {\doibase 10.1088/0305-4470/39/22/R01} {\bibfield
  {journal} {\bibinfo  {journal} {J. Phys. A. Math. Gen.}\ }\textbf {\bibinfo
  {volume} {39}},\ \bibinfo {pages} {R143} (\bibinfo {year}
  {2006})}\BibitemShut {NoStop}%
\bibitem [{\citenamefont {Dyson}(1953)}]{Dyson1953}%
  \BibitemOpen
  \bibfield  {author} {\bibinfo {author} {\bibfnamefont {F.~J.}\ \bibnamefont
  {Dyson}},\ }\href {\doibase 10.1103/PhysRev.92.1331} {\bibfield  {journal}
  {\bibinfo  {journal} {Phys. Rev.}\ }\textbf {\bibinfo {volume} {92}},\
  \bibinfo {pages} {1331} (\bibinfo {year} {1953})}\BibitemShut {NoStop}%
\bibitem [{\citenamefont {Miller}\ and\ \citenamefont
  {Abrahams}(1960)}]{Miller1960}%
  \BibitemOpen
  \bibfield  {author} {\bibinfo {author} {\bibfnamefont {A.}~\bibnamefont
  {Miller}}\ and\ \bibinfo {author} {\bibfnamefont {E.}~\bibnamefont
  {Abrahams}},\ }\href {\doibase 10.1103/PhysRev.120.745} {\bibfield  {journal}
  {\bibinfo  {journal} {Phys. Rev.}\ }\textbf {\bibinfo {volume} {120}},\
  \bibinfo {pages} {745} (\bibinfo {year} {1960})}\BibitemShut {NoStop}%
\bibitem [{\citenamefont {Kurkij{\"{a}}rvi}(1973)}]{Kurkijarvi1973}%
  \BibitemOpen
  \bibfield  {author} {\bibinfo {author} {\bibfnamefont {J.}~\bibnamefont
  {Kurkij{\"{a}}rvi}},\ }\href {\doibase 10.1103/PhysRevB.8.922} {\bibfield
  {journal} {\bibinfo  {journal} {Phys. Rev. B}\ }\textbf {\bibinfo {volume}
  {8}},\ \bibinfo {pages} {922} (\bibinfo {year} {1973})}\BibitemShut {NoStop}%
\bibitem [{Note8()}]{Note8}%
  \BibitemOpen
  \bibinfo {note} {In the language of the random resistors this means that the
  average resistance is finite.}\BibitemShut {Stop}%
\bibitem [{\citenamefont {Bernasconi}\ \emph {et~al.}(1978)\citenamefont
  {Bernasconi}, \citenamefont {Alexander},\ and\ \citenamefont
  {Orbach}}]{Bernasconi1978}%
  \BibitemOpen
  \bibfield  {author} {\bibinfo {author} {\bibfnamefont {J.}~\bibnamefont
  {Bernasconi}}, \bibinfo {author} {\bibfnamefont {S.}~\bibnamefont
  {Alexander}}, \ and\ \bibinfo {author} {\bibfnamefont {R.}~\bibnamefont
  {Orbach}},\ }\href {\doibase 10.1103/PhysRevLett.41.185} {\bibfield
  {journal} {\bibinfo  {journal} {Phys. Rev. Lett.}\ }\textbf {\bibinfo
  {volume} {41}},\ \bibinfo {pages} {185} (\bibinfo {year} {1978})}\BibitemShut
  {NoStop}%
\bibitem [{\citenamefont {Bernasconi}\ \emph {et~al.}(1980)\citenamefont
  {Bernasconi}, \citenamefont {Schneider},\ and\ \citenamefont
  {Wyss}}]{Bernasconi1980}%
  \BibitemOpen
  \bibfield  {author} {\bibinfo {author} {\bibfnamefont {J.}~\bibnamefont
  {Bernasconi}}, \bibinfo {author} {\bibfnamefont {W.~R.}\ \bibnamefont
  {Schneider}}, \ and\ \bibinfo {author} {\bibfnamefont {W.}~\bibnamefont
  {Wyss}},\ }\href {\doibase 10.1007/BF01365374} {\bibfield  {journal}
  {\bibinfo  {journal} {Zeitschrift f{\"{u}}r Phys. B Condens. Matterr Phys. B
  Condens. Matter Quanta}\ }\textbf {\bibinfo {volume} {37}},\ \bibinfo {pages}
  {175} (\bibinfo {year} {1980})}\BibitemShut {NoStop}%
\bibitem [{\citenamefont {Alexander}(1981)}]{Alexander1981a}%
  \BibitemOpen
  \bibfield  {author} {\bibinfo {author} {\bibfnamefont {S.}~\bibnamefont
  {Alexander}},\ }\href {\doibase 10.1103/PhysRevB.23.2951} {\bibfield
  {journal} {\bibinfo  {journal} {Phys. Rev. B}\ }\textbf {\bibinfo {volume}
  {23}},\ \bibinfo {pages} {2951} (\bibinfo {year} {1981})}\BibitemShut
  {NoStop}%
\bibitem [{\citenamefont {Zhang}\ \emph
  {et~al.}(2016{\natexlab{b}})\citenamefont {Zhang}, \citenamefont {Zhao},
  \citenamefont {Devakul},\ and\ \citenamefont {Huse}}]{Zhang2016}%
  \BibitemOpen
  \bibfield  {author} {\bibinfo {author} {\bibfnamefont {L.}~\bibnamefont
  {Zhang}}, \bibinfo {author} {\bibfnamefont {B.}~\bibnamefont {Zhao}},
  \bibinfo {author} {\bibfnamefont {T.}~\bibnamefont {Devakul}}, \ and\
  \bibinfo {author} {\bibfnamefont {D.~A.}\ \bibnamefont {Huse}},\ }\href
  {\doibase 10.1103/PhysRevB.93.224201} {\bibfield  {journal} {\bibinfo
  {journal} {Phys. Rev. B}\ }\textbf {\bibinfo {volume} {93}},\ \bibinfo
  {pages} {224201} (\bibinfo {year} {2016}{\natexlab{b}})}\BibitemShut
  {NoStop}%
\bibitem [{\citenamefont {Vosk}\ and\ \citenamefont
  {Altman}(2013)}]{Vosk2013a}%
  \BibitemOpen
  \bibfield  {author} {\bibinfo {author} {\bibfnamefont {R.}~\bibnamefont
  {Vosk}}\ and\ \bibinfo {author} {\bibfnamefont {E.}~\bibnamefont {Altman}},\
  }\href {\doibase 10.1103/PhysRevLett.110.067204} {\bibfield  {journal}
  {\bibinfo  {journal} {Phys. Rev. Lett.}\ }\textbf {\bibinfo {volume} {110}},\
  \bibinfo {pages} {067204} (\bibinfo {year} {2013})}\BibitemShut {NoStop}%
\bibitem [{\citenamefont {Dasgupta}\ and\ \citenamefont
  {Ma}(1980)}]{Dasgupta1980}%
  \BibitemOpen
  \bibfield  {author} {\bibinfo {author} {\bibfnamefont {C.}~\bibnamefont
  {Dasgupta}}\ and\ \bibinfo {author} {\bibfnamefont {S.-k.}\ \bibnamefont
  {Ma}},\ }\href {\doibase 10.1103/PhysRevB.22.1305} {\bibfield  {journal}
  {\bibinfo  {journal} {Phys. Rev. B}\ }\textbf {\bibinfo {volume} {22}},\
  \bibinfo {pages} {1305} (\bibinfo {year} {1980})}\BibitemShut {NoStop}%
\bibitem [{\citenamefont {Fisher}(1992)}]{Fisher1992}%
  \BibitemOpen
  \bibfield  {author} {\bibinfo {author} {\bibfnamefont {D.~S.}\ \bibnamefont
  {Fisher}},\ }\href {\doibase 10.1103/PhysRevLett.69.534} {\bibfield
  {journal} {\bibinfo  {journal} {Phys. Rev. Lett.}\ }\textbf {\bibinfo
  {volume} {69}},\ \bibinfo {pages} {534} (\bibinfo {year} {1992})}\BibitemShut
  {NoStop}%
\bibitem [{\citenamefont {Fisher}(1994)}]{Fisher1994}%
  \BibitemOpen
  \bibfield  {author} {\bibinfo {author} {\bibfnamefont {D.~S.}\ \bibnamefont
  {Fisher}},\ }\href {\doibase 10.1103/PhysRevB.50.3799} {\bibfield  {journal}
  {\bibinfo  {journal} {Phys. Rev. B}\ }\textbf {\bibinfo {volume} {50}},\
  \bibinfo {pages} {3799} (\bibinfo {year} {1994})}\BibitemShut {NoStop}%
\bibitem [{\citenamefont {Fisher}(1995)}]{Fisher1995}%
  \BibitemOpen
  \bibfield  {author} {\bibinfo {author} {\bibfnamefont {D.~S.}\ \bibnamefont
  {Fisher}},\ }\href {\doibase 10.1103/PhysRevB.51.6411} {\bibfield  {journal}
  {\bibinfo  {journal} {Phys. Rev. B}\ }\textbf {\bibinfo {volume} {51}},\
  \bibinfo {pages} {6411} (\bibinfo {year} {1995})}\BibitemShut {NoStop}%
\bibitem [{\citenamefont {Agarwal}\ \emph
  {et~al.}(2015{\natexlab{b}})\citenamefont {Agarwal}, \citenamefont {Demler},\
  and\ \citenamefont {Martin}}]{Agarwal2015}%
  \BibitemOpen
  \bibfield  {author} {\bibinfo {author} {\bibfnamefont {K.}~\bibnamefont
  {Agarwal}}, \bibinfo {author} {\bibfnamefont {E.~A.}\ \bibnamefont {Demler}},
  \ and\ \bibinfo {author} {\bibfnamefont {I.}~\bibnamefont {Martin}},\ }\href
  {\doibase 10.1103/PhysRevB.92.184203} {\bibfield  {journal} {\bibinfo
  {journal} {Phys. Rev. B}\ }\textbf {\bibinfo {volume} {92}},\ \bibinfo
  {pages} {184203} (\bibinfo {year} {2015}{\natexlab{b}})}\BibitemShut
  {NoStop}%
\bibitem [{\citenamefont {Nauts}\ and\ \citenamefont
  {Wyatt}(1983)}]{nauts_new_1983}%
  \BibitemOpen
  \bibfield  {author} {\bibinfo {author} {\bibfnamefont {A.}~\bibnamefont
  {Nauts}}\ and\ \bibinfo {author} {\bibfnamefont {R.~E.}\ \bibnamefont
  {Wyatt}},\ }\href {\doibase 10.1103/PhysRevLett.51.2238} {\bibfield
  {journal} {\bibinfo  {journal} {Physical Review Letters}\ }\textbf {\bibinfo
  {volume} {51}},\ \bibinfo {pages} {2238} (\bibinfo {year}
  {1983})}\BibitemShut {NoStop}%
\bibitem [{\citenamefont {Moler}\ and\ \citenamefont
  {Van~Loan}(2003)}]{moler_nineteen_2003}%
  \BibitemOpen
  \bibfield  {author} {\bibinfo {author} {\bibfnamefont {C.}~\bibnamefont
  {Moler}}\ and\ \bibinfo {author} {\bibfnamefont {C.}~\bibnamefont
  {Van~Loan}},\ }\href {\doibase 10.1137/S00361445024180} {\bibfield  {journal}
  {\bibinfo  {journal} {SIAM Review}\ }\textbf {\bibinfo {volume} {45}},\
  \bibinfo {pages} {3} (\bibinfo {year} {2003})}\BibitemShut {NoStop}%
\bibitem [{\citenamefont {Arnoldi}(1951)}]{arnoldi_principle_1951}%
  \BibitemOpen
  \bibfield  {author} {\bibinfo {author} {\bibfnamefont {W.~E.}\ \bibnamefont
  {Arnoldi}},\ }\href@noop {} {\bibfield  {journal} {\bibinfo  {journal}
  {Quarterly of Applied Mathematics}\ }\textbf {\bibinfo {volume} {9}},\
  \bibinfo {pages} {17} (\bibinfo {year} {1951})}\BibitemShut {NoStop}%
\bibitem [{\citenamefont {Suzuki}(1990)}]{Suzuki1990}%
  \BibitemOpen
  \bibfield  {author} {\bibinfo {author} {\bibfnamefont {M.}~\bibnamefont
  {Suzuki}},\ }\href {\doibase 10.1016/0375-9601(90)90962-N} {\bibfield
  {journal} {\bibinfo  {journal} {Phys. Lett. A}\ }\textbf {\bibinfo {volume}
  {146}},\ \bibinfo {pages} {319} (\bibinfo {year} {1990})}\BibitemShut
  {NoStop}%
\bibitem [{\citenamefont {Sornborger}\ and\ \citenamefont
  {Stewart}(1999)}]{Sornborger1999}%
  \BibitemOpen
  \bibfield  {author} {\bibinfo {author} {\bibfnamefont {A.~T.}\ \bibnamefont
  {Sornborger}}\ and\ \bibinfo {author} {\bibfnamefont {E.~D.}\ \bibnamefont
  {Stewart}},\ }\href {\doibase 10.1103/PhysRevA.60.1956} {\bibfield  {journal}
  {\bibinfo  {journal} {Phys. Rev. A}\ }\textbf {\bibinfo {volume} {60}},\
  \bibinfo {pages} {1956} (\bibinfo {year} {1999})}\BibitemShut {NoStop}%
\bibitem [{\citenamefont {{\v{Z}}nidari{\v{c}}}\ \emph
  {et~al.}(2008)\citenamefont {{\v{Z}}nidari{\v{c}}}, \citenamefont {Prosen},\
  and\ \citenamefont {Prelov{\v{s}}ek}}]{Znidaric2008}%
  \BibitemOpen
  \bibfield  {author} {\bibinfo {author} {\bibfnamefont {M.}~\bibnamefont
  {{\v{Z}}nidari{\v{c}}}}, \bibinfo {author} {\bibfnamefont {T.}~\bibnamefont
  {Prosen}}, \ and\ \bibinfo {author} {\bibfnamefont {P.}~\bibnamefont
  {Prelov{\v{s}}ek}},\ }\href {\doibase 10.1103/PhysRevB.77.064426} {\bibfield
  {journal} {\bibinfo  {journal} {Phys. Rev. B}\ }\textbf {\bibinfo {volume}
  {77}},\ \bibinfo {pages} {064426} (\bibinfo {year} {2008})}\BibitemShut
  {NoStop}%
\bibitem [{\citenamefont {Bardarson}\ \emph {et~al.}(2012)\citenamefont
  {Bardarson}, \citenamefont {Pollmann},\ and\ \citenamefont
  {Moore}}]{bardarson_unbounded_2012}%
  \BibitemOpen
  \bibfield  {author} {\bibinfo {author} {\bibfnamefont {J.~H.}\ \bibnamefont
  {Bardarson}}, \bibinfo {author} {\bibfnamefont {F.}~\bibnamefont {Pollmann}},
  \ and\ \bibinfo {author} {\bibfnamefont {J.~E.}\ \bibnamefont {Moore}},\
  }\href {\doibase 10.1103/PhysRevLett.109.017202} {\bibfield  {journal}
  {\bibinfo  {journal} {Physical Review Letters}\ }\textbf {\bibinfo {volume}
  {109}},\ \bibinfo {pages} {017202} (\bibinfo {year} {2012})}\BibitemShut
  {NoStop}%
\bibitem [{\citenamefont {Serbyn}\ \emph
  {et~al.}(2013{\natexlab{a}})\citenamefont {Serbyn}, \citenamefont
  {Papi{\'{c}}},\ and\ \citenamefont {Abanin}}]{Serbyn2013b}%
  \BibitemOpen
  \bibfield  {author} {\bibinfo {author} {\bibfnamefont {M.}~\bibnamefont
  {Serbyn}}, \bibinfo {author} {\bibfnamefont {Z.}~\bibnamefont {Papi{\'{c}}}},
  \ and\ \bibinfo {author} {\bibfnamefont {D.~A.}\ \bibnamefont {Abanin}},\
  }\href {\doibase 10.1103/PhysRevLett.110.260601} {\bibfield  {journal}
  {\bibinfo  {journal} {Phys. Rev. Lett.}\ }\textbf {\bibinfo {volume} {110}},\
  \bibinfo {pages} {260601} (\bibinfo {year} {2013}{\natexlab{a}})}\BibitemShut
  {NoStop}%
\bibitem [{\citenamefont {Deng}\ \emph {et~al.}(2017)\citenamefont {Deng},
  \citenamefont {Li}, \citenamefont {Pixley}, \citenamefont {Wu},\ and\
  \citenamefont {{Das Sarma}}}]{Deng2016a}%
  \BibitemOpen
  \bibfield  {author} {\bibinfo {author} {\bibfnamefont {D.-L.}\ \bibnamefont
  {Deng}}, \bibinfo {author} {\bibfnamefont {X.}~\bibnamefont {Li}}, \bibinfo
  {author} {\bibfnamefont {J.~H.}\ \bibnamefont {Pixley}}, \bibinfo {author}
  {\bibfnamefont {Y.-L.}\ \bibnamefont {Wu}}, \ and\ \bibinfo {author}
  {\bibfnamefont {S.}~\bibnamefont {{Das Sarma}}},\ }\href {\doibase
  10.1103/PhysRevB.95.024202} {\bibfield  {journal} {\bibinfo  {journal} {Phys.
  Rev. B}\ }\textbf {\bibinfo {volume} {95}},\ \bibinfo {pages} {024202}
  (\bibinfo {year} {2017})}\BibitemShut {NoStop}%
\bibitem [{\citenamefont {Vidal}(2003)}]{vidal_efficient_2003}%
  \BibitemOpen
  \bibfield  {author} {\bibinfo {author} {\bibfnamefont {G.}~\bibnamefont
  {Vidal}},\ }\href {\doibase 10.1103/PhysRevLett.91.147902} {\bibfield
  {journal} {\bibinfo  {journal} {Physical Review Letters}\ }\textbf {\bibinfo
  {volume} {91}},\ \bibinfo {pages} {147902} (\bibinfo {year}
  {2003})}\BibitemShut {NoStop}%
\bibitem [{\citenamefont {Vidal}(2004)}]{vidal_efficient_2004}%
  \BibitemOpen
  \bibfield  {author} {\bibinfo {author} {\bibfnamefont {G.}~\bibnamefont
  {Vidal}},\ }\href {\doibase 10.1103/PhysRevLett.93.040502} {\bibfield
  {journal} {\bibinfo  {journal} {Physical Review Letters}\ }\textbf {\bibinfo
  {volume} {93}},\ \bibinfo {pages} {040502} (\bibinfo {year}
  {2004})}\BibitemShut {NoStop}%
\bibitem [{\citenamefont
  {Schollw{\"o}ck}(2005)}]{schollwock_density-matrix_2005}%
  \BibitemOpen
  \bibfield  {author} {\bibinfo {author} {\bibfnamefont {U.}~\bibnamefont
  {Schollw{\"o}ck}},\ }\href {\doibase 10.1103/RevModPhys.77.259} {\bibfield
  {journal} {\bibinfo  {journal} {Reviews of Modern Physics}\ }\textbf
  {\bibinfo {volume} {77}},\ \bibinfo {pages} {259} (\bibinfo {year}
  {2005})}\BibitemShut {NoStop}%
\bibitem [{\citenamefont {Lindblad}(1976)}]{lindblad_generators_1976}%
  \BibitemOpen
  \bibfield  {author} {\bibinfo {author} {\bibfnamefont {G.}~\bibnamefont
  {Lindblad}},\ }\href {\doibase 10.1007/BF01608499} {\bibfield  {journal}
  {\bibinfo  {journal} {Communications in Mathematical Physics}\ }\textbf
  {\bibinfo {volume} {48}},\ \bibinfo {pages} {119} (\bibinfo {year}
  {1976})}\BibitemShut {NoStop}%
\bibitem [{\citenamefont {Prosen}\ and\ \citenamefont {{\v Z}nidari{\v
  c}}(2009)}]{prosen_matrix_2009}%
  \BibitemOpen
  \bibfield  {author} {\bibinfo {author} {\bibfnamefont {T.}~\bibnamefont
  {Prosen}}\ and\ \bibinfo {author} {\bibfnamefont {M.}~\bibnamefont {{\v
  Z}nidari{\v c}}},\ }\href {\doibase 10.1088/1742-5468/2009/02/P02035}
  {\bibfield  {journal} {\bibinfo  {journal} {Journal of Statistical Mechanics:
  Theory and Experiment}\ }\textbf {\bibinfo {volume} {2009}},\ \bibinfo
  {pages} {P02035} (\bibinfo {year} {2009})}\BibitemShut {NoStop}%
\bibitem [{\citenamefont {{\v Z}nidari{\v
  c}}(2010)}]{znidaric_dephasing-induced_2010}%
  \BibitemOpen
  \bibfield  {author} {\bibinfo {author} {\bibfnamefont {M.}~\bibnamefont {{\v
  Z}nidari{\v c}}},\ }\href {\doibase 10.1088/1367-2630/12/4/043001} {\bibfield
   {journal} {\bibinfo  {journal} {New Journal of Physics}\ }\textbf {\bibinfo
  {volume} {12}},\ \bibinfo {pages} {043001} (\bibinfo {year}
  {2010})}\BibitemShut {NoStop}%
\bibitem [{\citenamefont {Ledoux}(2001)}]{ledoux_concentration_2001}%
  \BibitemOpen
  \bibfield  {author} {\bibinfo {author} {\bibfnamefont {M.}~\bibnamefont
  {Ledoux}},\ }\href@noop {} {\emph {\bibinfo {title} {The concentration of
  measure phenomenon}}},\ \bibinfo {series} {Mathematical {Surveys} and
  {Monographs}}, Vol.~\bibinfo {volume} {89}\ (\bibinfo  {publisher} {American
  Mathematical Society, Providence, RI},\ \bibinfo {year} {2001})\BibitemShut
  {NoStop}%
\bibitem [{\citenamefont {Popescu}\ \emph {et~al.}(2006)\citenamefont
  {Popescu}, \citenamefont {Short},\ and\ \citenamefont
  {Winter}}]{Popescu2006}%
  \BibitemOpen
  \bibfield  {author} {\bibinfo {author} {\bibfnamefont {S.}~\bibnamefont
  {Popescu}}, \bibinfo {author} {\bibfnamefont {A.~J.}\ \bibnamefont {Short}},
  \ and\ \bibinfo {author} {\bibfnamefont {A.}~\bibnamefont {Winter}},\ }\href
  {\doibase 10.1038/nphys444} {\bibfield  {journal} {\bibinfo  {journal} {Nat.
  Phys.}\ }\textbf {\bibinfo {volume} {2}},\ \bibinfo {pages} {754} (\bibinfo
  {year} {2006})}\BibitemShut {NoStop}%
\bibitem [{\citenamefont {Goldstein}\ \emph {et~al.}(2006)\citenamefont
  {Goldstein}, \citenamefont {Lebowitz}, \citenamefont {Tumulka},\ and\
  \citenamefont {Zangh{\`i}}}]{goldstein_canonical_2006}%
  \BibitemOpen
  \bibfield  {author} {\bibinfo {author} {\bibfnamefont {S.}~\bibnamefont
  {Goldstein}}, \bibinfo {author} {\bibfnamefont {J.~L.}\ \bibnamefont
  {Lebowitz}}, \bibinfo {author} {\bibfnamefont {R.}~\bibnamefont {Tumulka}}, \
  and\ \bibinfo {author} {\bibfnamefont {N.}~\bibnamefont {Zangh{\`i}}},\
  }\href {\doibase 10.1103/PhysRevLett.96.050403} {\bibfield  {journal}
  {\bibinfo  {journal} {Physical Review Letters}\ }\textbf {\bibinfo {volume}
  {96}},\ \bibinfo {pages} {050403} (\bibinfo {year} {2006})}\BibitemShut
  {NoStop}%
\bibitem [{\citenamefont {Reimann}(2007)}]{Reimann2007}%
  \BibitemOpen
  \bibfield  {author} {\bibinfo {author} {\bibfnamefont {P.}~\bibnamefont
  {Reimann}},\ }\href {\doibase 10.1103/PhysRevLett.99.160404} {\bibfield
  {journal} {\bibinfo  {journal} {Phys. Rev. Lett.}\ }\textbf {\bibinfo
  {volume} {99}},\ \bibinfo {pages} {160404} (\bibinfo {year}
  {2007})}\BibitemShut {NoStop}%
\bibitem [{\citenamefont {Bartsch}\ and\ \citenamefont
  {Gemmer}(2009)}]{Bartsch2009}%
  \BibitemOpen
  \bibfield  {author} {\bibinfo {author} {\bibfnamefont {C.}~\bibnamefont
  {Bartsch}}\ and\ \bibinfo {author} {\bibfnamefont {J.}~\bibnamefont
  {Gemmer}},\ }\href {\doibase 10.1103/PhysRevLett.102.110403} {\bibfield
  {journal} {\bibinfo  {journal} {Phys. Rev. Lett.}\ }\textbf {\bibinfo
  {volume} {102}},\ \bibinfo {pages} {110403} (\bibinfo {year}
  {2009})}\BibitemShut {NoStop}%
\bibitem [{\citenamefont {Gelman}\ and\ \citenamefont
  {Kosloff}(2003)}]{gelman_simulating_2003}%
  \BibitemOpen
  \bibfield  {author} {\bibinfo {author} {\bibfnamefont {D.}~\bibnamefont
  {Gelman}}\ and\ \bibinfo {author} {\bibfnamefont {R.}~\bibnamefont
  {Kosloff}},\ }\href {\doibase 10.1016/j.cplett.2003.09.119} {\bibfield
  {journal} {\bibinfo  {journal} {Chemical Physics Letters}\ }\textbf {\bibinfo
  {volume} {381}},\ \bibinfo {pages} {129} (\bibinfo {year}
  {2003})}\BibitemShut {NoStop}%
\bibitem [{\citenamefont {Sugiura}\ and\ \citenamefont
  {Shimizu}(2012)}]{sugiura_thermal_2012}%
  \BibitemOpen
  \bibfield  {author} {\bibinfo {author} {\bibfnamefont {S.}~\bibnamefont
  {Sugiura}}\ and\ \bibinfo {author} {\bibfnamefont {A.}~\bibnamefont
  {Shimizu}},\ }\href {\doibase 10.1103/PhysRevLett.108.240401} {\bibfield
  {journal} {\bibinfo  {journal} {Physical Review Letters}\ }\textbf {\bibinfo
  {volume} {108}},\ \bibinfo {pages} {240401} (\bibinfo {year}
  {2012})}\BibitemShut {NoStop}%
\bibitem [{\citenamefont {Sugiura}\ and\ \citenamefont
  {Shimizu}(2013)}]{sugiura_canonical_2013}%
  \BibitemOpen
  \bibfield  {author} {\bibinfo {author} {\bibfnamefont {S.}~\bibnamefont
  {Sugiura}}\ and\ \bibinfo {author} {\bibfnamefont {A.}~\bibnamefont
  {Shimizu}},\ }\href {\doibase 10.1103/PhysRevLett.111.010401} {\bibfield
  {journal} {\bibinfo  {journal} {Physical Review Letters}\ }\textbf {\bibinfo
  {volume} {111}},\ \bibinfo {pages} {010401} (\bibinfo {year}
  {2013})}\BibitemShut {NoStop}%
\bibitem [{\citenamefont {Elsayed}\ and\ \citenamefont
  {Fine}(2013)}]{elsayed_regression_2013}%
  \BibitemOpen
  \bibfield  {author} {\bibinfo {author} {\bibfnamefont {T.~A.}\ \bibnamefont
  {Elsayed}}\ and\ \bibinfo {author} {\bibfnamefont {B.~V.}\ \bibnamefont
  {Fine}},\ }\href {\doibase 10.1103/PhysRevLett.110.070404} {\bibfield
  {journal} {\bibinfo  {journal} {Physical Review Letters}\ }\textbf {\bibinfo
  {volume} {110}},\ \bibinfo {pages} {070404} (\bibinfo {year}
  {2013})}\BibitemShut {NoStop}%
\bibitem [{\citenamefont {Steinigeweg}\ \emph {et~al.}(2014)\citenamefont
  {Steinigeweg}, \citenamefont {Khodja}, \citenamefont {Niemeyer},
  \citenamefont {Gogolin},\ and\ \citenamefont
  {Gemmer}}]{steinigeweg_pushing_2014}%
  \BibitemOpen
  \bibfield  {author} {\bibinfo {author} {\bibfnamefont {R.}~\bibnamefont
  {Steinigeweg}}, \bibinfo {author} {\bibfnamefont {A.}~\bibnamefont {Khodja}},
  \bibinfo {author} {\bibfnamefont {H.}~\bibnamefont {Niemeyer}}, \bibinfo
  {author} {\bibfnamefont {C.}~\bibnamefont {Gogolin}}, \ and\ \bibinfo
  {author} {\bibfnamefont {J.}~\bibnamefont {Gemmer}},\ }\href {\doibase
  10.1103/PhysRevLett.112.130403} {\bibfield  {journal} {\bibinfo  {journal}
  {Physical Review Letters}\ }\textbf {\bibinfo {volume} {112}},\ \bibinfo
  {pages} {130403} (\bibinfo {year} {2014})}\BibitemShut {NoStop}%
\bibitem [{Note9()}]{Note9}%
  \BibitemOpen
  \bibinfo {note} {This means typically states from the center of the spectrum
  where the density of states is exponentially large.}\BibitemShut {Stop}%
\bibitem [{\citenamefont {Cuppen}(1981)}]{cuppen_divide_1981}%
  \BibitemOpen
  \bibfield  {author} {\bibinfo {author} {\bibfnamefont {J.~J.~M.}\
  \bibnamefont {Cuppen}},\ }\href {\doibase 10.1007/BF01396757} {\bibfield
  {journal} {\bibinfo  {journal} {Numerische Mathematik}\ }\textbf {\bibinfo
  {volume} {36}},\ \bibinfo {pages} {177} (\bibinfo {year} {1981})}\BibitemShut
  {NoStop}%
\bibitem [{\citenamefont {Dhillon}(1998)}]{dhillon_new_1998}%
  \BibitemOpen
  \bibfield  {author} {\bibinfo {author} {\bibfnamefont {I.~S.}\ \bibnamefont
  {Dhillon}},\ }\emph {\bibinfo {title} {A {New} {O} ({N}(2)) {Algorithm} for
  the {Symmetric} {Tridiagonal} {Eigenvalue}/{Eigenvector} {Problem}}},\
  \href@noop {} {Ph.D. thesis},\ \bibinfo  {school} {University of California
  at Berkeley}, \bibinfo {address} {Berkeley, CA, USA} (\bibinfo {year}
  {1998}),\ \bibinfo {note} {uMI Order No. GAX98-03176}\BibitemShut {NoStop}%
\bibitem [{\citenamefont {Amestoy}\ \emph {et~al.}(2001)\citenamefont
  {Amestoy}, \citenamefont {Duff}, \citenamefont {L'Excellent},\ and\
  \citenamefont {Koster}}]{amestoy_fully_2001}%
  \BibitemOpen
  \bibfield  {author} {\bibinfo {author} {\bibfnamefont {P.}~\bibnamefont
  {Amestoy}}, \bibinfo {author} {\bibfnamefont {I.}~\bibnamefont {Duff}},
  \bibinfo {author} {\bibfnamefont {J.}~\bibnamefont {L'Excellent}}, \ and\
  \bibinfo {author} {\bibfnamefont {J.}~\bibnamefont {Koster}},\ }\href
  {\doibase 10.1137/S0895479899358194} {\bibfield  {journal} {\bibinfo
  {journal} {SIAM Journal on Matrix Analysis and Applications}\ }\textbf
  {\bibinfo {volume} {23}},\ \bibinfo {pages} {15} (\bibinfo {year}
  {2001})}\BibitemShut {NoStop}%
\bibitem [{\citenamefont {Amestoy}\ \emph {et~al.}(2006)\citenamefont
  {Amestoy}, \citenamefont {Guermouche}, \citenamefont
  {L{\textquoteright}Excellent},\ and\ \citenamefont
  {Pralet}}]{amestoy_hybrid_2006}%
  \BibitemOpen
  \bibfield  {author} {\bibinfo {author} {\bibfnamefont {P.~R.}\ \bibnamefont
  {Amestoy}}, \bibinfo {author} {\bibfnamefont {A.}~\bibnamefont {Guermouche}},
  \bibinfo {author} {\bibfnamefont {J.-Y.}\ \bibnamefont
  {L{\textquoteright}Excellent}}, \ and\ \bibinfo {author} {\bibfnamefont
  {S.}~\bibnamefont {Pralet}},\ }\href {\doibase 10.1016/j.parco.2005.07.004}
  {\bibfield  {journal} {\bibinfo  {journal} {Parallel Computing}\ }\bibinfo
  {series} {Parallel {Matrix} {Algorithms} and {Applications}
  ({PMAA}{\textquoteright}04)},\ \textbf {\bibinfo {volume} {32}},\ \bibinfo
  {pages} {136} (\bibinfo {year} {2006})}\BibitemShut {NoStop}%
\bibitem [{\citenamefont {Li}(2005)}]{li_overview_2005}%
  \BibitemOpen
  \bibfield  {author} {\bibinfo {author} {\bibfnamefont {X.~S.}\ \bibnamefont
  {Li}},\ }\href {\doibase 10.1145/1089014.1089017} {\bibfield  {journal}
  {\bibinfo  {journal} {ACM Trans. Math. Softw.}\ }\textbf {\bibinfo {volume}
  {31}},\ \bibinfo {pages} {302} (\bibinfo {year} {2005})}\BibitemShut
  {NoStop}%
\bibitem [{\citenamefont {Yu}\ \emph {et~al.}(2017)\citenamefont {Yu},
  \citenamefont {Pekker},\ and\ \citenamefont {Clark}}]{yu_finding_2017}%
  \BibitemOpen
  \bibfield  {author} {\bibinfo {author} {\bibfnamefont {X.}~\bibnamefont
  {Yu}}, \bibinfo {author} {\bibfnamefont {D.}~\bibnamefont {Pekker}}, \ and\
  \bibinfo {author} {\bibfnamefont {B.~K.}\ \bibnamefont {Clark}},\ }\href
  {\doibase 10.1103/PhysRevLett.118.017201} {\bibfield  {journal} {\bibinfo
  {journal} {Physical Review Letters}\ }\textbf {\bibinfo {volume} {118}},\
  \bibinfo {pages} {017201} (\bibinfo {year} {2017})}\BibitemShut {NoStop}%
\bibitem [{\citenamefont {Serbyn}\ \emph
  {et~al.}(2016{\natexlab{b}})\citenamefont {Serbyn}, \citenamefont
  {Michailidis}, \citenamefont {Abanin},\ and\ \citenamefont
  {Papi{\'c}}}]{serbyn_power-law_2016}%
  \BibitemOpen
  \bibfield  {author} {\bibinfo {author} {\bibfnamefont {M.}~\bibnamefont
  {Serbyn}}, \bibinfo {author} {\bibfnamefont {A.~A.}\ \bibnamefont
  {Michailidis}}, \bibinfo {author} {\bibfnamefont {D.~A.}\ \bibnamefont
  {Abanin}}, \ and\ \bibinfo {author} {\bibfnamefont {Z.}~\bibnamefont
  {Papi{\'c}}},\ }\href {\doibase 10.1103/PhysRevLett.117.160601} {\bibfield
  {journal} {\bibinfo  {journal} {Physical Review Letters}\ }\textbf {\bibinfo
  {volume} {117}},\ \bibinfo {pages} {160601} (\bibinfo {year}
  {2016}{\natexlab{b}})}\BibitemShut {NoStop}%
\bibitem [{\citenamefont {Khemani}\ \emph
  {et~al.}(2016{\natexlab{b}})\citenamefont {Khemani}, \citenamefont
  {Pollmann},\ and\ \citenamefont {Sondhi}}]{Khemani}%
  \BibitemOpen
  \bibfield  {author} {\bibinfo {author} {\bibfnamefont {V.}~\bibnamefont
  {Khemani}}, \bibinfo {author} {\bibfnamefont {F.}~\bibnamefont {Pollmann}}, \
  and\ \bibinfo {author} {\bibfnamefont {S.~L.}\ \bibnamefont {Sondhi}},\
  }\href {\doibase 10.1103/PhysRevLett.116.247204} {\bibfield  {journal}
  {\bibinfo  {journal} {Phys. Rev. Lett.}\ }\textbf {\bibinfo {volume} {116}},\
  \bibinfo {pages} {247204} (\bibinfo {year} {2016}{\natexlab{b}})}\BibitemShut
  {NoStop}%
\bibitem [{\citenamefont {Lim}\ and\ \citenamefont {Sheng}(2016)}]{Lim}%
  \BibitemOpen
  \bibfield  {author} {\bibinfo {author} {\bibfnamefont {S.~P.}\ \bibnamefont
  {Lim}}\ and\ \bibinfo {author} {\bibfnamefont {D.~N.}\ \bibnamefont
  {Sheng}},\ }\href {\doibase 10.1103/PhysRevB.94.045111} {\bibfield  {journal}
  {\bibinfo  {journal} {Phys. Rev. B}\ }\textbf {\bibinfo {volume} {94}},\
  \bibinfo {pages} {045111} (\bibinfo {year} {2016})}\BibitemShut {NoStop}%
\bibitem [{\citenamefont {Kennes}\ and\ \citenamefont
  {Karrasch}(2016)}]{Kennes2015}%
  \BibitemOpen
  \bibfield  {author} {\bibinfo {author} {\bibfnamefont {D.~M.}\ \bibnamefont
  {Kennes}}\ and\ \bibinfo {author} {\bibfnamefont {C.}~\bibnamefont
  {Karrasch}},\ }\href {\doibase 10.1103/PhysRevB.93.245129} {\bibfield
  {journal} {\bibinfo  {journal} {Phys. Rev. B}\ }\textbf {\bibinfo {volume}
  {93}},\ \bibinfo {pages} {245129} (\bibinfo {year} {2016})}\BibitemShut
  {NoStop}%
\bibitem [{\citenamefont {Pekker}\ and\ \citenamefont
  {Clark}(2017)}]{Pekker2014b}%
  \BibitemOpen
  \bibfield  {author} {\bibinfo {author} {\bibfnamefont {D.}~\bibnamefont
  {Pekker}}\ and\ \bibinfo {author} {\bibfnamefont {B.~K.}\ \bibnamefont
  {Clark}},\ }\href {\doibase 10.1103/PhysRevB.95.035116} {\bibfield  {journal}
  {\bibinfo  {journal} {Phys. Rev. B}\ }\textbf {\bibinfo {volume} {95}},\
  \bibinfo {pages} {035116} (\bibinfo {year} {2017})}\BibitemShut {NoStop}%
\bibitem [{\citenamefont {Chandran}\ \emph {et~al.}(2015)\citenamefont
  {Chandran}, \citenamefont {Carrasquilla}, \citenamefont {Kim}, \citenamefont
  {Abanin},\ and\ \citenamefont {Vidal}}]{chandran_spectral_2015}%
  \BibitemOpen
  \bibfield  {author} {\bibinfo {author} {\bibfnamefont {A.}~\bibnamefont
  {Chandran}}, \bibinfo {author} {\bibfnamefont {J.}~\bibnamefont
  {Carrasquilla}}, \bibinfo {author} {\bibfnamefont {I.~H.}\ \bibnamefont
  {Kim}}, \bibinfo {author} {\bibfnamefont {D.~A.}\ \bibnamefont {Abanin}}, \
  and\ \bibinfo {author} {\bibfnamefont {G.}~\bibnamefont {Vidal}},\ }\href
  {\doibase 10.1103/PhysRevB.92.024201} {\bibfield  {journal} {\bibinfo
  {journal} {Physical Review B}\ }\textbf {\bibinfo {volume} {92}},\ \bibinfo
  {pages} {024201} (\bibinfo {year} {2015})}\BibitemShut {NoStop}%
\bibitem [{\citenamefont {Pollmann}\ \emph {et~al.}(2016)\citenamefont
  {Pollmann}, \citenamefont {Khemani}, \citenamefont {Cirac},\ and\
  \citenamefont {Sondhi}}]{Pollmann2016}%
  \BibitemOpen
  \bibfield  {author} {\bibinfo {author} {\bibfnamefont {F.}~\bibnamefont
  {Pollmann}}, \bibinfo {author} {\bibfnamefont {V.}~\bibnamefont {Khemani}},
  \bibinfo {author} {\bibfnamefont {J.~I.}\ \bibnamefont {Cirac}}, \ and\
  \bibinfo {author} {\bibfnamefont {S.~L.}\ \bibnamefont {Sondhi}},\ }\href
  {\doibase 10.1103/PhysRevB.94.041116} {\bibfield  {journal} {\bibinfo
  {journal} {Phys. Rev. B}\ }\textbf {\bibinfo {volume} {94}},\ \bibinfo
  {pages} {041116} (\bibinfo {year} {2016})}\BibitemShut {NoStop}%
\bibitem [{\citenamefont {Wahl}\ \emph {et~al.}(2017)\citenamefont {Wahl},
  \citenamefont {Pal},\ and\ \citenamefont {Simon}}]{Wahl2016}%
  \BibitemOpen
  \bibfield  {author} {\bibinfo {author} {\bibfnamefont {T.~B.}\ \bibnamefont
  {Wahl}}, \bibinfo {author} {\bibfnamefont {A.}~\bibnamefont {Pal}}, \ and\
  \bibinfo {author} {\bibfnamefont {S.~H.}\ \bibnamefont {Simon}},\ }\href
  {\doibase 10.1103/PhysRevX.7.021018} {\bibfield  {journal} {\bibinfo
  {journal} {Phys. Rev. X}\ }\textbf {\bibinfo {volume} {7}},\ \bibinfo {pages}
  {021018} (\bibinfo {year} {2017})}\BibitemShut {NoStop}%
\bibitem [{\citenamefont {Inglis}\ and\ \citenamefont
  {Pollet}(2016)}]{inglis_accessing_2016}%
  \BibitemOpen
  \bibfield  {author} {\bibinfo {author} {\bibfnamefont {S.}~\bibnamefont
  {Inglis}}\ and\ \bibinfo {author} {\bibfnamefont {L.}~\bibnamefont
  {Pollet}},\ }\href {\doibase 10.1103/PhysRevLett.117.120402} {\bibfield
  {journal} {\bibinfo  {journal} {Physical Review Letters}\ }\textbf {\bibinfo
  {volume} {117}},\ \bibinfo {pages} {120402} (\bibinfo {year}
  {2016})}\BibitemShut {NoStop}%
\bibitem [{\citenamefont {Serbyn}\ \emph
  {et~al.}(2013{\natexlab{b}})\citenamefont {Serbyn}, \citenamefont
  {Papi{\'{c}}},\ and\ \citenamefont {Abanin}}]{Serbyn2013a}%
  \BibitemOpen
  \bibfield  {author} {\bibinfo {author} {\bibfnamefont {M.}~\bibnamefont
  {Serbyn}}, \bibinfo {author} {\bibfnamefont {Z.}~\bibnamefont {Papi{\'{c}}}},
  \ and\ \bibinfo {author} {\bibfnamefont {D.~A.}\ \bibnamefont {Abanin}},\
  }\href {\doibase 10.1103/PhysRevLett.111.127201} {\bibfield  {journal}
  {\bibinfo  {journal} {Phys. Rev. Lett.}\ }\textbf {\bibinfo {volume} {111}},\
  \bibinfo {pages} {127201} (\bibinfo {year} {2013}{\natexlab{b}})}\BibitemShut
  {NoStop}%
\bibitem [{\citenamefont {Huse}\ \emph {et~al.}(2014)\citenamefont {Huse},
  \citenamefont {Nandkishore},\ and\ \citenamefont {Oganesyan}}]{Huse2013}%
  \BibitemOpen
  \bibfield  {author} {\bibinfo {author} {\bibfnamefont {D.~A.}\ \bibnamefont
  {Huse}}, \bibinfo {author} {\bibfnamefont {R.}~\bibnamefont {Nandkishore}}, \
  and\ \bibinfo {author} {\bibfnamefont {V.}~\bibnamefont {Oganesyan}},\ }\href
  {\doibase 10.1103/PhysRevB.90.174202} {\bibfield  {journal} {\bibinfo
  {journal} {Phys. Rev. B}\ }\textbf {\bibinfo {volume} {90}},\ \bibinfo
  {pages} {174202} (\bibinfo {year} {2014})}\BibitemShut {NoStop}%
\bibitem [{\citenamefont {Stan}\ \emph {et~al.}(2009)\citenamefont {Stan},
  \citenamefont {Dahlen},\ and\ \citenamefont {van Leeuwen}}]{Stan2009}%
  \BibitemOpen
  \bibfield  {author} {\bibinfo {author} {\bibfnamefont {A.}~\bibnamefont
  {Stan}}, \bibinfo {author} {\bibfnamefont {N.~E.}\ \bibnamefont {Dahlen}}, \
  and\ \bibinfo {author} {\bibfnamefont {R.}~\bibnamefont {van Leeuwen}},\
  }\href {\doibase 10.1063/1.3127247} {\bibfield  {journal} {\bibinfo
  {journal} {J. Chem. Phys.}\ }\textbf {\bibinfo {volume} {130}},\ \bibinfo
  {pages} {224101} (\bibinfo {year} {2009})}\BibitemShut {NoStop}%
\bibitem [{\citenamefont {Kadanoff}\ and\ \citenamefont
  {Baym}(1994)}]{Kadanoff1994}%
  \BibitemOpen
  \bibfield  {author} {\bibinfo {author} {\bibfnamefont {L.~P.}\ \bibnamefont
  {Kadanoff}}\ and\ \bibinfo {author} {\bibfnamefont {G.}~\bibnamefont
  {Baym}},\ }\href {http://www.amazon.com/dp/020141046X} {\emph {\bibinfo
  {title} {{Quantum Statistical Mechanics}}}},\ edited by\ \bibinfo {editor}
  {\bibfnamefont {D.}~\bibnamefont {Pines}}\ (\bibinfo  {publisher} {Westview
  Press},\ \bibinfo {year} {1994})\ p.\ \bibinfo {pages} {224}\BibitemShut
  {NoStop}%
\bibitem [{\citenamefont {{\v{S}}pi{\v{c}}ka}\ \emph
  {et~al.}(2005{\natexlab{a}})\citenamefont {{\v{S}}pi{\v{c}}ka}, \citenamefont
  {Velick{\'{y}}},\ and\ \citenamefont {Kalvov{\'{a}}}}]{Spicka2005}%
  \BibitemOpen
  \bibfield  {author} {\bibinfo {author} {\bibfnamefont {V.}~\bibnamefont
  {{\v{S}}pi{\v{c}}ka}}, \bibinfo {author} {\bibfnamefont {B.}~\bibnamefont
  {Velick{\'{y}}}}, \ and\ \bibinfo {author} {\bibfnamefont {A.}~\bibnamefont
  {Kalvov{\'{a}}}},\ }\href {\doibase 10.1016/j.physe.2005.05.014} {\bibfield
  {journal} {\bibinfo  {journal} {Phys. E Low-dimensional Syst.
  Nanostructures}\ }\textbf {\bibinfo {volume} {29}},\ \bibinfo {pages} {154}
  (\bibinfo {year} {2005}{\natexlab{a}})}\BibitemShut {NoStop}%
\bibitem [{\citenamefont {{\v{S}}pi{\v{c}}ka}\ \emph
  {et~al.}(2005{\natexlab{b}})\citenamefont {{\v{S}}pi{\v{c}}ka}, \citenamefont
  {Velick{\'{y}}},\ and\ \citenamefont {Kalvov{\'{a}}}}]{Spicka2005a}%
  \BibitemOpen
  \bibfield  {author} {\bibinfo {author} {\bibfnamefont {V.}~\bibnamefont
  {{\v{S}}pi{\v{c}}ka}}, \bibinfo {author} {\bibfnamefont {B.}~\bibnamefont
  {Velick{\'{y}}}}, \ and\ \bibinfo {author} {\bibfnamefont {A.}~\bibnamefont
  {Kalvov{\'{a}}}},\ }\href {\doibase 10.1016/j.physe.2005.05.016} {\bibfield
  {journal} {\bibinfo  {journal} {Phys. E Low-dimensional Syst.
  Nanostructures}\ }\textbf {\bibinfo {volume} {29}},\ \bibinfo {pages} {196}
  (\bibinfo {year} {2005}{\natexlab{b}})}\BibitemShut {NoStop}%
\bibitem [{\citenamefont {Latini}\ \emph {et~al.}(2014)\citenamefont {Latini},
  \citenamefont {Perfetto}, \citenamefont {Uimonen}, \citenamefont {van
  Leeuwen},\ and\ \citenamefont {Stefanucci}}]{Latini2013}%
  \BibitemOpen
  \bibfield  {author} {\bibinfo {author} {\bibfnamefont {S.}~\bibnamefont
  {Latini}}, \bibinfo {author} {\bibfnamefont {E.}~\bibnamefont {Perfetto}},
  \bibinfo {author} {\bibfnamefont {A.-M.}\ \bibnamefont {Uimonen}}, \bibinfo
  {author} {\bibfnamefont {R.}~\bibnamefont {van Leeuwen}}, \ and\ \bibinfo
  {author} {\bibfnamefont {G.}~\bibnamefont {Stefanucci}},\ }\href {\doibase
  10.1103/PhysRevB.89.075306} {\bibfield  {journal} {\bibinfo  {journal} {Phys.
  Rev. B}\ }\textbf {\bibinfo {volume} {89}},\ \bibinfo {pages} {075306}
  (\bibinfo {year} {2014})}\BibitemShut {NoStop}%
\bibitem [{\citenamefont {Bera}\ \emph {et~al.}(2017)\citenamefont {Bera},
  \citenamefont {{De Tomasi}}, \citenamefont {Weiner},\ and\ \citenamefont
  {Evers}}]{Bera2016}%
  \BibitemOpen
  \bibfield  {author} {\bibinfo {author} {\bibfnamefont {S.}~\bibnamefont
  {Bera}}, \bibinfo {author} {\bibfnamefont {G.}~\bibnamefont {{De Tomasi}}},
  \bibinfo {author} {\bibfnamefont {F.}~\bibnamefont {Weiner}}, \ and\ \bibinfo
  {author} {\bibfnamefont {F.}~\bibnamefont {Evers}},\ }\href {\doibase
  10.1103/PhysRevLett.118.196801} {\bibfield  {journal} {\bibinfo  {journal}
  {Phys. Rev. Lett.}\ }\textbf {\bibinfo {volume} {118}},\ \bibinfo {pages}
  {196801} (\bibinfo {year} {2017})}\BibitemShut {NoStop}%
\bibitem [{\citenamefont {Roscilde}\ \emph {et~al.}(2016)\citenamefont
  {Roscilde}, \citenamefont {Naldesi},\ and\ \citenamefont
  {Ercolessi}}]{Naldesi2016a}%
  \BibitemOpen
  \bibfield  {author} {\bibinfo {author} {\bibfnamefont {T.}~\bibnamefont
  {Roscilde}}, \bibinfo {author} {\bibfnamefont {P.}~\bibnamefont {Naldesi}}, \
  and\ \bibinfo {author} {\bibfnamefont {E.}~\bibnamefont {Ercolessi}},\ }\href
  {\doibase 10.21468/SciPostPhys.1.1.010} {\bibfield  {journal} {\bibinfo
  {journal} {SciPost Phys.}\ }\textbf {\bibinfo {volume} {1}},\ \bibinfo
  {pages} {010} (\bibinfo {year} {2016})}\BibitemShut {NoStop}%
\bibitem [{\citenamefont {L{\"u}schen}\ \emph {et~al.}(2016)\citenamefont
  {L{\"u}schen}, \citenamefont {Bordia}, \citenamefont {Scherg}, \citenamefont
  {Alet}, \citenamefont {Altman}, \citenamefont {Schneider},\ and\
  \citenamefont {Bloch}}]{luschen_evidence_2016}%
  \BibitemOpen
  \bibfield  {author} {\bibinfo {author} {\bibfnamefont {H.~P.}\ \bibnamefont
  {L{\"u}schen}}, \bibinfo {author} {\bibfnamefont {P.}~\bibnamefont {Bordia}},
  \bibinfo {author} {\bibfnamefont {S.}~\bibnamefont {Scherg}}, \bibinfo
  {author} {\bibfnamefont {F.}~\bibnamefont {Alet}}, \bibinfo {author}
  {\bibfnamefont {E.}~\bibnamefont {Altman}}, \bibinfo {author} {\bibfnamefont
  {U.}~\bibnamefont {Schneider}}, \ and\ \bibinfo {author} {\bibfnamefont
  {I.}~\bibnamefont {Bloch}},\ }\href@noop {} {\enquote {\bibinfo {title}
  {Evidence for {Griffiths}-{Type} {Dynamics} near the {Many}-{Body}
  {Localization} {Transition} in {Quasi}-{Periodic} {Systems}},}\ } (\bibinfo
  {year} {2016}),\ \Eprint {http://arxiv.org/abs/1612.07173} {arXiv:1612.07173}
  \BibitemShut {NoStop}%
\bibitem [{\citenamefont {Bar~Lev}\ \emph {et~al.}(2017)\citenamefont
  {Bar~Lev}, \citenamefont {Kennes}, \citenamefont {Kl{\"o}ckner},
  \citenamefont {Reichman},\ and\ \citenamefont
  {Karrasch}}]{bar_lev_transport_2017}%
  \BibitemOpen
  \bibfield  {author} {\bibinfo {author} {\bibfnamefont {Y.}~\bibnamefont
  {Bar~Lev}}, \bibinfo {author} {\bibfnamefont {D.~M.}\ \bibnamefont {Kennes}},
  \bibinfo {author} {\bibfnamefont {C.}~\bibnamefont {Kl{\"o}ckner}}, \bibinfo
  {author} {\bibfnamefont {D.~R.}\ \bibnamefont {Reichman}}, \ and\ \bibinfo
  {author} {\bibfnamefont {C.}~\bibnamefont {Karrasch}},\ }\href@noop {}
  {\enquote {\bibinfo {title} {Transport in quasiperiodic interacting systems:
  from superdiffusion to subdiffusion},}\ } (\bibinfo {year} {2017}),\ \Eprint
  {http://arxiv.org/abs/1702.04349} {arXiv:1702.04349} \BibitemShut {NoStop}%
\bibitem [{\citenamefont {Huveneers}(2017)}]{Huveneers2017}%
  \BibitemOpen
  \bibfield  {author} {\bibinfo {author} {\bibfnamefont {F.}~\bibnamefont
  {Huveneers}},\ }\href {http://arxiv.org/abs/1701.05755} {\  (\bibinfo {year}
  {2017})},\ \Eprint {http://arxiv.org/abs/1701.05755} {arXiv:1701.05755}
  \BibitemShut {NoStop}%
\bibitem [{\citenamefont {Fischer}\ and\ \citenamefont
  {Hertz}(1993)}]{Fischer1993spin}%
  \BibitemOpen
  \bibfield  {author} {\bibinfo {author} {\bibfnamefont {K.~H.}\ \bibnamefont
  {Fischer}}\ and\ \bibinfo {author} {\bibfnamefont {J.~A.}\ \bibnamefont
  {Hertz}},\ }\href@noop {} {\emph {\bibinfo {title} {{Spin Glasses}}}},\
  Cambridge Studies in Magnetism\ (\bibinfo  {publisher} {Cambridge University
  Press},\ \bibinfo {year} {1993})\BibitemShut {NoStop}%
\bibitem [{\citenamefont {Schiulaz}\ and\ \citenamefont
  {M{\"{u}}ller}(2014)}]{Schiulaz2013}%
  \BibitemOpen
  \bibfield  {author} {\bibinfo {author} {\bibfnamefont {M.}~\bibnamefont
  {Schiulaz}}\ and\ \bibinfo {author} {\bibfnamefont {M.}~\bibnamefont
  {M{\"{u}}ller}},\ }in\ \href {\doibase 10.1063/1.4893505} {\emph {\bibinfo
  {booktitle} {AIP Conf. Proc.}}}\ (\bibinfo {year} {2014})\ pp.\ \bibinfo
  {pages} {11--23}\BibitemShut {NoStop}%
\bibitem [{\citenamefont {Yao}\ \emph {et~al.}(2016)\citenamefont {Yao},
  \citenamefont {Laumann}, \citenamefont {Cirac}, \citenamefont {Lukin},\ and\
  \citenamefont {Moore}}]{Yao2014}%
  \BibitemOpen
  \bibfield  {author} {\bibinfo {author} {\bibfnamefont {N.~Y.}\ \bibnamefont
  {Yao}}, \bibinfo {author} {\bibfnamefont {C.~R.}\ \bibnamefont {Laumann}},
  \bibinfo {author} {\bibfnamefont {J.~I.}\ \bibnamefont {Cirac}}, \bibinfo
  {author} {\bibfnamefont {M.~D.}\ \bibnamefont {Lukin}}, \ and\ \bibinfo
  {author} {\bibfnamefont {J.~E.}\ \bibnamefont {Moore}},\ }\href {\doibase
  10.1103/PhysRevLett.117.240601} {\bibfield  {journal} {\bibinfo  {journal}
  {Phys. Rev. Lett.}\ }\textbf {\bibinfo {volume} {117}},\ \bibinfo {pages}
  {240601} (\bibinfo {year} {2016})}\BibitemShut {NoStop}%
\bibitem [{\citenamefont {Schiulaz}\ \emph {et~al.}(2015)\citenamefont
  {Schiulaz}, \citenamefont {Silva},\ and\ \citenamefont
  {M{\"{u}}ller}}]{Schiulaz2014}%
  \BibitemOpen
  \bibfield  {author} {\bibinfo {author} {\bibfnamefont {M.}~\bibnamefont
  {Schiulaz}}, \bibinfo {author} {\bibfnamefont {A.}~\bibnamefont {Silva}}, \
  and\ \bibinfo {author} {\bibfnamefont {M.}~\bibnamefont {M{\"{u}}ller}},\
  }\href {\doibase 10.1103/PhysRevB.91.184202} {\bibfield  {journal} {\bibinfo
  {journal} {Phys. Rev. B}\ }\textbf {\bibinfo {volume} {91}},\ \bibinfo
  {pages} {184202} (\bibinfo {year} {2015})}\BibitemShut {NoStop}%
\bibitem [{\citenamefont {Hickey}\ \emph {et~al.}(2016)\citenamefont {Hickey},
  \citenamefont {Genway},\ and\ \citenamefont {Garrahan}}]{Hickey2014}%
  \BibitemOpen
  \bibfield  {author} {\bibinfo {author} {\bibfnamefont {J.~M.}\ \bibnamefont
  {Hickey}}, \bibinfo {author} {\bibfnamefont {S.}~\bibnamefont {Genway}}, \
  and\ \bibinfo {author} {\bibfnamefont {J.~P.}\ \bibnamefont {Garrahan}},\
  }\href {\doibase 10.1088/1742-5468/2016/05/054047} {\bibfield  {journal}
  {\bibinfo  {journal} {J. Stat. Mech. Theory Exp.}\ }\textbf {\bibinfo
  {volume} {2016}},\ \bibinfo {pages} {054047} (\bibinfo {year}
  {2016})}\BibitemShut {NoStop}%
\bibitem [{\citenamefont {{De Roeck}}\ and\ \citenamefont
  {Huveneers}(2014{\natexlab{a}})}]{DeRoeck2014}%
  \BibitemOpen
  \bibfield  {author} {\bibinfo {author} {\bibfnamefont {W.}~\bibnamefont {{De
  Roeck}}}\ and\ \bibinfo {author} {\bibfnamefont {F.}~\bibnamefont
  {Huveneers}},\ }\href {\doibase 10.1007/s00220-014-2116-8} {\bibfield
  {journal} {\bibinfo  {journal} {Commun. Math. Phys.}\ }\textbf {\bibinfo
  {volume} {332}},\ \bibinfo {pages} {1017} (\bibinfo {year}
  {2014}{\natexlab{a}})}\BibitemShut {NoStop}%
\bibitem [{\citenamefont {{De Roeck}}\ and\ \citenamefont
  {Huveneers}(2014{\natexlab{b}})}]{DeRoeck2014a}%
  \BibitemOpen
  \bibfield  {author} {\bibinfo {author} {\bibfnamefont {W.}~\bibnamefont {{De
  Roeck}}}\ and\ \bibinfo {author} {\bibfnamefont {F.}~\bibnamefont
  {Huveneers}},\ }\href {\doibase 10.1103/PhysRevB.90.165137} {\bibfield
  {journal} {\bibinfo  {journal} {Phys. Rev. B}\ }\textbf {\bibinfo {volume}
  {90}},\ \bibinfo {pages} {165137} (\bibinfo {year}
  {2014}{\natexlab{b}})}\BibitemShut {NoStop}%
\bibitem [{\citenamefont {Papi{\'{c}}}\ \emph {et~al.}(2015)\citenamefont
  {Papi{\'{c}}}, \citenamefont {Stoudenmire},\ and\ \citenamefont
  {Abanin}}]{Papic2015}%
  \BibitemOpen
  \bibfield  {author} {\bibinfo {author} {\bibfnamefont {Z.}~\bibnamefont
  {Papi{\'{c}}}}, \bibinfo {author} {\bibfnamefont {E.~M.}\ \bibnamefont
  {Stoudenmire}}, \ and\ \bibinfo {author} {\bibfnamefont {D.~A.}\ \bibnamefont
  {Abanin}},\ }\href {\doibase 10.1016/j.aop.2015.08.024} {\bibfield  {journal}
  {\bibinfo  {journal} {Ann. Phys. (N. Y).}\ }\textbf {\bibinfo {volume}
  {362}},\ \bibinfo {pages} {714} (\bibinfo {year} {2015})}\BibitemShut
  {NoStop}%
\bibitem [{\citenamefont {Roeck}\ and\ \citenamefont
  {Huveneers}(2015)}]{Roeck2015a}%
  \BibitemOpen
  \bibfield  {author} {\bibinfo {author} {\bibfnamefont {W.~D.}\ \bibnamefont
  {Roeck}}\ and\ \bibinfo {author} {\bibfnamefont {F.}~\bibnamefont
  {Huveneers}},\ }in\ \href {\doibase 10.1007/978-3-319-16637-7} {\emph
  {\bibinfo {booktitle} {From Part. Syst. to Partial Differ. Equations II}}},\
  \bibinfo {series} {Springer Proceedings in Mathematics {\&} Statistics},
  Vol.\ \bibinfo {volume} {129},\ \bibinfo {editor} {edited by\ \bibinfo
  {editor} {\bibfnamefont {P.}~\bibnamefont {Gon{\c{c}}alves}}\ and\ \bibinfo
  {editor} {\bibfnamefont {A.~J.}\ \bibnamefont {Soares}}}\ (\bibinfo
  {publisher} {Springer International Publishing},\ \bibinfo {address} {Cham},\
  \bibinfo {year} {2015})\ pp.\ \bibinfo {pages} {173--192}\BibitemShut
  {NoStop}%
\bibitem [{\citenamefont {van Horssen}\ \emph {et~al.}(2015)\citenamefont {van
  Horssen}, \citenamefont {Levi},\ and\ \citenamefont
  {Garrahan}}]{Horssen2015}%
  \BibitemOpen
  \bibfield  {author} {\bibinfo {author} {\bibfnamefont {M.}~\bibnamefont {van
  Horssen}}, \bibinfo {author} {\bibfnamefont {E.}~\bibnamefont {Levi}}, \ and\
  \bibinfo {author} {\bibfnamefont {J.~P.}\ \bibnamefont {Garrahan}},\ }\href
  {\doibase 10.1103/PhysRevB.92.100305} {\bibfield  {journal} {\bibinfo
  {journal} {Phys. Rev. B}\ }\textbf {\bibinfo {volume} {92}},\ \bibinfo
  {pages} {100305} (\bibinfo {year} {2015})}\BibitemShut {NoStop}%
\bibitem [{\citenamefont {Garrison}\ \emph {et~al.}(2016)\citenamefont
  {Garrison}, \citenamefont {Mishmash},\ and\ \citenamefont
  {Fisher}}]{Garrison2016}%
  \BibitemOpen
  \bibfield  {author} {\bibinfo {author} {\bibfnamefont {J.~R.}\ \bibnamefont
  {Garrison}}, \bibinfo {author} {\bibfnamefont {R.~V.}\ \bibnamefont
  {Mishmash}}, \ and\ \bibinfo {author} {\bibfnamefont {M.~P.~A.}\ \bibnamefont
  {Fisher}},\ }\href {http://arxiv.org/abs/1606.05650} {\  (\bibinfo {year}
  {2016})},\ \Eprint {http://arxiv.org/abs/1606.05650} {arXiv:1606.05650}
  \BibitemShut {NoStop}%
\bibitem [{\citenamefont {Antipov}\ \emph {et~al.}(2016)\citenamefont
  {Antipov}, \citenamefont {Javanmard}, \citenamefont {Ribeiro},\ and\
  \citenamefont {Kirchner}}]{Antipov2016}%
  \BibitemOpen
  \bibfield  {author} {\bibinfo {author} {\bibfnamefont {A.~E.}\ \bibnamefont
  {Antipov}}, \bibinfo {author} {\bibfnamefont {Y.}~\bibnamefont {Javanmard}},
  \bibinfo {author} {\bibfnamefont {P.}~\bibnamefont {Ribeiro}}, \ and\
  \bibinfo {author} {\bibfnamefont {S.}~\bibnamefont {Kirchner}},\ }\href
  {\doibase 10.1103/PhysRevLett.117.146601} {\bibfield  {journal} {\bibinfo
  {journal} {Phys. Rev. Lett.}\ }\textbf {\bibinfo {volume} {117}},\ \bibinfo
  {pages} {146601} (\bibinfo {year} {2016})}\BibitemShut {NoStop}%
\bibitem [{\citenamefont {Bouchaud}\ \emph {et~al.}(1996)\citenamefont
  {Bouchaud}, \citenamefont {Cugliandolo}, \citenamefont {Kurchan},\ and\
  \citenamefont {M{\'{e}}zard}}]{Alamos1971}%
  \BibitemOpen
  \bibfield  {author} {\bibinfo {author} {\bibfnamefont {J.-P.}\ \bibnamefont
  {Bouchaud}}, \bibinfo {author} {\bibfnamefont {L.}~\bibnamefont
  {Cugliandolo}}, \bibinfo {author} {\bibfnamefont {J.}~\bibnamefont
  {Kurchan}}, \ and\ \bibinfo {author} {\bibfnamefont {M.}~\bibnamefont
  {M{\'{e}}zard}},\ }\href {\doibase 10.1016/0378-4371(95)00423-8} {\bibfield
  {journal} {\bibinfo  {journal} {Phys. A Stat. Mech. its Appl.}\ }\textbf
  {\bibinfo {volume} {226}},\ \bibinfo {pages} {243} (\bibinfo {year}
  {1996})}\BibitemShut {NoStop}%
\bibitem [{\citenamefont {Steinigeweg}\ \emph {et~al.}(2009)\citenamefont
  {Steinigeweg}, \citenamefont {Wichterich},\ and\ \citenamefont
  {Gemmer}}]{Steinigeweg2009a}%
  \BibitemOpen
  \bibfield  {author} {\bibinfo {author} {\bibfnamefont {R.}~\bibnamefont
  {Steinigeweg}}, \bibinfo {author} {\bibfnamefont {H.}~\bibnamefont
  {Wichterich}}, \ and\ \bibinfo {author} {\bibfnamefont {J.}~\bibnamefont
  {Gemmer}},\ }\href {\doibase 10.1209/0295-5075/88/10004} {\bibfield
  {journal} {\bibinfo  {journal} {EPL}\ }\textbf {\bibinfo {volume} {88}},\
  \bibinfo {pages} {10004} (\bibinfo {year} {2009})}\BibitemShut {NoStop}%
\bibitem [{\citenamefont {Yan}\ \emph {et~al.}(2015)\citenamefont {Yan},
  \citenamefont {Jiang},\ and\ \citenamefont {Zhao}}]{Yan2015}%
  \BibitemOpen
  \bibfield  {author} {\bibinfo {author} {\bibfnamefont {Y.}~\bibnamefont
  {Yan}}, \bibinfo {author} {\bibfnamefont {F.}~\bibnamefont {Jiang}}, \ and\
  \bibinfo {author} {\bibfnamefont {H.}~\bibnamefont {Zhao}},\ }\href {\doibase
  10.1140/epjb/e2014-50797-4} {\bibfield  {journal} {\bibinfo  {journal} {Eur.
  Phys. J. B}\ }\textbf {\bibinfo {volume} {88}},\ \bibinfo {pages} {11}
  (\bibinfo {year} {2015})}\BibitemShut {NoStop}%
\bibitem [{\citenamefont {Steinigeweg}\ \emph {et~al.}(2017)\citenamefont
  {Steinigeweg}, \citenamefont {Jin}, \citenamefont {Schmidtke}, \citenamefont
  {{De Raedt}}, \citenamefont {Michielsen},\ and\ \citenamefont
  {Gemmer}}]{Steinigeweg2017}%
  \BibitemOpen
  \bibfield  {author} {\bibinfo {author} {\bibfnamefont {R.}~\bibnamefont
  {Steinigeweg}}, \bibinfo {author} {\bibfnamefont {F.}~\bibnamefont {Jin}},
  \bibinfo {author} {\bibfnamefont {D.}~\bibnamefont {Schmidtke}}, \bibinfo
  {author} {\bibfnamefont {H.}~\bibnamefont {{De Raedt}}}, \bibinfo {author}
  {\bibfnamefont {K.}~\bibnamefont {Michielsen}}, \ and\ \bibinfo {author}
  {\bibfnamefont {J.}~\bibnamefont {Gemmer}},\ }\href {\doibase
  10.1103/PhysRevB.95.035155} {\bibfield  {journal} {\bibinfo  {journal} {Phys.
  Rev. B}\ }\textbf {\bibinfo {volume} {95}},\ \bibinfo {pages} {035155}
  (\bibinfo {year} {2017})}\BibitemShut {NoStop}%
\end{thebibliography}%

\end{document}